\documentclass[a4paper, 11pt]{article}

\usepackage[]{natbib}
\usepackage{setspace}
\usepackage{graphicx}
\usepackage{amsmath, amsthm, amssymb, amsfonts, bm}
\usepackage{url}
\usepackage{bm}
\usepackage{multirow}
\usepackage{geometry}
\usepackage{enumitem}
\usepackage{setspace}
\usepackage[ruled, vlined]{algorithm2e}

\usepackage{xcolor}                    
\usepackage{hyperref}
\hypersetup{
    colorlinks=true,       
    linkcolor=magenta,      
    citecolor=cyan,        
    filecolor=red,      
    urlcolor=green       
}

\def\spacingset#1{\renewcommand{\baselinestretch}%
{#1}\small\normalsize} \spacingset{1}

\makeatletter
\def\url@leostyle{%
  \@ifundefined{selectfont}{\def\UrlFont{\sf}}{\def\UrlFont{\small\ttfamily}}}
\makeatother 

\theoremstyle{definition}
\newtheorem{thm}{Theorem}
\theoremstyle{definition}

\theoremstyle{definition}
\newtheorem{lem}{Lemma}
\theoremstyle{definition}
\newtheorem{prop}{Proposition}
\theoremstyle{definition}

\theoremstyle{remark}
\newtheorem{rem}{Remark}
\theoremstyle{definition}

\DeclareFontFamily{U}{mathx}{\hyphenchar\font45}
\DeclareFontShape{U}{mathx}{m}{n}{
      <5> <6> <7> <8> <9> <10>
      <10.95> <12> <14.4> <17.28> <20.74> <24.88>
      mathx10
      }{}
\DeclareSymbolFont{mathx}{U}{mathx}{m}{n}
\DeclareFontSubstitution{U}{mathx}{m}{n}
\DeclareMathAccent{\widecheck}{0}{mathx}{"71}

\newcommand{\E}{\mathbb{E}}
\newcommand{\pr}{\mathbb{P}}
\newcommand{\var}{\text{var}}

\newcommand{\iid}{\text{\scriptsize{i.i.d}}}
\newcommand{\R}{\mathbb{R}}
\newcommand{\Z}{\mathbb{Z}}

\newcommand{\mbf}{\mathbf}
\newcommand{\mc}{\mathcal}

\renewcommand{\l}{\left}
\renewcommand{\r}{\right}

\newcommand{\vep}{\varepsilon}
\newcommand{\heta}{\wh{\eta}}

\newcommand{\bvep}{\bm{\varepsilon}}
\newcommand{\bOm}{\bm{\Omega}}
\newcommand{\bTh}{\bm{\Theta}}
\newcommand{\bbI}{\mathbb{I}}

\newcommand{\uDel}{\underline{\Delta}}
\newcommand{\ubV}{\underline{\mathbf{V}}}
\newcommand{\ubW}{\underline{\mathbf{W}}}
\newcommand{\ubX}{\underline{\mathbf{X}}}
\newcommand{\ubZ}{\underline{\mathbf{Z}}}

\newcommand{\ubh}{\underline{\mathbf{h}}}

\newcommand{\VaR}{\text{VaR}}
\newcommand{\sVaR}{\text{sVaR}}

\def\wh{\widehat}
\def\wt{\widetilde}

\title{\Large High-dimensional GARCH process segmentation \\
with an application to Value-at-Risk}
\author{}
\author{\normalsize Haeran Cho$^1$ \hskip 1cm Karolos K. Korkas$^2$}\vskip -3cm
\date{\small{\today}}

\begin{document}
\maketitle

\footnotetext[1]{School of Mathematics, University of Bristol, UK. Email: \url{haeran.cho@bristol.ac.uk}
\\
Supported by the Engineering and Physical Sciences Research Council grant no. EP/N024435/1.}

\footnotetext[2]{School of Computing and Mathematical Sciences, University of Greenwich, UK.
Email: \url{kkorkas@yahoo.co.uk}}

\begin{abstract}
\noindent 
Models for financial risk often assume that underlying asset returns are stationary. However, there is strong evidence that multivariate financial time series entail changes not only in their within-series dependence structure, but also in the cross-sectional dependence among them. In particular, the stressed Value-at-Risk of a portfolio, a popularly adopted measure of market risk, cannot be gauged adequately unless such structural breaks are taken into account in its estimation. We propose a method for consistent detection of multiple change points in high-dimensional panel data set where both conditional variance of individual time series and their correlations are allowed to change over time. We prove its consistency in multiple change point estimation, and demonstrate its good performance through simulation studies and an application to the Value-at-Risk problem on a real dataset. Our methodology is implemented in the R package \verb+segMGarch+, available from CRAN. \par
\smallskip

 \par
\end{abstract}

\spacingset{1.45} 
\onehalfspacing

\section{Introduction}
\label{sec:intro}

The increased financial uncertainty during the recent global economic crisis 
has confirmed the close volatility linkage between asset markets. 
For example, there is strong evidence that the economy and oil prices \citep{hamilton2003}, 
foreign exchange rates \citep{Baillie1991}, equity markets \citep{Baele2003} and
crude oil and agricultural commodities \citep{du2011} are related.
These co-movements are naturally expected since the rate of information 
influences the volatility in asset returns and, therefore, 
the information flow from one asset market can be
incorporated into another related market \citep{ross1989}. 
Hence, good understanding of the correlations among multiple markets is
crucial for policy makers, financial institutions and investors.

The joint modelling of financial returns as a multivariate 
generalised autoregressive conditional heteroscedastic (GARCH) process
has attracted considerable attention in the literature.
A partial list includes the vectorised multivariate GARCH model \citep{bollerslev1988}, 
Baba-Engle-Kraft-Kroner (BEKK) model \citep{engle1995}, 
constant conditional correlation (CCC) model \citep{bollerslev1990}, 
dynamic conditional correlation (DCC) model \citep{engle2002}, 
generalised orthogonal GARCH model \citep{van2002},  
full-factor multivariate GARCH model \citep{vrontos2003} 
and conditionally uncorrelated components-based multivariate volatility processes \citep{fan2008}; 
for a survey of multivariate GARCH modelling and inference, see \cite{bauwens2006}.

The assumption that the underlying dynamics remain unchanged is restrictive 
considering that the fundamentals driving an economy, 
the asset markets in particular, exhibit sudden changes or regimes switches. 
Empirical evidence of change points (a.k.a.\ structural breaks or breakpoints)
in various macroeconomic and financial time series is well documented, 
such as those in exchange rates \citep{pesaran2007},
equity and bond returns \citep{cappiello2006},
commodity \citep{ewing2013} and share \citep{barigozzi2018} prices.
Further, \cite{diebold1999} and \cite{mikosch2004} noted that 
stochastic regime switching may be confused with long-range dependence. 
In this paper, we observe the importance of accounting for structural breaks 
in the volatilities and correlations of a multi-asset portfolio. 
In particular, we show that the stressed Value-at-Risk, 
a popular measure of market risk widely adopted by financial institutions,
under-estimates the exposure of a portfolio when 
a pre-selected period of a fixed length is used as the stress period,
compared to when the most volatile period is identified by change point analysis.


The problem of testing for at most a {\em single} structural break in univariate
conditional heteroscedastic models has been studied in 
\cite{kokoszka2000}, \cite{kokoszka2002}, \cite{lee2003}, \cite{berkes2004} and \cite{de2004},
to name a few.
For multiple change point detection,
\cite{piotr2013} proposed the BASTA (binary segmentation for transformed ARCH),
a two-stage procedure for detecting change points in the conditional variance of univariate series,
while \cite{andreou2003} studied change point detection in the co-movement of bivariate returns. 
More recent change point methods for (conditional) covariance structure of multivariate data include \cite{aue2009}, \cite{dette2018} and \cite{barassi2018} on testing for a single change point
and \cite{wang2018} on multiple change point detection. 
We also note some recent works on homogeneity pursuit for panel data with factor structures \citep{ke2016},
where the focus lies in identifying a group structure in the panel data
by casting the problem as that of change point detection.

In this paper, we propose a methodology for
multiple change point detection in multivariate, possibly high-dimensional GARCH processes.
It simultaneously segments high-dimensional GARCH processes 
by identifying `common' change points,
each of which can be shared by a subset or all of the component time series
as a change point in their within-series and/or cross-sectional dependence structure.
The methodology consists of two stages: 
The first stage transforms the $N$-dimensional time series
into the $N(N + 1)/2$-dimensional panel data consisting of 
empirical residual series and their cross-products,
whereby change points in the complex structure (conditional variance and covariance) 
are made detectable as change points in the level of the transformed data,
at the price of the increased dimensionality.
A number of methodologies have been investigated
for change point analysis in the mean of high-dimensional panel data,
such as \cite{horvath2012}, \cite{jirak2014}, \cite{cho2013} and \cite{wang2018s}.
Among many, we adopt the double CUSUM Binary Segmentation procedure 
proposed in \cite{cho2016} for the second stage, 
which achieves consistency in multiple change point estimation
while permitting within-series and cross-sectional correlations.
Extending the mixing property originally derived for univariate, 
time-varying ARCH processes in \cite{piotr2011}  
to that of time-varying bivariate GARCH processes (Proposition~\ref{prop:one}),
we establish the consistency of the combined methodology (Theorem~\ref{thm:dcbs}).

The rest of the paper is organised as follows. 
Section~\ref{sec:model} introduces a time-varying multivariate GARCH model 
which provides a framework for the theoretical treatment of our methodology. 
Section~\ref{sec:method} describes the proposed two-stage methodology
and investigates its theoretical properties, 
and simulation studies in Section~\ref{sec:sim}
confirm its good finite sample performance. 
In Section~\ref{sec:real}, we apply the methodology to a real financial data set 
and show the importance of accounting for structural breaks 
in a portfolio of assets in risk management. 
Section~\ref{sec:conc} concludes the paper,
and the Appendix contains 
the proofs of theoretical results
and additional simulation results. 
Our methodology is implemented in the R package \verb+segMGarch+, available from CRAN.

\subsubsection*{Notation}

By $\R_+$, we denote the set of positive real numbers.
For any set $\Pi \subset \{1, \ldots, N\}$, we denote its cardinality by $|\Pi|$.
For given observations $\mbf r_t \in \R^N$,
we denote by $\mc F_t$ the $\sigma$-algebra $\sigma\{\mbf r_s, \, s \le t\}$. 
Also, we use the notations $a \vee b = \max(a, b)$ and $a \wedge b = \min(a, b)$.
The notation $a_n \asymp b_n$ indicates that $a_n$ is of the order of $b_n$, 
i.e.\ $a_n = O(b_n)$ and $b_n = O(a_n)$,
and $a_n \gg b_n$ indicates that $a_n^{-1} b_n \to 0$.
We denote a vector of zeros by $\mbf 0$ whose dimension should be clear from the context.

\section{Time-varying multivariate GARCH model}
\label{sec:model}

We consider the following time-varying multivariate GARCH
(tv-MGARCH) model denoted by $\mbf r_t = (r_{1, t}, \ldots, r_{N, t})^\top$, $t=1, \ldots, T$:
\begin{align}
r_{i, t} &= \sqrt{h_{i, t}}\vep_{i, t} \quad \text{where} \quad
h_{i, t} = \omega_i(t) + \sum_{j=1}^p \alpha_{i, j}(t)r_{i, t-j}^2 + \sum_{k=1}^q \beta_{i, k}(t)h_{i, t-k}.
\label{eq:garch}
\end{align}
The independent innovations $\bvep_t = (\vep_{1, t}, \ldots, \vep_{N, t})^\top$
satisfy $\E(\bvep_t) = \mbf 0_N$ and 
$\var(\bvep_t) = \bm\Sigma_\vep(t) = [\sigma_{i, i'}(t)]_{i, i'=1}^N$ 
with $\sigma_{i, i}(t) = 1$ 
and $\sigma_{i, i'}(t) = \sigma_{i', i}(t)$ for all $1 \le i, i^\prime \le N$ and $t$. 
We denote the vector of parameters involved in modelling 
the within-panel conditional variance of $r_{i, t}$ by
\begin{align*}
\bOm_i(t) = (\omega_i(t), \alpha_{i, 1}(t), \ldots, \alpha_{i, p}(t), 
\beta_{i, 1}(t), \ldots, \beta_{i, q}(t))^\top \in \R^{1+p+q},
\end{align*}
and that involved in modelling the cross-correlations of 
$\vep_{i, t}$ with $\vep_{i^\prime, t}, \, i^\prime \ne i$, by
\begin{align*}
\bTh_i(t) = (\sigma_{i, 1}(t), \ldots, \sigma_{i, i-1}(t), 
\sigma_{i, i+1}(t), \ldots, \sigma_{i, N}(t))^\top \in \R^{N-1}.
\end{align*}
Then, the following assumption specifies how change points enter into the structure of $\mbf r_t$.
\begin{enumerate}[label=(A\arabic*), start = 1]
\setlength\itemsep{0em}
\item\label{eq:a0} 
The functions $\bOm_i(t)$ and $\bTh_i(t)$ are {\it piecewise constant} in $t$
and share $B \equiv B_T$ change points $\eta_b \equiv \eta_{b, T}, \, b = 1, \ldots, B$
(satisfying $0 \equiv \eta_0 < \eta_1 < \ldots < \eta_B < \eta_{B+1} \equiv T$)
across $i = 1, \ldots, N$, such that at any $\eta_b$, 
there exists $\Pi_b \subset \{(i, i'):\, 1 \le i \le i' \le N\}$ with $|\Pi_b| \ge 1$,
where
$(i, i') \in \Pi_b$ iff 
either $\bOm_i(\eta_b) \ne \bOm_i(\eta_b+1)$ (then $(i, i) \in \Pi_b$)
or $\sigma_{i, i'}(\eta_b) \ne \sigma_{i, i'}(\eta_b+1)$.
\end{enumerate}
Under Assumption~\ref{eq:a0}, the tv-MGARCH process $\mbf r_t$
is approximately piecewise stationary over each segment 
$[\eta_b + 1, \eta_{b + 1}], \, b = 0, \ldots, B$, with boundary effects around the change points.
Assumption~\ref{eq:a0} does not rule out that $|\Pi_b| < N(N+1)/2$,
i.e.\ it is permitted that only a {\em subset} of the components of $\bOm_i(t)$ and $\bTh_i(t)$ 
undergoes a change at each $\eta_b$.

Under model~\eqref{eq:garch}, it is assumed that
the (unconditional) correlations across the components of $\mbf r_t$ are
attributed to those of $\bvep_t$, 
and that the conditional variance of each component series is
separately modelled as a univariate GARCH process,
and the model is reduced to the CCC model of \cite{bollerslev1990} 
over each stationary segment $[\eta_b+1, \eta_{b+1}]$. 
We adopt this simplistic approach in order to specify 
the parameters ($\bOm_i(t)$ and $\bTh_i(t)$) that introduce structural changes 
and to motivate the data transformation introduced later for 
change point analysis in high-dimensional panel data,
rather than to provide a complete model for 
time-varying cross-sectional conditional correlations of the panel data;
the latter task can be accomplished 
once stationary segments are identified by the estimated change points. 
We note that the literature on multivariate GARCH processes, as those cited in Introduction,
considers a relatively lower-dimensional applications ($N \le 8$),
whereas we consider GARCH modelling of both simulated and 
real datasets of higher dimensions ($N$ up to $100$) in this paper,
which renders the simplistic approach essential for computational feasibility.


We further assume the following assumption on the model~\eqref{eq:garch}.

\begin{enumerate}[label=(A\arabic*), start = 2]
\setlength\itemsep{0em}
\item\label{eq:a1}  The dimensionality $N$ satisfies $N \asymp T^\theta$ for
some $\theta \in [0, \infty)$.

\item\label{eq:a2} For some $\epsilon_1 > 0$, $\Xi_1 < \infty$ and all
$T$, we have
\begin{align*}
\min_{1 \le i \le N}\inf_{t\in\Z}\omega_i(t) > \epsilon_1 \text{ and } 
\max_{1 \le i \le N}\sup_{t\in\Z}\omega_i(t) \le \Xi_1 < \infty.
\end{align*}

\item\label{eq:a3} For some $\epsilon_2 \in (0, 1)$ and all $T$, we have
\begin{align*}
\max_{1 \le i \le N} \sup_{t\in\Z} 
\l( \sum_{j=1}^p\alpha_{i, j}(t)+\sum_{k=1}^q\beta_{i, k}(t) \r) \le 1-\epsilon_2.
\end{align*}
\end{enumerate}
Assumption~\ref{eq:a1} indicates that the dimensionality can either be
fixed or increase with $T$ at a polynomial rate. Assumptions~\ref{eq:a2}--\ref{eq:a3} 
guarantee that between any two consecutive change points, 
each $r_{i, t}$ admits a well-defined solution a.s. and is weakly stationary
(see e.g.\ Theorem~4.35 of \cite{douc2014}).

\cite{boussama2011} shows that 
there exists a unique, strictly stationary and strongly mixing solution
for the BEKK multivariate GARCH models introduced in \cite{engle1995}, see their Theorem~2.4.
Relatively few results exist on mixing for non-stationary processes,
but such results are essential in investigating the properties of 
the change point detection methodology developed under~\eqref{eq:garch}.
Adopting the tools developed in \cite{piotr2011}, 
who investigate the mixing rate of univariate, time-varying ARCH processes,
we derive that any pair of the components of a tv-MGARCH process,
$\mbf r_{ii', t} = (r_{i, t}, r_{i', t})^\top$, $1 \le i < i' \le N$,
is strong mixing at a geometric rate under the following Lipschitz-type condition on 
the joint density of $(\vep_{i, t}^2, \vep_{i', t}^2)^\top$.

\begin{enumerate}[label=(A\arabic*), start = 5]
\setlength\itemsep{0em}
\item\label{eq:a4} The joint distribution of $\vep_{i, t}^2$ and $\vep_{i', t}^2$, 
denoted by $f_{i, i'}(u, v)$, satisfies the following: 
for any $a>0$, there exists fixed $K > 0$ independent of $a$ such that 
\begin{align*} 
\l( \int | f_{i, i'}(u, v) - f_{i, i'}(u(1+a), v) | dudv \r) \vee 
\l( \int | f_{i, i'}(u, v) - f_{i, i'}(u, v(1+a)) | dudv \r) \le Ka 
\end{align*} 
uniformly over $i, i'=1, \ldots, N, \, i \ne i'$.
\end{enumerate}

\begin{prop}
\label{prop:one} Under Assumptions~\ref{eq:a0} and~\ref{eq:a2}--\ref{eq:a4}, there exists some $\alpha \in
(0, 1)$ such that
\begin{align*}
\sup_{1 \le i < i' \le N}\sup_{\substack{G \in \sigma(\mbf r_{i, i', u}: \, u \ge t+k),
\\ H \in \sigma(\mbf r_{i, i', u}: \, u \le t)}} | \pr(G \cap H) - \pr(G)\pr(H) | \le M\alpha^k,
\end{align*}
where $M$ is a finite constant independent of $t$ and $k$.
\end{prop}
See Appendix~\ref{sec:mixing} for the proof,
which relies on the Markovian nature of the time-varying bivariate GARCH processes.
The strong mixing property of the tv-MGARCH process derived in Proposition~\ref{prop:one}
proves useful in controlling the behaviour of empirical residuals 
from the data transformation proposed in Section~\ref{sec:transform} below, 
see Proposition~\ref{prop:two}~(ii), and
this in turn allows us to rigorously establish the consistency of the proposed change point detection methodology.

\section{Two-stage change point detection methodology}
\label{sec:method}

In this section, we describe the proposed two-stage change point detection methodology
under the tv-MGARCH model in~\eqref{eq:garch}.
More specifically, Section~\ref{sec:transform} introduces 
a data transformation of the $N$-dimensional tv-MGARCH process 
into a panel of $d \equiv d_N = N(N + 1)/2$ series,
which makes any change in its conditional variance and correlation structure as defined in~\ref{eq:a0},
detectable as that in the mean of the transformed data. 
Then in Section~\ref{sec:dcbs}, we outline the application of
the double CUSUM binary segmentation algorithm \citep{cho2016} 
to the transformed panel data for the detection and localisation of the multiple change points.
Section~\ref{sec:theor} establishes the consistency of 
the combined two-stage procedure in 
estimating both the total number and the locations of the change points $\eta_b, \, b = 1, \ldots, B$,
and Section~\ref{sec:choice} discusses the selection of 
any tuning parameters involved in the methodology.

\subsection{Stage 1: Transformation of tv-MGARCH processes}
\label{sec:transform}

We first describe a data transformation for detecting change points 
in $\bOm_i(t)$ for each $i = 1, \ldots, N$,
which involves a function $g_0: \R^{1+p+q} \to \R$ that takes
$\mbf r_{i, t}^{t-p} = (r_{i, t}, \ldots, r_{i, t-p})^\top$ and
$\mbf h_{i, t-1}^{t-q} = (h_{i, t-1}, \ldots, h_{i, t-q})^\top$ as an input, and generates:
\begin{align}
U_{i, t} &= g_0(\mbf r_{i, t}^{t-p}, \mbf h_{i, t-1}^{t-q}) =
\frac{r_{i, t}}{\sqrt{\widecheck{h}_{i, t}}}, \quad \text{where}
\label{eq:gone}
\\
\widecheck{h}_{i, t} &= C_{i, 0} + \sum_{j=1}^pC_{i, j}r_{i, t-j}^2 
+ \sum_{k=1}^qC_{i, p+k} h_{i, t-k}+\epsilon r_{i, t}^2,
\nonumber
\end{align}
where $C_{i, 0} \in \R_+$, $C_{i, j} \in \R_+ \cup \{0\}, \, j = 1, \ldots, p + q$,
and $\epsilon$ denotes a small positive constant. 
For now, we ignore that the conditional variance $\mbf h_{i, t-1}^{t-q}$ is unobservable
and that the GARCH orders $p$ and $q$ are unknown; 
we address their selection and the choice of $C_{i, j}$ 
in Section~\ref{sec:choice}.

While it is possible to adopt an alternative transformation as $g_0$, 
the proposed transformation in~\eqref{eq:gone} is closely related to empirical residuals
which have been popularly adopted for 
change point analysis in time-varying and conditionally heteroscedastic univariate processes,
see \cite{kokoszka2002}, \cite{lee2003}, \cite{de2004}, \cite{piotr2013} and \cite{barassi2018}. 
For stationary processes, empirical residuals 
approximates a series of i.i.d.\ innovations and, even in the presence of change points, 
such transformation tends to reduce the autocorrelations.
Moreover, $U_{i, t}$ `encodes' the presence and the locations of 
any change point in $\bOm_i(t)$. 
More specifically,
\begin{align}
U_{i, t}^2 = \frac{h_{i, t}}{\widecheck{h}_{i, t}} \cdot \frac{r_{i, t}^2}{h_{i, t}}  
= \frac{h_{i, t}}{\widecheck{h}_{i, t}} \vep_{i, t}^2 
= \frac{h_{i, t}}{\widecheck{h}_{i, t}} + \frac{h_{i, t}}{\widecheck{h}_{i, t}}(\vep_{i, t}^2 - 1), \label{eq:motiv:u}
\end{align}
so that $U^2_{i, t}$ contains any change in $\bOm_i(t)$ as a change in its `level',
recalling that $\E(\vep_{i, t}^2) = 1$.
For notational convenience, we define
\begin{align*}
g_1(\mbf r_{i, t}^{t-p}, \mbf h_{i, t-1}^{t-q}) \equiv \l\{ g_0(\mbf r_{i, t}^{t-p}, \mbf h_{i, t-1}^{t-q}) \r\}^2 =
U_{i, t}^2 = \frac{r_{i, t}^2}{\widecheck{h}_{i, t}}.
\end{align*}

To detect any changes in the cross-sectional dependence structure of $\mbf r_t$, 
we adopt the transformation $g_2: \R^{2+2p+2q} \to \R$:
\begin{align}
\label{eq:gtwo}
U_{ii', t} = g_2(\mbf r_{i, t}^{t-p}, \mbf h_{i, t-1}^{t-q}, \mbf r_{i', t}^{t-p}, \mbf h_{i', t-1}^{t-q}) 
= (U_{i, t} + s_{i, i'} U_{i', t})^2
\end{align}
where $s_{i, i'} \in \{1, -1\}$.
Similarly to~\eqref{eq:motiv:u}, $U_{ii', t}$ admits the decomposition
\begin{align*}
U_{ii', t} 
=& \, \frac{h_{i, t}}{\widecheck{h}_{i, t}} + 
\frac{h_{i, t}}{\widecheck{h}_{i, t}}(\vep_{i, t}^2-1) + 
\frac{h_{i', t}}{\widecheck{h}_{i', t}} + \frac{h_{i', t}}{\widecheck{h}_{i', t}}(\vep_{i', t}^2-1) 
\\
&+ 2s_{i, i'} \sqrt{\frac{h_{i, t}h_{i', t}}{\widecheck{h}_{i, t}\widecheck{h}_{i', t}}}\;
\l\{ \sigma_{i, i'}(t) + (\vep_{i, t}\vep_{i', t} - \sigma_{i, i'}(t)) \r\},
\end{align*}
from which we conclude that a change in $\sigma_{i, i'}(t)$
are detectable from $U_{ii', t}$ as that in its level.

Regarding $U_{i, t}$ as empirical residuals obtained by applying volatility filters,
\cite{andreou2003} propose to examine
$U_{1, t} U_{2, t}$, $U_{1, t}^2 U_{2, t}^2$ or $|U_{1, t} U_{2, t}|$
for detecting structural changes in the co-movement of a pair of series $(r_{1, t}, r_{2, t})^\top$. 
Instead, we adopt $U_{ii', t}$, whose formulation is
motivated by the observation made in \cite{cho2013}: 
For given $(a_t, b_t)$, any changes in the second-order dependence structure,
$\E(a_t^2)$, $\E(b_t^2)$ and $\E(a_tb_t)$,
are detectable by jointly examining $\E(a_t^2)$, $\E(b_t^2)$ and 
$\E\{(a_t + s_{a, b} b_t)^2\}$ for any $s_{a, b} \in \{1, -1\}$.
Moreover, $U_{i, t}^2$, $U_{i', t}^2$ and $U_{ii^\prime, t}$ are in the same scale
in the sense that if $U_{i, t}$ and $U_{i', t}$ are Gaussian random variables,
$U_{i, t}^2$, $U_{i', t}^2$ and $U_{ii^\prime, t}$ are distributed as 
random variables following scaled $\chi^2_1$ distributions.
This enables us to regard $U_{ii', t}$ on an equal footing with $U_{i, t}^2$ and $U_{i', t}^2$, 
which is essential in {\it simultaneous} segmentation of $\mbf r_t$
via joint consideration of all the coordinate series of the panel data 
\begin{align}
\label{eq:input} 
\{U_{i, t}^2, \, 1 \le i \le N, \, U_{ii', t}, \, 1
\le i < i' \le N; \, 1 \le t \le T\}.
\end{align}

\begin{rem}[Choice of $s_{i, i'}$]
Our theoretical results do not depend on the choice of $s_{i, i'}$.
In practice, we set
$s_{i, i'} = - \text{sign}(\wh{\text{cor}}(U_{i, t}, U_{i', t}))$
where $\wh{\text{cor}}(U_{i, t}, U_{i', t})$ denotes the sample correlation between 
$U_{i, t}$ and $U_{i', t}$ over $t = 1, \ldots, T$,
in an attempt to better bring out any change in $\sigma_{i, i'}(t)$
as that in the level of $U_{ii', t}$;
a similar choice has also been considered in \cite{cho2013} and \cite{barigozzi2018}.
\end{rem}

In summary, we transform the $N$-dimensional time series $\mbf r_t$ to
the $d$-dimensional panel data with $d = N(N+1)/2$ in~\eqref{eq:input},
where any changes in $\bOm_i(t)$ and $\bTh_i(t)$, $i = 1, \ldots, N$ are
detectable as changes in the level of the coordinate time series. 
In doing so, the problem of detecting multiple change points 
in the more complex dependence structure of $\mbf r_t$,
is transformed into a relatively easier problem of detecting change points 
in the level of the panel data in~\eqref{eq:input}.
While this brings in the increase of dimensionality,
the panel data segmentation algorithm adopted in the second stage 
(detailed in Section~\ref{sec:dcbs}) handles the high dimensionality well 
under a mild assumption on the dimensionality in Assumption~\ref{eq:a1}.

\subsection{Stage 2: Double CUSUM binary segmentation}
\label{sec:dcbs}

The double CUSUM binary segmentation (DCBS) algorithm.
first introduced in \cite{cho2016},
is applicable to high-dimensional panel data
for the detection and estimation of multiple change points in its level.
The $d$-dimensional panel in~\eqref{eq:input} 
obtained from transforming the original time series $\mbf r_t$,
encodes the change points under~\eqref{eq:garch} in its level
as described in Section~\ref{sec:transform},
and thus can serve as an input to the DCBS algorithm.
For ease of notation, we introduce the DCBS algorithm
with each element time series of the panel data in~\eqref{eq:input}
denoted by 
\begin{align}
x_{j, t} = U_{ii', t} \quad \text{with} \quad 
j \equiv j(i, i') = (N - i/2)(i-1)+i', \, 1 \le i \le i' \le N,
\label{eq:xj}
\end{align}
(where $U_{ii, t} \equiv U_{i, t}^2$) such that $j \in \{1, \ldots, d\}$.

We first describe the double cumulative sum (CUSUM) statistics
computed on a generic segment $[s, e]$ for some $1 \le s < e \le T$,
and then provide a full description of the DCBS algorithm.
CUSUM statistics have been widely adopted for
change point detection in both univariate and multivariate data.
A series of (weighted) CUSUM statistics, calculated on $x_{j, t}, \, s \le t \le e$, is given by
\begin{align}
\label{eq:single_cusum}
\mc X^j_{s, c, e} =
\sqrt{\frac{(c-s+1)(e-c)}{e-s+1}}\left(\frac{1}{c-s+1}\sum_{t=s}^c
x_{j, t} - \frac{1}{e-c}\sum_{t=c+1}^e x_{j, t} \right)
\quad \text{for} \quad 
s \le c < e.
\end{align}
In the context of testing the null hypothesis of no change point 
against the at-most-one-change alternative, 
asymptotic properties of the CUSUM-based test have been extensively studied, 
see e.g.\ \cite{csorgo1997} and \cite{aue2013}.
A large value of the CUSUM statistic $|\mc X^j_{s, c, e}|$ indicates the presence of a change point
in the level of $x_{j, t}$ in the vicinity of $t = c$.
Combined with the binary segmentation algorithm, 
the CUSUM statistic has frequently been adopted for univariate time series segmentation;
see \cite{vostrikova1981} and \cite{venkatraman1992} 
for the theoretical treatment of its application to the canonical additive model with independent noise, 
and \cite{piotr2013} the case of piecewise stationary ARCH processes.


\begin{algorithm}[htbp]
\caption{{\tt DCBinSeg} (Double CUSUM Binary Segmentation algorithm)}
\label{alg:dcbs}
\DontPrintSemicolon
\SetAlgoLined
\KwIn{Panel data $\{x_{j, t}, \, 1 \le t \le T, \, 1 \le j \le d\}$, the threshold $\pi_{d, T}$, 
the start and the end of a given segment $(s, e)$, the set of change point estimators $\wh{\mc B}$}
\BlankLine
{\bf Step 1:} Compute $\mc D_{s, e}(c, m)$ for $s \le c < e$ and $1 \le m \le d$\;
\BlankLine

{\bf Step 2:} Set 
$\mc T_{s, e} \leftarrow \max_{s \le c < e}\max_{1 \le m \le d} \mc D_{s, e}(c, m)$ and
$\wh\eta \leftarrow \arg\max_{s \le c < e}\max_{1 \le m \le d} \mc D_{s, e}(c, m)$\;
\BlankLine

{\bf Step 3:} \If{$\mc T_{s, e} > \pi_{n, T}$}{
$\wh{\mc B} \leftarrow \wh{\mc B} \cup \{\wh\eta\}$\;

$\wh{\mc B} \leftarrow$ {\tt DCBinSeg}($\{x_{j, t}\}$, $\pi_{d, T}$, $s$, $\wh\eta$, $\wh{\mc B}$)\;

$\wh{\mc B} \leftarrow$ {\tt DCBinSeg}($\{x_{j, t}\}$, $\pi_{d, T}$, $\wh\eta+1$, $e$, $\wh{\mc B}$)\;
}
\BlankLine
\KwOut{$\wh{\mc B}$}
\end{algorithm}

\cite{cho2016} propose the DCBS algorithm for simultaneous segmentation of
multivariate, possibly high-dimensional time series panel data.
It guarantees consistency in estimating both the total number and locations
of multiple change points, while permitting both serial and cross-sectional correlations in $x_{j, t}$,
which is highly relevant to the time series setting considered in this paper.
Over a given segment $[s, e]$, 
we aggregate the CUSUM statistics $\mc X^j_{s, c, e}$ over $1 \le j \le d$,
and generates a two-dimensional array of double CUSUM (DC) statistics
\begin{align}
\mc D_{s, e}(c, m) = \sqrt{\frac{m(2d-m)}{2d}} 
\left(\frac{1}{m} \sum_{j=1}^m |\mc X^{(j)}_{s, c, e}| - 
\frac{1}{2d-m} \sum_{j=n+1}^d |\mc X^{(j)}_{s, c, e}|\right) \label{double:cusum:one}
\end{align}
for $s \le c  < e$ and $1 \le m \le d$, where
$|\mc X^{(j)}_{s, c, e}|$ denote the ordered CUSUM statistics at each $c$ such that 
$|\mc X^{(1)}_{s, c, e}| \ge |\mc X^{(2)}_{s, c, e}| \ge \ldots \ge |\mc X^{(d)}_{s, c, e}|$. 
Then, the test statistic is derived by maximising the two-dimensional array 
over both time and cross-sectional indices, as
\begin{align}
\mc T_{s, e} = \max_{s \le c < e} \max_{1 \le m \le d} \mc D_{s, e}(c, m), \label{eq:test:stat} 
\end{align}
which is compared against a threshold, $\pi_{d, T}$, for determining
the presence of a change point over the interval $[s, e]$. 
If $\mc T_{s, e} > \pi_{d, T}$, we regard that there exists
at least one change point in $[s, e]$ and identify the location of a change point as
\begin{align*}
\heta = \arg\max_{s \le c < e} \max_{1 \le m \le d} \mc D_{s, e}(c, m).
\end{align*}

\begin{rem}
We briefly remark upon the choice of scaling for the DC statistics in \eqref{double:cusum:one},
namely $\sqrt{m(2d - m)/(2d)}$ in place of $\sqrt{m(d - m)/d}$.
While the latter is in line with the scaling adopted for the weighted CUSUM statistic in \eqref{eq:single_cusum},
when applied to the cross-sectional dimension,
it does not favour a change point shared by more than $d/2$ rows of the panel data 
and in fact, acts as a penalty when all $d$ rows share a change point, which is counter-intuitive. 
The choice of the former scaling resolves this issue, and 
favourably regards the `density' of a change point in considering its cross-sectional magnitude.
\end{rem}

The DCBS algorithm recursively applies 
the above steps of testing and locating a single change point
over the segments determined by the previously detected change points,
until they are no longer partitioned according to a threshold $\pi_{d, T}$.
A pseudocode of the DCBS algorithm is provided in Algorithm~\ref{alg:dcbs}.
The call of {\tt DCBinSeg} with $s = 1$, $e = T$ and $\wh{\mc B} = \emptyset$
as the initial arguments returns the set of change point estimators
$\wh{\mc B} = \{\wh\eta_b:\, b = 1, \ldots, \wh{B}\}$.
We discuss the choice of $\pi_{d, T}$ in Section~\ref{sec:choice}.

\subsection{Theoretical properties}
\label{sec:theor}

Theoretical investigation into the properties of the two-stage procedure
is divided into two parts;
firstly, we show that the transformed panel data in~\eqref{eq:input}
admits a decomposition into piecewise constant signals
and the residuals meeting some desirable properties.
Then, we establish the consistency of the DCBS algorithm 
applied with such panel data as an input.

We introduce the following notations for 
presenting the first part of our results. 
For each $b = 1, \ldots, B$, let $\{\mbf r^{b}_t\}$ denote 
a stationary multivariate GARCH($p, q$) process defined 
with the parameters $\bOm_i(\eta_b+1)$ and $\bTh_i(\eta_b + 1)$, 
and the innovations coinciding with $\bvep_t$ over the associated segment $[\eta_b+1, \eta_{b+1}]$; 
we denote the corresponding conditional variance by $\{\mbf h^b_t\}$.
Then, let
$\wt{U}^{b}_{i, t} = g_0(\mbf r_{i, t}^{b, t-p}, \mbf h_{i, t-1}^{b, t-q})$ and
$\wt{U}^{b}_{ii', t} = g_2(\mbf r_{i, t}^{b, t-p}, \mbf h_{i, t-1}^{b, t-q}, 
\mbf r_{i', t}^{b, t-p}, \mbf h_{i', t-1}^{b, t-q})$,
which are constructed analogously as $U_{i, t}$ and $U_{ii', t}$
with stationary $\mbf r^b_t$ and $\mbf h^b_t$ replacing their nonstationary counterparts
in~\eqref{eq:gone} and~\eqref{eq:gtwo}, respectively.
Finally, we denote the index of the change point strictly to the left of and nearest to $t$ by 
$v(t) =\max\{0 \le b \le B: \, \eta_b < t\}$,
with which piecewise stationary processes $\mbf r^{v(t)}_t$, $\mbf h^{v(t)}_t$,
$\wt{U}^{v(t)}_{i, t}$ and $\wt{U}^{v(t)}_{ii', t}$ are defined.

\begin{prop}
\label{prop:two}
Suppose that~\ref{eq:a0}--\ref{eq:a4} hold, and 
recall the definition of $\{x_{j, t}, \, 1 \le j \le d, \, 1 \le t \le T\}$ from~\eqref{eq:xj}.
Then, the following decomposition holds:
\begin{align}
\label{eq:panel} 
x_{j, t} = f_{j, t} + z_{j, t}, \quad 1 \le j \le d, \quad 1 \le t \le T.
\end{align}
\begin{enumerate}[label = (\roman*)]
\item Setting $f_{j, t} = \wt{g}_{ii', t}$ (with $\wt{g}_{ii, t} \equiv \wt{g}_{i, t}$)
where $\wt{g}_{i, t} = \E\{(\wt{U}^{v(t)}_{i, t})^2\}$ and $\wt{g}_{ii', t} = \E(\wt{U}^{v(t)}_{ii', t})$,
we have $f_{j, t}$ piecewise constant with all change points belong to $\mc B = \{\eta_1, \ldots, \eta_B\}$;
conversely, at each $\eta_b \in \mc B$, there exists a non-empty set
$\wt\Pi_b \subset \{1, \ldots, d\}$ where
\begin{align}
\label{eq:f}
\wt\Pi_b = \Big\{j: \, \Delta_{j, b} := |f_{j, \eta_b+1} - f_{j, \eta_b}| \ne 0\Big\}.
\end{align}

\item The residuals $z_{j, t} = U_{ii', t} - \E(\wt{U}^{v(t)}_{ii', t})$ satisfy
\begin{align*}
\max_{1 \le j \le d} \max_{1 \le s < e \le T}
\frac{1}{\sqrt{e-s+1}}\left\vert \sum_{t=s}^e z_{j, t} \right\vert
 = O_p(\sqrt{\log(T)}).
\end{align*}
\end{enumerate}
\end{prop}

The proof of Proposition~\ref{prop:two} can be found in Appendix~\ref{pf:lem:one}.
Unlike $\E(U_{i, t}^2)$ or $\E(U_{ii', t})$, 
we have $\wt{g}_{i, t}$ and $\wt{g}_{ii', t}$
{\em exactly} piecewise constant without any boundary effects.
By its construction, $z_{j, t}$ do not necessarily satisfy $\E(z_{j, t}) = 0$.
However, due to the mixing properties of the tv-MGARCH processes (Proposition~\ref{prop:one}), 
scaled partial sums of $z_{j, t}$ can be appropriately bounded.


In order to establish the theoretical consistency of the DCBS algorithm, 
we impose the following conditions on the detectability of each change point $\eta_b$.

\begin{enumerate}[label = (B\arabic*), start = 1]
\setlength\itemsep{0em}
\item\label{eq:b2} There exists a fixed constant $c > 0$ such that
$\min_{0 \le b \le B} (\eta_{b+1} - \eta_b) \ge cT^\gamma$ for
some $\gamma \in (6/7, 1]$ (recalling that $\eta_0=0$ and $\eta_{B+1} = T$).
\item\label{eq:b3} The number of change points, $B \equiv B_T$, satisfies $B = o(\log(T))$.
\item\label{eq:b4} Recall the definitions of $\wt\Pi_b$ and $\Delta_{j, b}$ given in~\eqref{eq:f}. Then,
\begin{align*}
\frac{T^{7\gamma/4-3/2} \uDel_{d, T}}{\sqrt{d \, \log(T)}} \to \infty \text{ as } T \to \infty,
\quad \text{where} \quad
\uDel_{d, T} := \min_{1 \le b \le B} |\wt\Pi_b|^{-1/2} \sum_{j \in \wt\Pi_b} \Delta_{j, b}.
\end{align*}
\end{enumerate}

In conjunction with \ref{eq:b2}, Assumption~\ref{eq:b3} imposes 
a bound on the total number of change points $B$, 
which is permitted to grow slowly with $T$. 
Assumption~\ref{eq:b4} specifies the minimum requirement on the {\em cross-sectional size of the change},
quantified by $\uDel_{d, T}$, for all change points to be detected as well as being located with accuracy. 
The quantity $\uDel_{d, T}$ combines both the cross-sectional `density' of each change point, $|\wt\Pi_b|$, 
and the magnitude of jumps, $\sum_{j \in \wt\Pi_b} \Delta_{j, b}$, over all $b = 1, \ldots, B$;
for example, if $\Delta_{j, b} \equiv \Delta_b$ for all $j \in \wt\Pi_b$,
we have $|\wt\Pi_b|^{-1/2} \sum_{j \in \wt\Pi_b} \Delta_{j, b} = |\wt\Pi_b|^{1/2} \Delta_b$,
which increases with both $|\wt\Pi_b|$ and $\Delta_b$.
Typically, change point detection becomes more challenging 
as the distance between two adjacent change points decreases (with decreasing $\gamma$),
and also as the dimensionality $d$ increases relative to the density of the change point $\vert \wt\Pi_b \vert$,
which is reflected in~\ref{eq:b4}.
We highlight that our methodology does not require each change point to be 
common to all the cross-sections of $\{x_{j, t}\}$
(and, consequently, in all $\bOm_i(t)$ and $\bTh_i(t)$) provided that Assumption~\ref{eq:b4} is met.
In other words, a relatively sparse change point $\eta_b$ (due to small $d^{-1}|\wt\Pi_b|$)
is detectable with accuracy as long as $\Delta_{j, b}$ are sufficiently large.

It is not trivial to relate $\Delta_{j, b}$, the magnitude of a jump in
$f_{j, t} = \wt{g}_{ii', t} = \E\{(U^{v(t)}_{ii'})^2\}$ 
(with $j = j(i, i')$ as defined in~\eqref{eq:xj})
and the changes in $\bOm_i(t)$ or $\bTh_i(t)$ due to the presence of the complex transformation.
However, the density or sparsity of a change point, 
measured by $|\Pi_b|$ (defined in Assumption~\ref{eq:a0} ),
is preserved by $|\wt\Pi_b|$ and, in fact, $(i, i') \in \Pi_b$ iff $j = j(i, i') \in \wt\Pi_b$.
For further discussion on the high-dimensional efficiency of DC test statistic, 
we refer to Remark~3.1 of \cite{cho2016}.

\begin{thm}
\label{thm:dcbs} 
Suppose that Assumptions~\ref{eq:a0}--\ref{eq:a4} and \ref{eq:b2}--\ref{eq:b4} hold.
Let $\heta_b, \ b=1, \ldots, \wh{B}$ 
(with $1 < \heta_1 < \ldots < \heta_{\wh{B}} < T$),
denote the change point estimators returned by the DCBS algorithm 
with a threshold $\pi_{d, T}$ satisfying
$C'd\uDel_{d, T}^{-1} T^{5(1-\gamma)/2}\sqrt{\log(T)} < \pi_{d, T} < C''\uDel_{d, T} T^{\gamma-1/2}$ 
for some constants $C', C'' > 0$.
Then, there exists $c_0 > 0$ such that
\begin{align*}
\pr\left\{\wh{B}=B; \, |\heta_b - \eta_b| < c_0\rho_{d, T} 
\text{ \ for \ } b=1, \ldots, B\right\} \to 1
\end{align*}
as $T \to \infty$, where $\rho_{d, T} = d\uDel_{d, T}^{-2}T^{5(1-\gamma)}\log(T)$.
\end{thm}

For the proof of Theorem~\ref{thm:dcbs}, see Appendix~\ref{sec:pf:one}.
From the condition imposed on the rate of $\uDel_{d, T}$ in~\ref{eq:b4}, 
it is easily seen that $\rho_{d, T} /T^\gamma \to 0$ as $T \to \infty$. 
That is, in the re-scaled time interval $[0, 1]$,
the change point estimators satisfy 
$T^{-1}|\heta_b-\eta_b| \le T^{-\gamma}|\heta_b-\eta_b| \to 0$ for all $b = 1, \ldots, B$.
Defining the optimality in change point detection as when 
each of the true change points and the corresponding estimated
change point are within the distance of $O_p(1)$ (see e.g.\ \cite{korostelev1987}), 
it is attained up to a logarithmic factor
when the change points are maximally spread ($\gamma = 1$),
and the jumps are dense ($|\wt\Pi_b| \asymp d$) and of large magnitude 
($\sum_{j \in \wt\Pi_b} \Delta_{j, b} \asymp d$). 

\subsection{Selection of tuning parameters}
\label{sec:choice}

\subsubsection{Choice of parameters for transformation}
\label{sec:choice:one}

Empirical performance of the two-stage methodology, its power in particular, 
is influenced by the choice of the transformation function $g_0$
determined by the coefficients $C_{i, j}, \, j = 0, \ldots, p+q$, in~\eqref{eq:gone}.
In most references given at the beginning of Section~\ref{sec:transform}, 
$C_{i, j}$ are set as the maximum likelihood estimates (MLEs) of
the GARCH parameters obtained from the whole sample assuming the stationarity, 
say $\wh\omega_i, \wh\alpha_{i, j}, \, 1 \le j \le p$ and $\wh\beta_{i, k}, \, 1 \le k \le q$.

The BASTA--res algorithm proposed by \cite{piotr2013} 
performs change point detection in the univariate GARCH process
by analysing the transformation of the input time series
obtained similarly to $U_{i, t}^2$. 
They recommend the use of `dampened' versions of the GARCH parameter estimates.
In our setting, this leads to the choice of $C_{i, 0} = \wh\omega_i$,
$C_{i, j} = \wh\alpha_{i, j}/F_i, \, 1 \le j \le p$, and $C_{i, p+k} = \wh\beta_{i, k}/F_i, \, 1 \le k \le q$, 
with within-series dampening parameters $F_i \ge 1$. 

Empirically, the motivation behind the introduction of $F_i$ is as follows. 
For $r_{i, t}$ with time-varying parameters, we often observe that 
the GARCH parameters are over-estimated such that
$\sum_{j = 1}^p \wh\alpha_{i, j} + \sum_{k = 1}^q \wh\beta_{i, k}$ is close to, or even exceeds, one. 
Therefore, using the raw estimates in place of $C_{i, j}$'s in~\eqref{eq:gone} 
leads to change points not well-detectable in the resultant transformed panel data.
Thus we adopt the dampening parameter $F_i$ and select it as
\begin{align*}
F_i = \max\left[1, \frac{\min(0.99, \sum_{j = 1}^p \wh\alpha_{i, j} + \sum_{k = 1}^q \wh\beta_{i, k})} 
{\max\big\{0.01, 1 - (\sum_{j = 1}^p \wh\alpha_{i, j} + \sum_{k = 1}^q \wh\beta_{i, k})\big\}}\right].
\end{align*}
By construction, $F_i$ is bounded as $F_i \in [1, 99]$ and 
approximately brings $\wh\omega_i$ and $\sum_{j=1}^p\wh\alpha_{i, j} + \sum_{k=1}^q\wh\beta_{i, k}$ 
to the same scale.

The transformation $g_0$ also involves the unobservable conditional variance $h_{i, t}$, 
which we propose to replace with the empirical estimates
\begin{align}
\label{eq:cond:var} \wh h_{i, t} = \wh\omega_i + 
\sum_{j=1}^p \wh\alpha_{i, j} r_{i, t-j}^2 + \sum_{k=1}^q \wh\beta_{i, k}\wh h_{i, t-k}
\end{align}
obtained with the MLEs of the GARCH parameters.

Typically, the GARCH orders $p$ and $q$ are unknown and may even vary over time.
We propose to use $(p, q) = (1, 1)$.
The GARCH($1$, $1$) model is simple yet known to provide a good fit 
to a wide range of datasets (see e.g.\ \cite{hansen2005}). 
Besides, the model~\eqref{eq:garch} is adopted for the purpose of change point analysis
rather than for describing the time series themselves.
In simulation studies reported in Section~\ref{sec:sim}, 
we study the effect of mis-specifying the GARCH orders on change point analysis, 
which shows that the choice of $(p, q) = (1, 1)$ works reasonably well 
even when it under-specifies the true GARCH orders.


\subsubsection{Choice of threshold for DCBS algorithm}
\label{sec:choice:thr}

Theorem~\ref{thm:dcbs} provides a range of the threshold $\pi_{d, T}$ 
that guarantees the consistency of the proposed methodology.
However, the theoretical range involves typically unattainable
knowledge on the quantities such as $\gamma$ or $\uDel_{d, T}$.
Moreover, even when such knowledge is available, finite sample performance
may be affected by the choice of the multiplicative constant to a given rate. 
Instead, we propose a parametric resampling procedure,
which enables us to approximate the distribution of the DC test statistic 
in the absence of any change point.  
A similar approach has been widely adopted in the change point literature 
including \cite{kokoszka2002} in the context of testing the presence of a change point 
in univariate GARCH processes. 

\begin{algorithm}[htb]
\caption{Bootstrap algorithm for threshold selection}
\label{alg:boot}
\DontPrintSemicolon
\SetAlgoLined
\KwIn{Empirical residuals $\{\wh\vep_{i, t}, \, 1 \le i \le N, \, 1 \le t \le T\}$, 
GARCH parameter estimators $\wh\omega_i$, $\wh\alpha_{i, j}, \, 1 \le j \le p$,
and $\wh\beta_{i, k}, \, 1 \le k \le q$,
start and the end of a given segment $(s, e)$, 
bootstrap sample size $R$,  
the level of significance $\alpha \in [0, 1]$}
\BlankLine

{\bf Step 1:} Compute the empirical residuals $\wh\vep_{i, t} \leftarrow \wh{h}_{i, t}^{-1/2} r_{i, t}$\;

{\bf Step 2:} \For{$\ell = 1, \ldots, R$}{
{\bf Step 2.1:}
Generate bootstrap samples $\{\bm\vep^\ell_t\}_{t = 1}^T$
of $\{\wh{\bm\vep}_t = (\wh\vep_{1, t}, \ldots, \wh\vep_{N, t})^\top\}_{t = 1}^T$\;

{\bf Step 2.2:} Simulate an MGARCH process
\begin{align*}
r^\ell_{i, t} = (h^\ell_{i, t})^{1/2}\vep^\ell_{i, t}, \quad
\text{where} \quad h^\ell_{i, t} = \wh{\omega}_i +
\sum_{j=1}^p\wh{\alpha}_{i, j}(r^\ell_{i, t-j})^2 + \sum_{k=1}^q \wh{\beta}_{i, k} h^\ell_{i, t-k}
\end{align*}

{\bf Step 2.3:} Generate $\{x^\ell_{j, t}\}$ as 
\begin{align*}
\Big\{g_1(\mbf r^{\ell, t - p}_{i, t}, \mbf h^{\ell, t-q}_{i, t-1}), \, 1 \le i \le N, \, 
g_2(\mbf r^{\ell, t - p}_{i, t}, \mbf h^{\ell, t-q}_{i, t-1}, \mbf r^{\ell, t - p}_{i', t}, \mbf h^{\ell, t-q}_{i', t-1}), 
\, 1 \le i < i' \le N; \, 1 \le t \le T \Big\}
\end{align*}

{\bf Step 2.3:} Calculate $\mc T^\ell_{s, e}$ from $\{x^\ell_{j, t}\}$ as in \eqref{eq:test:stat}\;
}

{\bf Step 3:} Select $\pi^{(s, e)}_{d, T}$ as the $100(1-\alpha)$\%-percentile 
of $\mc T^\ell_{s, e}, \ \ell=1, \ldots, R$\;
\BlankLine

\KwOut{$\pi^{(s, e)}_{d, T}$}
\end{algorithm}

Algorithm~\ref{alg:boot} outlines the proposed resampling scheme,
where we derive the segment-dependent threshold $\pi^{(s, e)}_{d, T}$
for each segment $[s, e]$ considered at some iteration of the DCBS algorithm (see Algorithm~\ref{alg:dcbs}).
It takes the empirical residuals $\wh\vep_{i, t} = \wh{h}_{i, t}^{-1/2} r_{i, t}$
where the conditional variance is estimated as in~\eqref{eq:cond:var}
with the MLE of the GARCH parameters $\wh\omega_i$, $\wh\alpha_{i, j}$ and $\wh\beta_{i, k}$.
While theoretical investigation into the validity of the bootstrap procedure
is beyond the scope of this paper, 
we verify its good performance on simulated datasets in Section~\ref{sec:sim}.
In all our numerical studies, we set $\alpha = 0.05$ and $R = 100$.

\section{Simulation study}
\label{sec:sim}

\subsection{Models}
\label{sec:sim:model}

We investigate the numerical performance of the two-stage data segmentation methodology
proposed in Section~\ref{sec:method}
on the datasets simulated from the models below.
We are not aware of another change point detection methodology
applicable to the segmentation of high-dimensional GARCH processes,
which makes a comparative numerical study challenging.
Instead, we adopt the simulation models from the literature on 
change point analysis in univariate GARCH processes;
Model \ref{m:0} is motivated by the simulation models
considered in \cite{piotr2013}, 
Models~\ref{m:2}--\ref{m:3} by those in \cite{kokoszka2002}.
We note that under~\ref{m:0}, there does not exist any change point, i.e.\ $B = 0$.

\begin{enumerate}[label = (M\arabic*), start = 0]
\setlength\itemsep{0em} 
\item \label{m:0} \textbf{Stationary MGARCH $(1, 1)$ processes.} Let
$\omega_i = \omega + \delta_{\omega, i}$, 
$\alpha_{i, 1} = \alpha + \delta_{\alpha, i}$ and 
$\beta_{i, 1} =  \beta  + \delta_{\beta, i}$, 
where $\delta_{\cdot, 1} \sim_{\iid} \mc U(-\Delta, \Delta)$ for some small $\Delta > 0$ 
is added to each GARCH
parameter so that every $r_{i, t}$ has a slightly different set of
GARCH parameters. The innovations are generated from two different
distributions, namely (i) $\bvep_t \sim_{\iid} \mc N(\mbf 0, \bm\Sigma_\vep)$ 
where $\sigma_{i, i'} =  \varrho^{|i-i'|}$ with $\varrho = -0.75$ and 
(ii) $\vep_{i, t} \sim_{\iid} t_{10}$ for each $i$ and $t$.  
We consider $T = 1000$ and $N \in \{50,100\}$.
\begin{enumerate}[label = (M0.\arabic*), start = 1]
\setlength\itemsep{0em} 
\item \label{m:0:1} $(\omega, \alpha_1, \beta_1) = (0.4, 0.1, 0.5)$.
\item \label{m:0:2} $(\omega, \alpha_1, \beta_1) = (0.1, 0.1, 0.8)$.
\end{enumerate}

\item \label{m:2} \textbf{tv-MGARCH $(1, 1)$ processes with two change points.} 
We introduce the first change point $\eta_1 = [T/4]$ to the GARCH parameters
such that for a randomly chosen $\mathcal{S}_1 \subset \{1, \ldots, N\}$,
GARCH parameters $\omega_i(t), \alpha_{i, 1}(t)$ and $\beta_{i, 1}(t)$ for 
$i \in \mathcal{S}_1$ change at $t = \eta_1$ as
$\omega_i(t) = \omega^{(1)}\bbI(t \le \eta_1)+ \omega^{(2)}\bbI(t > \eta_1) + \delta_{\omega, i}$, 
$\alpha_{i, 1}(t) = \alpha^{(1)}_1\bbI(t \le \eta_1)+ \alpha^{(2)}_1\bbI(t > \eta_1) + \delta_{\alpha, i}$ 
and $\beta_{i, 1}(t) = \beta^{(1)}_1\bbI(t \le \eta_1)+ \beta^{(2)}_1\bbI(t > \eta_1)  + \delta_{\beta, i}$,
where $\delta_{\cdot, i}$ is as in~\ref{m:0},
and the parameters before and after the change point are chosen as below.
\begin{enumerate}[label = (M1.\arabic*), start = 1]
\setlength\itemsep{0em} 
\item \label{m:2:1} $(\omega, \alpha_1, \beta_1):$ $(0.1, 0.3, 0.3)$ $\to$ $(0.15, 0.25, 0.65)$. 
\item \label{m:2:2} $(\omega, \alpha_1, \beta_1):$ $(0.1, 0.3, 0.3)$ $\to$ $(0.125, 0.1, 0.6)$.
\item \label{m:2:3} $(\omega, \alpha_1, \beta_1):$ $(0.1, 0.3, 0.3)$
$\to$ $(0.15, 0.15, 0.25)$.
\end{enumerate}
The second change point $\eta_2 = [3T/5]$ is introduced to the cross-correlation structure.
Initially, $\bm\Sigma_\vep(t) = \bm\Sigma_\vep$ defined in~\ref{m:0} for $t \le \eta_2$. 
Then, for a randomly chosen $\mathcal{S}_2 \subset \{1, \ldots, N\}$, 
the rows of $\bm\Sigma_\vep(t)$ corresponding to $\vep_{i, t}, \, i \in \mc S_2$ 
swap their locations arbitrarily.
For the generation of $\bm\vep_t = (\bm\Sigma_\vep(t))^{1/2} \mbf v_t$, we consider two settings:
(i) $v_{i, t} \sim_{\iid} \mc N(0, 1)$ for all $i$ and $t$, and 
(ii) $v_{i, t} \sim_{\iid} t_{10}$ for all $i$ and $t$.
We set $|\mathcal{S}_1| = |\mathcal{S}_2| = [\varrho N]$ with 
$\varrho \in \{1, 0.75, 0.5, 0.25\}$ controlling the `sparsity' of the change points,
and consider $N \in \{50, 100\}$ and the sample size $T = 500$ for (i) and $T = 2500$ for (ii); 
the large sample size under~(ii) is to ensure the stability of the quasi-MLE
of the GARCH coefficients for the transformation.

\item \label{m:3} \textbf{Mis-specification of the orders $p$ and $q$.}
\begin{enumerate}[label = (M2.\arabic*), start = 1]
\setlength\itemsep{0em} 
\item \label{m:3:1}  {\bf Over-specification.} The two change points are
introduced to tv-MGARCH$(1, 1)$ processes with Gaussian innovations as in \ref{m:2:1}--\ref{m:2:2}
(referred to as (M2.1.1)--(M2.1.2)), but the GARCH orders are
mis-specified in the transformation $g_0$ (see~\eqref{eq:gone}) as $(p, q) = (2,  2)$. 
\item \label{m:3:2} {\bf Under-specification.} The two change points are introduced to
tv-MGARCH$(2, 2)$ processes with Gaussian innovations
as in \ref{m:2}, with the GARCH parameters change at $\eta_1$ as:
\begin{enumerate}[label = (M2.2.\arabic*), start = 1]
\setlength\itemsep{0em} 
\item \label{m:3:2:1} $(\omega, \alpha_1, \alpha_2, \beta_1, \beta_2)$: 
$(0.1, 0.1, 0.2, 0.1, 0.2)$ $\to$ $(0.15, 0.15, 0.1, 0.35, 0.3)$;
or 
\item \label{m:3:2:2} $(\omega, \alpha_1, \alpha_2, \beta_1, \beta_2)$: 
$(0.1, 0.1, 0.2, 0.1, 0.2)$ $\to$ $(0.125, 0.1, 0, 0.3, 0.3)$.
\end{enumerate}
The covariance matrix of the innovations change at $\eta_2$ as in \ref{m:2}. 
The GARCH orders are mis-specified in the transformation $g_0$ as $(p, q) = (1, 1)$.
\end{enumerate}

\item \label{m:4} \textbf{Full-factor multivariate GARCH($1$, $1$) model with time-varying factors and loadings.} 
Proposed in \cite{vrontos2003}, each $r_{i, t}$ is generated as a linear
combination of the independent factors $f_{j, t}$, $j = 1, \ldots, N$ which are GARCH$(1, 1)$ processes.
\begin{align*}
\mbf r_t &= \mathbf{W} \mathbf{f}_t, \quad \mathbf{f}_t|\mc F_{t-1} \sim_{\iid}
\mc N_N(\mbf 0, \mathbf{H}_t), \quad \mathbf{H}_t =
\text{diag}(h_{1, t}, \ldots, h_{N, t}),
\\
\text{where } h_{i, t} &= \omega_i + \alpha_{i, 1} f_{i, t-1}^2 +
\beta_{i, 1} h_{i, t-1}, \ t = 1, \ldots, T; \, i = 1, \ldots, N,
\end{align*}
with $w_{i, i'} \sim_{\iid} \mc N(1, 1)$ for the loading matrix $\mathbf{W}$. 
For a randomly chosen $\mathcal{S}_1 \subset \{1, \ldots, N\}$, 
GARCH parameters of $\mathbf{f}$ change at $\eta_1 = [T/4]$ as
\begin{enumerate}[label = (M3.\arabic*), start = 1]
\setlength\itemsep{0em} 
\item \label{m:4:1} $(\omega, \alpha_1, \beta_1):$ $(0.1, 0.3, 0.3)$ $\to$ $(0.15, 0.25, 0.65)$. 
\item \label{m:4:2} $(\omega, \alpha_1, \beta_1):$ $(0.1, 0.3, 0.3)$ $\to$ $(0.125, 0.1, 0.6)$.
\end{enumerate}
Another change point is introduced to the loading matrix at $\eta_2 = [3T/5]$, 
by swapping the rows of $\mathbf{W}$ corresponding to a randomly chosen 
$\mathcal{S}_2 \subset \{1, \ldots, N\}$, 
which brings in a change in the conditional cross-correlations as well as within-series conditional variance.
The cardinality of $\mathcal{S}_1$ and $\mathcal{S}_2$ is
controlled as in \ref{m:2} with $\varrho \in \{1, 0.75, 0.5, 0.25\}$,
and we consider $T = 500$ and $N \in \{50, 100\}$.
\end{enumerate}

In all follows, we set $p = q = 1$ as suggested in Section~\ref{sec:choice:one} for
the transformation $g_0$ unless specifically chosen otherwise (see \ref{m:3:2}),
and set $\alpha = 0.05$ and $R = 100$ for threshold selection
as described in Section~\ref{sec:choice:thr}.
All simulation results reported are based on $100$ realisations.

\subsection{Results}
\label{sec:results}

Firstly, we perform at-most-one-change test on the data simulated from \ref{m:0}
by conducting only a single iteration of the DCBS algorithm in Algorithm~\ref{alg:dcbs}, 
which enables us to investigate its size behaviour, see Tables~\ref{table:sim:zero}.
In Appendix~\ref{sec:sim:add}, we investigate the power and localisation accuracy
of such a test in various single change point scenarios motivated by \cite{piotr2013}.

Overall, we observe that the DC-based test performs well in size control.
When the innovations are generated from a Gaussian distribution
with cross-correlations, the test manages to keep the size below
the significance level $\alpha = 0.05$ when $N = 50$, while
more spurious false alarms are observed as the dimensionality grows for~\ref{m:0:2}. 
We note that, although not directly comparable, 
Table~1 of \cite{piotr2013} observed similar size behaviour from their
procedure as well as the change point test from \cite{andreou2002} 
applied to univariate GARCH processes generated with the same GARCH parameters as in~\ref{m:0:2}.  
When the innovations are drawn from a $t_{10}$-distribution,
the parameter configuration of~\ref{m:0:2} brings in greater size distortion.

\begin{table}[htbp]
\caption{\ref{m:0} Size of the change point test at $\alpha = 0.05$ when $T = 1000$.} 
\label{table:sim:zero} \centering
\begin{tabular}{c|cc|cc}
\hline
&   \multicolumn{2}{c}{Gaussian $\vep_{i, t}$} &        
\multicolumn{2}{c}{$t_{10}$-distributed $\vep_{i, t}$}      \\
$N$ &   \ref{m:0:1} &    \ref{m:0:2} &    \ref{m:0:1} &    \ref{m:0:2}  \\ 
\hline\hline
$50$ &  0.01 &  0.05 &  0.03 &  0.13    \\
$100$ & 0.02 &  0.09 &  0.02 &  0.3 \\  \hline
\end{tabular}
\end{table}

Table~\ref{table:sim:two:one}--\ref{table:sim:two:two} report the
results from applying the proposed methodology to 
multiple change point detection from tv-MGARCH processes generated as described in~\ref{m:2},
with Gaussian and heavy-tailed $\bm\vep_t$;
Figures~\ref{fig:sim:two:50}--\ref{fig:sim:two:100}
in Appendix illustrate
the locations of the estimated change points. 
For the change point $\eta_1$ attributed solely to the change in GARCH parameters,
when it results in time-varying unconditional variance $\var(r_{i, t}), \, i \in \mc S_1$,
as in \ref{m:2:1}--\ref{m:2:2} (with \ref{m:2:1} bringing a larger jump in the variance),
the change point is more easily detected than 
when the unconditional variance is kept approximately constant as in~\ref{m:2:3}.
Between the two types of change points considered in~\ref{m:2}, 
the detection of $\eta_1$ is more challenging as it becomes sparser and $N$ increases,
compared to $\eta_2$ where the cross-sectional correlations undergo a change.
This is explained by the fact that, the sparsity of $\eta_1$,
measured by the number of series containing $\eta_1$ scaled by $d$ ($\approx N^2$),
is in the order of $\varrho N^{-1}$ whereas that of $\eta_2$ is in the order of $\varrho^2$.
With heavier-tailed innovations, 
both the detection and the localisation accuracy deteriorates,
possibly as the quasi-MLE procedure for the first-stage transformation 
suffers due to the heavy-tailedness
and in particular, the detection of $\eta_1$ is the most affected between the two.

\begin{table}[htbp]
\caption{\ref{m:2}~(i) The number of estimated change points (\%) and 
the accuracy in change point location  (\% of $|\heta_1 - \eta_1| < \log^2T$) 
when $\alpha = 0.05$, $T = 500$,
$N = 50$ (left) and $N = 100$ (right) with Gaussian innovations.} 
\label{table:sim:two:one}
\centering \small{\begin{tabular}{cc|ccccc|cc||ccccc|cc}
\hline
&   &   \multicolumn{5}{c}{$\wh{B}$} &                  \multicolumn{2}{c}{accuracy (\%)} &     \multicolumn{5}{c}{$\wh{B}$} &                  \multicolumn{2}{c}{accuracy (\%)}       \\
&   $\varrho$ & 0 & 1 & 2 & 3 & $\ge 4$ &   $\eta_1$ &  $\eta_2$ &
0 & 1 & 2 & 3 & $\ge 4$ &   $\eta_1$ &  $\eta_2$    \\ 
\hline\hline
\multirow{4}{*}{\ref{m:2:1}} &   1 & 0  & 0  & 100  &   0  & 0  & 100  &   100  &   0  & 0  & 100  &   0  & 0 & 90 &   99 \\
&   0.75 &  0 & 0  & 100  &   0  & 0 & 100  &   100  &   0  & 0  & 100  &   0  & 0  & 100  &   100  \\
&   0.5 &   0  & 0  & 100  & 0  & 0  & 100  &   100  &   0  & 0   & 100  &   0  & 0   & 97 &   99 \\
&   0.25 &  0 & 7 & 92 &    1 & 0 & 100 &   93 &    0 & 91 &    9
& 0 & 0 & 100 &   6   \\  
\hline
\multirow{4}{*}{\ref{m:2:2}} &   1 & 0 & 0 & 100 &   0 & 0 & 100 &   100 &   0 & 0 & 100 &   0 & 0 & 100 &   100 \\
&   0.75 &  0 & 2 & 98 &    0 & 0 & 96 &    100 &   0 & 0 & 100 &   0 & 0 & 100 &   100 \\
&   0.5 &   0 & 38 &    62 &    0 & 0 & 60 &    100 &   0 & 2 & 98 &    0 & 0 & 98 &    100 \\
&   0.25 &  0 & 95 &    5 & 0 & 0 & 2 & 100 &   0 & 80 &    20 & 0
& 0 & 32 &    84  \\  
\hline
\multirow{4}{*}{\ref{m:2:3}} &   1 & 0 & 89 &    11 &    0 & 0 & 3 & 100 &   0 & 68 &    31 &    1 & 0 & 13 &    100 \\
&   0.75 &  0 & 89 &    11 &    0 & 0 & 3 & 100 &   0 & 85 &    15 &    0 & 0 & 3 & 100 \\
&   0.5 &   0 & 94 &    6 & 0 & 0 & 0 & 100 &   0 & 88 &    12 &    0 & 0 & 3 & 100 \\
&   0.25 &  0 & 97 &    3 & 0 & 0 & 0 & 100 &   1 & 96 &    3 & 0
& 0 & 0 & 98  \\  
\hline
\end{tabular}}
\end{table}

\begin{table}[htbp]
\caption{\ref{m:2}~(ii) The number of estimated change points (\%) and 
the accuracy in change point location when $\alpha = 0.05$, $T = 500$,
$N = 50$ (left) and $N = 100$ (right) with $t_{10}$ innovations.} 
\label{table:sim:two:two}
\centering \small{\begin{tabular}{cc|ccccc|cc||ccccc|cc}
\hline
&   &   \multicolumn{5}{c}{$\wh{B}$} &                  \multicolumn{2}{c}{accuracy (\%)} &     \multicolumn{5}{c}{$\wh{B}$} &                  \multicolumn{2}{c}{accuracy (\%)}       \\
&   $\varrho$ & 0 & 1 & 2 & 3 & $\ge 4$ &   $\eta_1$ &  $\eta_2$ &
0 & 1 & 2 & 3 & $\ge 4$ &   $\eta_1$ &  $\eta_2$    \\ 
\hline\hline
\multirow{4}{*}{\ref{m:2:1}} & 1 & 0 & 5 & 74 & 12 & 8 & 92 & 100 & 0 & 7 & 72 & 11 & 10 & 90 & 100 \\
& 0.75 & 0 & 2 & 86 & 8 & 4 & 98 & 100 & 1 & 1 & 67 & 20 & 11 & 96 & 99\\
& 0.5 & 0 & 2 & 78 & 14 & 6 & 98 & 100 & 3 & 1 & 72 & 12 & 12 & 97 & 100 \\
& 0.25 & 0 & 28 & 58 & 6 & 8 & 100 & 70 & 1 & 47 & 39 & 8 & 5 & 99 & 40 \\  
\hline
\multirow{4}{*}{\ref{m:2:2}}& 1 & 0 & 31 & 67 & 2 & 0 & 67 & 100 & 0 & 0 & 100 & 0 & 0 & 90 & 99 \\
& 0.75 & 0 & 11 & 86 & 3 & 0 & 86 & 100 & 0 & 0 & 100 & 0 & 0 & 100 & 100 \\
& 0.5 & 0 & 5 & 93 & 2 & 0 & 100 & 100 & 0 & 5 & 93 & 2 & 0 & 94 & 100 \\
& 0.25 & 0 & 3 & 94 & 2 & 0 & 94 & 99 & 0 & 91 & 9 & 0 & 0 & 100 & 6   \\   
\hline
\multirow{4}{*}{\ref{m:2:3}} & 0 & 0 & 99 & 1 & 0 & 0 & 0 & 100 & 0 & 100 & 0 & 0 & 0 & 0 &    100 \\
& 0.75 & 0 & 99 & 1 & 0 & 0 & 0 & 100 & 0 & 100 & 0 & 0 & 0 & 0 & 100 \\
& 0.5 &   0 & 100 & 0 & 0 & 0 & 0 & 100 & 0 & 100 & 0 & 0 & 0 & 0 & 100 \\
& 0.25 &  0 & 99 & 1 & 0 & 0 & 0 & 100 & 1 & 100 & 0 & 0 & 0 & 0 & 100  \\  
\hline
\end{tabular}}
\end{table}

Comparing Tables~\ref{table:sim:three:one}--\ref{table:sim:three:two}
to Table~\ref{table:sim:two:one},
we observe that mis-specifying the GARCH orders $(p, q)$ does not
noticeably worsen the performance of our methodology,
confirming the robustness of our methodology to the choice of $p$ and $q$.

\begin{table}[htbp]
\caption{\ref{m:3:1} The number of estimated change points (\%) and 
the accuracy in change point location when $\alpha = 0.05$, $T = 500$,
$N = 50$ (left) and $N = 100$ (right).}
\label{table:sim:three:one} \centering
\small{\begin{tabular}{cc|ccccc|cc||ccccc|cc}
\hline
&   &   \multicolumn{5}{c}{$\wh{B}$} &                  \multicolumn{2}{c}{accuracy (\%)} &     \multicolumn{5}{c}{$\wh{B}$} &                  \multicolumn{2}{c}{accuracy (\%)}       \\
&   $\varrho$ & 0 & 1 & 2 & 3 & $\ge 4$ &   $\eta_1$ &  $\eta_2$ &
0 & 1 & 2 & 3 & $\ge 4$ &   $\eta_1$ &  $\eta_2$    \\  \hline\hline
\multirow{4}{*}{(M2.1.1)} & 1 & 0 & 0 & 98 &    2 & 0 & 100 &   100 &   0 & 0 & 100 &   0 & 0 & 100 &   100 \\
&   0.75 &  0 & 0 & 100 &   0 & 0 & 100 &   100 &   0 & 0 & 99 &    1 & 0 & 100 &   100 \\
&   0.5 &   0 & 0 & 95 &    5 & 0 & 100 &   100 &   0 & 0 & 100 &   0 & 0 & 100 &   100 \\
&   0.25 &  0 & 18 &    73 &    8 & 1 & 100 &   80 &    0 & 97 & 3
& 0 & 0 & 100 &   1   \\  \hline
\multirow{4}{*}{(M2.1.2)} & 1 & 0 & 1 & 97 &    2 & 0 & 99 &    100 &   0 & 0 & 99 &    1 & 0 & 100 &   100 \\
&   0.75 &  0 & 8 & 90 &    2 & 0 & 92 &    100 &   0 & 0 & 99 &    1 & 0 & 100 &   100 \\
&   0.5 &   0 & 50 &    47 &    3 & 0 & 46 &    100 &   0 & 7 & 85 &    8 & 0 & 93 &    100 \\
&   0.25 &  1 & 88 &    10 &    1 & 0 & 5 & 99 &    3 & 79 &    18 &    0 & 0 & 27 &    78  \\
\hline
\end{tabular}}
\end{table}

\begin{table}[htbp]
\caption{\ref{m:3:2} The number of estimated change points (\%) and 
the accuracy in change point location when $\alpha = 0.05$, $T = 500$,
$N = 50$ (left) and $N = 100$ (right).}
\label{table:sim:three:two} \centering
\small{\begin{tabular}{cc|ccccc|cc||ccccc|cc}
\hline
&   &   \multicolumn{5}{c}{$\wh{B}$} &                  \multicolumn{2}{c}{accuracy (\%)} &     \multicolumn{5}{c}{$\wh{B}$} &                  \multicolumn{2}{c}{accuracy (\%)}       \\
&   $\varrho$ & 0 & 1 & 2 & 3 & $\ge 4$ &   $\eta_1$ &  $\eta_2$ &
0 & 1 & 2 & 3 & $\ge 4$ &   $\eta_1$ &  $\eta_2$    \\  \hline\hline
\multirow{4}{*}{\ref{m:3:2:1}} & 1 & 0 & 0 & 100 &   0 & 0 & 100 &   100 &   0 & 0 & 100 &   0 & 0 & 100 &   100 \\
&   0.75 &  0 & 0 & 98 &    2 & 0 & 100 &   100 &   0 & 0 & 98 &    2 & 0 & 100 &   100 \\
&   0.5 &   0 & 0 & 99 &    1 & 0 & 100 &   100 &   0 & 0 & 99 &    1 & 0 & 100 &   100 \\
&   0.25 &  0 & 5 & 92 &    2 & 1 & 95 &    100 &   0 & 15 &    74
&    11 &    0 & 100 &   83  \\  \hline
\multirow{4}{*}{\ref{m:3:2:2}} & 1 & 0 & 0 & 99 &    1 & 0 & 100 &   100 &   0 & 0 & 100 &   0 & 0 & 100 &   100 \\
&   0.75 &  0 & 0 & 97 &    3 & 0 & 100 &   100 &   0 & 0 & 99 &    1 & 0 & 100 &   100 \\
&   0.5 &   0 & 0 & 98 &    2 & 0 & 100 &   100 &   0 & 0 & 98 &    2 & 0 & 100 &   100 \\
&   0.25 &  0 & 45 &    51 &    4 & 0 & 48 &    100 &   0 & 12 &    79 &    9 & 0 & 92 &    93  \\  
\hline
\end{tabular}}
\end{table}

For the full-factor MGARCH model in \ref{m:4}, 
Table~\ref{table:sim:four} shows that our methodology manages
to estimate the two change points with high accuracy,
but it also tends to return spurious estimators
(see also Figures~\ref{fig:sim:four:50}--\ref{fig:sim:four:100} in Appendix).
In this scenario, the cross-sectional dependence is strong
due to the presence of common factors, which renders the re-sampling procedure
for threshold selection discussed in Section~\ref{sec:choice:thr} less reliable.

\begin{table}[htbp]
\caption{\ref{m:4} The number of estimated change points (\%) and 
the accuracy in change point location when $\alpha = 0.05$, $T = 500$,
$N = 50$ (left) and $N = 100$ (right).}
\label{table:sim:four}
\centering 
\small{\begin{tabular}{cc|ccccc|cc||ccccc|cc}
\hline
&   &   \multicolumn{5}{c}{$\wh{B}$} &                  \multicolumn{2}{c}{accuracy (\%)} &     \multicolumn{5}{c}{$\wh{B}$} &                  \multicolumn{2}{c}{accuracy (\%)}       \\
&   $\varrho$ & 0 & 1 & 2 & 3 & $\ge 4$ &   $\eta_1$ &  $\eta_2$ &
0 & 1 & 2 & 3 & $\ge 4$ &   $\eta_1$ &  $\eta_2$    \\  \hline\hline
\multirow{4}{*}{\ref{m:4:1}} &   1 & 0 & 5 & 34 &    52 &    9 & 91 &    97 &    0 & 0 & 45 &    44 &    11 &    95 &    97  \\
&   0.75 &  0 & 3 & 36 &    49 &    12 &    95 &    90 &    0 & 0 & 33 &    60 &    7 & 95 &    96  \\
&   0.5 &   0 & 4 & 37 &    52 &    7 & 89 &    89 &    0 & 3 & 17 &    64 &    16 &    96 &    88  \\
&   0.25 &  0 & 14 &    42 &    39 &    5 & 77 &    53 &    0 & 6
& 32 &    48 &    14 &    92 &    70  \\  \hline
\multirow{4}{*}{\ref{m:4:2}} &   1 & 0 & 13 &    87 &    0 & 0 & 87 &    99 &    0 & 4 & 96 &    0 & 0 & 96 &    98  \\
&   0.75 &  0 & 22 &    78 &    0 & 0 & 79 &    96 &    0 & 7 & 93 &    0 & 0 & 93 &    97  \\
&   0.5 &   0 & 74 &    24 &    2 & 0 & 29 &    93 &    0 & 38 &    61 &    1 & 0 & 64 &    97  \\
&   0.25 &  6 & 89 &    5 & 0 & 0 & 8 & 70 &    0 & 93 &    7 & 0
& 0 & 8 & 92  \\  
\hline
\end{tabular}}
\end{table}

\section{Application to financial risk management}
\label{sec:real}

In this section, we consider the application of the methodology proposed in Section~\ref{sec:method}
to risk management using the Value-at-Risk (VaR) of a portfolio,
a widely used measure of market risk embraced by financial institutions  
for regulatory and other internal purposes.
In this exercise, we demonstrate the peril of ignoring
the change points in the volatility and the cross-sectional correlations of a multi-asset portfolio,
and show that the stress period identified by our proposed methodology
is effective in providing robust risk management. 

\subsection{Background: VaR, stressed VaR and its backtests}

VaR measures the extreme loss (change in value) of an asset or a portfolio of assets 
with a prescribed probability level during a given holding period.
It has been criticised due to unrealistic assumptions (linearity and normality), 
parameter sensitivity (to estimation and holding periods) and its inadequacy during crises 
especially when correlations between assets are observed to vary over time \citep{persaud2000, danielsson2002}. 
The last point is of particular interest
since, compared to the period of market stability,
correlations are observed to be significantly higher when markets are falling \citep{li2017}. 
\cite{rebonato2001} noted that a risk manager `would greatly over-estimate the degree of diversification 
in his portfolio in the event of a crash if he used the [correlation] matrix 
estimated during normal periods'. 
Works to address the criticism on VaR exist: \cite{valentinyi2004} examined 
whether detecting and taking into account change points improves upon VaR forecast. 
Similarly, \cite{spoikony2009} proposed to perform local change point analysis 
to detect regions of volatility homogeneity as an alternative to stationary GARCH modelling, 
and found that the local volatility estimator performed well in the application to VaR.

The Basel Accord (1996 Amendment) requires the use of stressed VaR (sVaR) 
which is based on a covariance matrix from a crisis period in the past. 
The accord does not specify the exact time period to be used but instead,
proposes the \emph{judgement-based} and the \emph{formulaic} approaches \citep{eba}. 
The former relies on a high-level analysis of the risks related to the holding portfolio, 
while the latter is a more systematic, quantitative approach
where our proposed methodology can contribute
by supplying the information about the latest stress periods. 

The VaR metric is defined as follows:
\begin{align*}
\VaR_t(\alpha) = -F^{-1}(\alpha|\mathcal{G}_t),
\end{align*}
where $F^{-1}(\cdot | \mathcal{G}_t)$ is the quantile function of the loss and profit distribution
with market- and portfolio-specific conditions contained in $\mc G_t$.
To backtest VaR by means of statistical tests,
we adopt two tests from \cite{kupiec1995} using
the Proportion of Failure (PoF) and the Time until First Failure (TFF). 
For a sample of $T$ observations, the test statistics take the form of 
a likelihood ratio test statistic:
\begin{align}
LR_{\text{PoF}} = -2 \log\left(\frac{(1-\alpha)^{T-x_f}\alpha^{x_f}}
{\left(1-\frac{x_f}{T}\right)^{T-x_f}\left(\frac{x_f}{T}\right)^{x_f}}\right)
\text{ and }
LR_{\text{TFF}} = -2 \log\left
(\frac{\alpha(1-\alpha)^{t_f -1}} {\left ( \frac{1}{t_f}\right
)\left ( 1- \frac{1}{t_f}\right )^{t_f-1}}\right)
\label{eq:kupiec}
\end{align}
for some $\alpha \in (0, 1)$,
where $x_f$ denotes the number of failures (losses in excess of the reported VaR) occurred 
and $t_f$ the number of days until the first failure within the $T$ observations occurs.
Under suitable assumptions,
both $LR_{\text{PoF}}$ and $LR_{\text{TFF}}$ follow a $\chi^2_1$-distribution asymptotically,
which can be used to test whether the VaR model is adequate. 
These tests do not control for the dependence in the failures,
i.e.\ excess losses beyond the reported VaR may cluster 
while the overall (unconditional) number of failures is not significantly different from $\alpha T$. 
To address this limitation, we adopt the dynamic conditional quantile (DQ) test 
by \cite{engle2004caviar} whose test statistic is defined as 
\begin{align}
\text{DQ} = \frac{\mbf H^\top\mbf Z(\mbf Z^\top\mbf Z)^{-1}\mbf Z^\top \mbf H}{T\alpha(1-\alpha)}
\sim_{H_0} \chi^2_{\bar{q}},
\label{eq:dq}
\end{align}
where $\mbf Z$ is the matrix of explanatory variables (e.g.\ raw and squared past returns)
with $\bar{q} = \text{rank}(\mbf Z)$,
and $\mbf H$ the vector collecting 
$\text{Hit}_t(\alpha) := \bbI(r_t < - \VaR_t(\alpha)) - \alpha$ 
over time with $r_t$ being the time series of portfolio returns.
Later in Sections~\ref{sec:svar:hist}--\ref{sec:svar:dcc},
we use the three backtests in~\eqref{eq:kupiec} and~\eqref{eq:dq} when
assessing the performance of our proposed methodology in the applications to stressed VaR.



\subsection{Change point detection in a multi-asset portfolio}

We collect the daily log-returns of the stocks composing the S\&P100 index from 1 January 2007 to 31 December 2020 (available from Yahoo! Finance). 
We winsorize the time series in order to restrict the influence of outliers on the MLE of the GARCH parameters for individual log-return series. In addition, identifying the pairs of stocks whose unconditional correlation is above $0.8$, one of each of the pair is randomly removed,
which results in $N = 79$ time series ($d_N = 3160$).
We take this step solely for the purpose of ensuring that
the estimation of multivariate GARCH processes we consider later in Section~\ref{sec:oos} for out-of-sample validation of the sVaR, is feasible and free of convergence issues; such high correlations are often attributed to the shares of the same company in different classes
(e.g.\ GOOG and GOOGL). We set 75\% of the total sample (from 1 January 2007 to 28 April 2016, $T = 2347$) as the training data
that serves as an input to the proposed change point detection methodology. The remaining observations are used for out-of-sample validation of the sVaR from different stress periods, which is discussed in the next section.

Our method, using the default parameters described in Section~\ref{sec:choice}, 
detects three change points from the training data, 
which form the four periods reported in Table~\ref{table:cpdetected} (Periods~1--4).
The table also reports Period~5, which covers the 12-month period 
following the bankruptcy of Lehman Brothers in September 2008 as the crisis period, 
with its length chosen in accordance with Basel~2.5 \citep{BABS193}. 
The results indicate that the highest VaR 
(obtained as quantiles of an equally weighted portfolio of the 79 stocks in modulus) 
was obtained from Period~2 which spans from 8 September 2008 to 5 May 2009, 
a rather expected outcome given the high volatility 
the market experienced after the bankruptcy of Lehman Brothers. 
We note that this period spans less than 12 months, 
and is shorter than the maximum allowed duration by Basel~2.5. 
In addition, it coincides with Bank of England's view on the historical periods per region 
with the worst market moves \citep{boe2018}. 
Period~4 also exhibits high stress characteristics 
but less severe than those observed in the aftermath of the Lehman Brothers bankruptcy. 
If a bank selects Period~5 (the 95\% VaR of which is 4.41\%) to calibrate its sVaR model, 
it will likely under-estimate the measure compared to 
that calibrated with Period~2 (the 95\% VaR of which is 5.32\%). 
In Section~\ref{sec:oos}, we explore this argument in further detail.

Out of independent interest, we applied the proposed change point detection methodology 
to the same dataset covering the recent pandemic crisis, 
i.e.\ from 6 August 2009 (after the latest change point identified from the training data) 
to 31 December 2020. 
It detects a change point at the beginning of the pandemic located around the end of February 2020, 
a few days before the Federal Reserve System decided to cut interest rates 
in order to reduce the economic impacts from the Covid-19 outbreak, 
see Period~6 and~7 in Table~\ref{table:cpdetected}. 
From the same table, we observe that the current pandemic crisis is not, 
measured by either the 95\% VaR or the 95\% VaR, 
as `stressed' as Period~2 coinciding with global financial crisis. 
One reason might be the extremely fast recovery of the capital markets
which likely offset the initial shocks. 
However, without further data it is not trivial to make a conclusive assessment.

\begin{table}[htbp]
\centering
\caption{Three change points detected by our methodology (in bold) 
from the training data spanning from 1 January 2007 to 28 April 2016, 
and the corresponding periods of stationarity (Periods~1--4). 
Period~5 covers the 12-month period following the bankruptcy of Lehman Brothers
with its length chosen in accordance with Basel 2.5. 
It also presents the single change point detected by our methodology (in italic)
from the data panning from 6 August 2009 to 31 December 2020 
and the corresponding periods of stationarity (Periods~6 and~7).
For each period, we report the 95\% and 99\% VaR of an equally weighted portfolio of the $N = 79$ stocks.} 
\label{table:cpdetected}
    \begin{tabular}{lrr|rr}
   	\hline
    & \multicolumn{2}{c|}{Period range (mm/dd/yyyy)} & \multicolumn{2}{c}{Value-at-Risk} \\
    & \multicolumn{1}{c}{From} & \multicolumn{1}{c|}{To} & \multicolumn{1}{c}{95\%} & \multicolumn{1}{c}{99\%} \\
   \hline\hline
    Period 1 & 01/01/2007 & {\bf 09/07/2008}        & 0.01836 & 0.02600\\
    Period 2 & 09/08/2008 & {\bf 05/04/2009}        & 0.05326 & 0.07508\\
    Period 3 & 05/05/2009 & {\bf 08/05/2009}        & 0.01904 & 0.02765\\
    Period 4 & 08/06/2009 & 04/28/2016              & 0.03217 & 0.04554\\
    \hline
    Period 5 & 09/08/2008 & 09/08/2009   & 0.04416 & 0.06254\\
    \hline
    Period~6 & 08/06/2009 & {\it 02/24/2020} & 0.02523 & 0.03880\\
    Period~7 & 02/25/2020 & 12/31/2020        & 0.03711 & 0.05560\\    
    \hline
    \end{tabular}
\end{table}%

%


\subsection{Out-of-sample performance of sVaR}
\label{sec:oos}

Our proposed methodology supplies a means 
to segment the data and identify the period of most volatility,
but does not automatically provide an sVaR model. 
After the change points are detected from a given dataset, 
a user should decide on the model according to regulatory or internal requirements.
In this section, we form optimal portfolios of the S\&P100 constituent stocks
(Section~\ref{sec:Optim_portf}) and assess which period from Table~\ref{table:cpdetected}
is the most appropriate for an sVaR purpose using two sVaR models;
in order to ensure that we have enough test data for the out-of-sample exercises,
only Periods~1--5 have been used for model calibration.
One based on the historical simulation (unconditional covariance modelling, Section~\ref{sec:svar:hist}) 
and the other based on the DCC model of \cite{engle2002} (conditional, Section~\ref{sec:svar:dcc}). 
The former is the most popular VaR model used by banks
and the latter is a dynamic extension of the CCC model \citep{bollerslev1990},
which in turn is closely related to the time-varying model~\eqref{eq:garch} adopted in this paper.

\subsubsection{Optimal portfolio formulation}
\label{sec:Optim_portf}

We form portfolios of the $N$ stocks over a given period
for testing the out-of-sample performance of the sVaR.
In particular, we calculate the portfolio return from the vector of the returns $\mbf r_t$ as
$r^{\text{opt}}_{t} = \mbf w_t^\top \bm\mu_t$,
where $\mbf w_t \in \R^N$ denotes a vector of the portfolio weights and 
$\bm\mu_t \in \R^N$ a vector of the mean returns of the assets at time $t$. 
There are many different methods for constructing $\bm\mu_t$, 
but we simply calculate the sample mean return for each stock using 
the observations from the most recent $T_{\text{sv}} \in \{250, 500, 650\}$ days,
which forms a rolling window that moves forward one day at a time. 
To obtain the optimal weights $\mbf w_t$, we solve the following Markowitz optimisation problem
\begin{align}
\min_{\mbf w \in \R^N} \mbf w^\top \wh{\bm\Sigma}_r(t) \mbf w \quad \text{such that} \quad
\mbf w^\top \bm\mu_t = \mu_{\text{target}} \mbf 1 \; \; \text{and} \; \; 
\mbf w^\top \mathbf{1} = 1, \label{eq:markowitz_opt}
\end{align}
where $\wh{\bm\Sigma}_r(t)$ is an estimator of the covariance matrix of $\mbf r_t$; 
to ensure that it is well-conditioned, we obtain $\wh{\bm\Sigma}_r(t)$
via a non-linear shrinkage estimation technique proposed by \cite{ledoit2012}
using the same $T_{\text{sv}}$-day rolling window.
The quantity $\mu_{\text{target}}$ denotes 
a target expected portfolio return that takes its value from
$\{0.25\%, 0.50\%, 0.75\%\}$. 
We repeatedly solve the optimisation problem every $h_{\text{reb}} \in \{5, 20\}$ days
keeping the portfolio weights $\mbf w_t$ unchanged in between,
which implies a weekly or monthly portfolio rebalancing. 
Following the above steps, we create an optimal portfolio for every triplet
$(T_{\text{sv}}, \mu_{\text{target}}, h_{\text{reb}})$
and use it to evaluate the adequacy of the sVaR  
as described in Sections~\ref{sec:svar:hist} and~\ref{sec:svar:dcc}.

\subsubsection{Historical simulation}
\label{sec:svar:hist}

Let the five periods identified in Table~\ref{table:cpdetected} be indexed with $b = 1, \ldots, 5$.
A financial institution typically uses historical simulation over one or two years 
($T_{\text{sv}} \in \{250,500\}$), to produce a one-day ahead forecast of sVaR as
\begin{align}
\widehat{\sVaR}^{(b)}_{t+1} &= - \Big[\text{upper $100 \times \alpha$-th percentile of } 
\{R^{(b)}_{\tau}\}_{\tau=t-T_{\text{sv}}+1}^t\Big] \label{svar}
\\
\text{where} \quad R^{(b)}_t &= \mbf r_t^\top (\mathbf{L}^{(b)} \mathbf{L}(t))^{-1} \mbf w_t, \label{cho_deco}
\end{align}
with $\mathbf{L}^{(b)}$ being the Cholesky decomposition of $\wh{\bm\Sigma}^{(b)}_r$,
an estimator of the (unconditional) covariance matrix for Period~$b$
(also obtained using the non-linear shrinkage estimation method of \cite{ledoit2012}, 
see Section~\ref{sec:Optim_portf}),
$\mathbf{L}(t)$ the Cholesky decomposition of $\wh{\bm\Sigma}_r(t)$
and $\mbf w_t$ the optimal weights obtained from \eqref{eq:markowitz_opt}. 
Equation~\eqref{cho_deco} transforms the return vector $\mbf r_t$ 
with covariance matrix $\wh{\bm\Sigma}_r(t)$,
into the portfolio return $R^{(b)}_t$ of a return vector with a (stressed) covariance $\wh{\bm\Sigma}^{(b)}_r$, 
for given $b = 1, \ldots, 5$ \citep{duffie1997}. 
For each $b$, we repeatedly forecast the one-day ahead sVaR of an optimal portfolio 
using $\widehat{\sVaR}_{t+1}^{(b)}$,
until we reach the end of the test data on 31 December 2020.
We repeat this exercise for each of the optimal portfolios formed 
with the triplets $(T_{\text{sv}}, \mu_{\text{target}}, h_{\text{reb}}) \in 
\{250, 500\} \times \{0.25\%, 0.50\%, 0.75\%\} \times \{5, 20\}$.

The results from sVaR backtests are provided in Table~\ref{tab:HS_res}.
In particular, the table indicates that Period~2 can safely be used to calibrate sVaR:
The corresponding number of failures is below the expected number of violations 
for the out-of-sample period at 99\% levels 
(approximately 5 and 7 for $500$ and $750$ days, respectively),
and the first failure occurred only after 
at least 173 days at the 99\% level depending on $\mu_{\text{target}}$. 
Note that the alternative hypothesis in the likelihood ratio tests is two-side and 
a small number of failures also reject the adequacy of an sVaR model. 
The sVaR model calibrated using Period~2 yields a favourable, albeit conservative, result and 
is more likely to be accepted by risk managers. 
Besides, it passes the traffic light test \citep{BABS1996} in all cases, 
whereby a VaR model is deemed valid (green zone) 
if the probability of observing up to $x_f$ failures 
is less than $0.95$ under the binomial distribution with $T_{\text{sv}}$ and $\alpha$ as the parameters.
Assessing the adequacy of stress periods using the DQ test, 
Period~2 is the best candidate for the sVaR calibration with $p$-value $> 0.1$ in most scenarios
except when $\mu_{\text{target}}$ is large and $T_{\text{sv}} = 500$.

We mention that Period~5 also exhibits a small number of failures, 
rather expected behaviour considering that it overlaps with Period~2 for eight months. 
It is a good candidate for a stress period as it passes almost all tests (except for the DQ test). 
However, the results could be significantly worse if the actual stress period were shorter, 
in which case a twelve-month period would be an equal mix of a stress period and a much less stressed period.

\begin{table}[htb]
\centering
\caption{Results from sVaR backtesting 
on the test data (from 1 May 2016 to 31 December 2020)
using the historical simulation approach 
calibrated on five different stress periods $b = 1, \ldots, 5$ from Table~\ref{table:cpdetected}.
From left to right, the table reports: The number of sVaR failures (PoF) 
and the number of days until the first failure in sVaR (TFF), 
with the corresponding $p$-values in brackets from the respective {\it two-sided} tests in~\eqref{eq:kupiec};
traffic light test (traffic); $p$-values from the DQ test (DQ) in~\eqref{eq:dq}.}
\resizebox{\columnwidth}{!}{
\begin{tabular}{ccc | lll r| lll r}
\hline
\multicolumn{3}{c|}{Scenario}    & \multicolumn{4}{c|}{5-day rebalance ($h_{\text{reb}} = 5$)} & \multicolumn{4}{c}{20-day rebalance ($h_{\text{reb}} = 20$)} \\
$\mu_{\text{target}}$ & $T_{\text{sv}}$ & Period & PoF & TFF & traffic & \multicolumn{1}{l|}{DQ} & PoF & TFF & traffic & \multicolumn{1}{l}{DQ} \\
\hline\hline
0.25\% & 250 &  1 & 17 (0.003) & 14 (0.132) & \colorbox{yellow}{yellow}(0.014) & 0     & 13 (0.244) & 195 (0.451) & \colorbox{green}{green}(0.143) & 0 \\
& & 2 & 0 (1) & NA (1) & \colorbox{green}{green}(1) & 0.157 & 0 (1) & NA (1) & \colorbox{green}{green}(1) & 0.157 \\
& & 3 & 20 (0.002) & 195 (0.451) & \colorbox{yellow}{yellow}(0.001) & 0     & 17 (0.022) & 195 (0.451) & \colorbox{yellow}{yellow}(0.014) & 0 \\
& & 4 & 7 (0.435) & 198 (0.439) & \colorbox{green}{green}(0.817) & 0     & 3 (0.016) & 723 (0.003) & \colorbox{green}{green}(0.995) & 0 \\
& & 5 & 1 (0) & 723 (0.003) & \colorbox{green}{green}(1) & 0.284     & 2 (0.004) & 723 (0.003) & \colorbox{green}{green}(0.999) & 0 \\
0.25\% & 500 &  1 & 20 (0) & 	117 (0.871) & \colorbox{red}{red}(0) & 0     & 20 (0) & 117 (0.871) & \colorbox{red}{red}(0) & 0 \\
& & 2 & 2 (0.03) & 480 (0.034) & \colorbox{green}{green}(0.991) & 0.513 & 3 (0.102) & 475 (0.036) & \colorbox{green}{green}(0.965) & 0.764 \\
& & 3 & 23 (0) & 117 (0.871)& \colorbox{red}{red}(0) & 0     & 24 (0) & 117 (0.871) & \colorbox{red}{red}(0) & 0 \\
& & 4 & 7 (0.927)& 473 (0.036) & \colorbox{green}{green}(0.515) & 0     & 8 (0.641) & 470 (0.037) & \colorbox{green}{green}(0.365) & 0 \\
& & 5 & 5 (0.476) & 473 (0.036) & \colorbox{green}{green}(0.805) & 0     & 4 (0.248) & 475 (0.036) & \colorbox{green}{green}(0.906) & 0 \\
0.5\% & 250 &  1 & 37 (0) & 14 (0.132) & \colorbox{red}{red}(0) & 0     & 34 (0) & 14 (0.132) & \colorbox{red}{red}(0) & 0 \\
& & 2 & 2 (0.004) & 732 (0.003) & \colorbox{green}{green}(0.999) & 0.352 & 1 (0) & 891 (0.001) & \colorbox{green}{green}(1) & 0.272 \\
& & 3 & 39 (0) & 14 (0.132) & \colorbox{red}{red}(0) & 0     & 35 (0) & 14 (0.132) & \colorbox{red}{red}(0) & 0 \\
& & 4 & 14 (0.146) & 195 (0.451) & \colorbox{yellow}{yellow}(0.087) & 0     & 12 (0.387) & 198 (0.439) & \colorbox{green}{green}(0.222) & 0 \\
& & 5 & 7 (0.435) & 596 (0.011) & \colorbox{green}{green}(0.817) & 0 & 4 (0.050) & 596 (0.011) & \colorbox{green}{green}(0.983) & 0.594 \\
0.5\% & 500 & 1 & 27 (0) & 117 (0.871) & \colorbox{red}{red}(0) & 0     & 37 (0) & 117 (0.871) & \colorbox{red}{red}(0) & 0 \\
& & 2 & 4 (0.248) & 480 (0.034) & \colorbox{green}{green}(0.906) & 0     & 4 (0.248) & 475 (0.034) & \colorbox{green}{green}(0.906) & 0 \\
& & 3 & 29 (0.022) & 117 (0.871) & \colorbox{red}{red}(0) & 0     & 33 (0) & 66 (0.696) & \colorbox{red}{red}(0) & 0 \\
& & 4 & 17 (0.001) & 173 (0.545) & \colorbox{yellow}{yellow}(0.001) & 0     & 15 (0.006) & 173 (0.545) & \colorbox{yellow}{yellow}(0.004) & 0 \\
& & 5 & 5 (0.476) & 480 (0.034) & \colorbox{green}{green}(0.805) & 0     & 7 (0.927) & 346 (0.117) & \colorbox{green}{green}(0.515) & 0.003 \\
0.75\% & 250 &  1 & 43 (0) & 14 (0.132) & \colorbox{red}{red}(0) & 0     & 44 (0) & 14 (0.132) & \colorbox{red}{red}(0) & 0 \\
& & 2 & 7 (0.435) & 596 (0.011) & \colorbox{green}{green}(0.817) & 0 & 4 (0.05) & 596 (0.011) & \colorbox{green}{green}(0.983) & 0.498 \\
& & 3 & 44 (0) & 14 (0.132) & \colorbox{red}{red}(0) & 0     & 38 (0) & 14 (0.132) & \colorbox{red}{red}(0) & 0 \\
& & 4 & 23 (0.007) & 14 (0.132) & \colorbox{red}{red}(0) & 0     & 21 (0.001) & 14 (0.132) & \colorbox{yellow}{yellow}(0.001) & 0 \\
& & 5 & 10 (0.809) & 596 (0.011) & \colorbox{green}{green}(0.447) & 0     & 6 (0.250) & 596 (0.011) & \colorbox{green}{green}(0.900) & 0.001 \\
0.75\% & 500 &  1 & 37 (0) & 66 (0.696) & \colorbox{red}{red}(0) & 0     & 43 (0) & 66 (0.696) & \colorbox{red}{red}(0) & 0 \\
& & 2 & 6 (0.765) & 173 (0.545) & \colorbox{green}{green}(0.669) & 0.002     & 6 (0.765) & 345 (0.118) & \colorbox{green}{green}(0.669) & 0.001 \\
& & 3 & 31 (0.000) & 66 (0.696) & \colorbox{red}{red}(0) & 0     & 34 (0) & 66 (0.696) & \colorbox{red}{red}(0) & 0 \\
& & 4 & 18 (0.00) & 173 (0.545) &\colorbox{red}{red}(0)& 0     & 21 (0) & 173 (0.545) &  \colorbox{red}{red}(0) & 0 \\
& & 5 & 10 (0.242) & 173 (0.545) & \colorbox{green}{green}(0.145) & 0     & 10 (0.242) & 345 (0.118) & \colorbox{green}{green}(0.145) & 0 \\
\hline
\end{tabular}}
\label{tab:HS_res}%
\end{table}%


\subsubsection{Dynamic conditional correlation model}
\label{sec:svar:dcc}

In the second exercise, we adopt the DCC model of \cite{engle2002} 
to model each stationary segment identified by the proposed change point detection methodology.
We restrict our analysis to $N = 30$ stocks randomly selected from the $79$ stocks in order to avoid convergence issues with the estimation of the DCC model. 

Under the DCC, the conditional covariance matrix is factorised as
$\mbf H_t = \mbf D_t \mbf R_t \mbf D_t$,
where $\mbf D_t = \text{diag}(h_{1,t}^{1/2}, \ldots, h_{N,t}^{1/2})$
and the conditional variance $h_{i, t}$ follows the GARCH($p, q$) model~\eqref{eq:garch} 
with time-invariant parameters. We consider the GARCH models with $p = q = 1$ and normally distributed innovations for their estimation; the results were similar with other model orders $p, q$ or innovation distributions. 
\cite{engle2002} propose the following dynamic correlation structure for $\mbf R_t$:
\begin{align}
\mbf R_t &= \text{diag}(\bm\Sigma_t)^{-1/2} \; \bm\Sigma_t \; \text{diag}(\bm\Sigma_t)^{-1/2}, 
\quad \text{with} \nonumber
\\
\bm\Sigma_t &= (1 - a_{\text{dcc}} - b_{\text{dcc}})\bar{\bm\Sigma}+ a_{\text{dcc}} \mbf v_{t-1}
\mbf v_{t-1}^\top + b_{\text{dcc}}\bm\Sigma_{t-1}, \label{eq:garch_cov}
\end{align}
where $\mbf v_t = \mbf D_t^{-1} \mbf r_t$ is the standardised residual vector
and $\bar{\bm\Sigma} = [\wh\varrho_{i, i'}]_{i, i' = 1}^N$ with 
$\wh\varrho_{i, i'}$ denoting the unconditional sample correlations between $v_{i, t}$ and $v_{i', t}$. 


Similarly to the approach taken in Section~\ref{sec:svar:hist}, we consider a
portfolio of assets with the return vector $\mbf r_t$ and 
the vector of portfolio weights $\mbf w_t$ obtained as in~\eqref{eq:markowitz_opt}
with the triplets $(T_{\text{sv}}, \mu_{\text{target}}, h_{\text{reb}}) \in 
\{650\} \times \{0.25\%, 0.50\%, 0.75\%\} \times \{5, 20\}$.
For each Period~$b$, we estimate the parameters
$a_{\text{dcc}}, b_{\text{dcc}}$ along with $\omega_i, \alpha_{i, 1}$ and $\beta_{i, 1}$
(their dependence on $b$ suppressed).
Then from the estimated DCC model for the corresponding period,
we obtain the one-day ahead forecast of the conditional covariance matrix, 
$\wh{\mbf H}^{(b)}_{t+1}$, via~\eqref{eq:garch_cov},
produce a forecast of the portfolio volatility as
$\wh{\sigma}^{(b)}_{t+1} = \sqrt{\mbf w_t^\top \wh{\mbf H}^{(b)}_{t+1} \mbf w_t}$,
and compare  $- 2.33 \times \wh{\sigma}^{(b)}_{t+1}$ (99\% sVaR up to its sign) to the actual portfolio returns,
until we reach the end of the test data on 31 December 2020.
The thus-obtained backtesting results are given in Table~\ref{tab:DCC_res}. 

Using the estimated parameters from Period~2 
yields the smallest number of failures compared with the rest:
At the 99\% level, the first failure occurred after 295 to 383 days, 
compared with 12 days taken for the rest of the stress periods. 
The DQ test also indicates that Period~2 is the most suitable 
for calibrating the 99\% sVaR metric with $p$-values $>0.10$ in all scenarios. 
Period~5 fails to pass all the tests as the corresponding numbers of failures exceeded the expected ones. 
This reinforces our argument that, since Period~5 includes 
the three month period (May to August 2009) of low stress characteristics,
overlapping with Period~3 in Table~\ref{table:cpdetected}, 
it is not suitable for calibrating the 99\% sVaR metric,
see also that Period~3 yields a large number of failures compared to Period~2.

\begin{table}[htbp]
\centering
\caption{Results from sVaR backtesting using the DCC models
on the test data (from 1 May 2016 to 31 December 2020)
calibrated using five different stress periods $b = 1, \ldots, 5$ from Table~\ref{table:cpdetected},
with $T_{\text{sv}} = 650$ used throughout.
From left to right, the table reports: The number of sVaR failures (PoF) 
and the number of days until the first failure in sVaR (TFF), 
with the corresponding $p$-values in brackets from the respective {\it two-sided} tests in~\eqref{eq:kupiec};
traffic light test (traffic); $p$-values from the DQ test (DQ) in~\eqref{eq:dq}.}
\resizebox{\columnwidth}{!}{
\begin{tabular}{cc | lll r| lll r}
\hline
\multicolumn{2}{c|}{Scenario}    & \multicolumn{4}{c|}{5-day rebalance ($h_{\text{reb}} = 5$)} & \multicolumn{4}{c}{20-day rebalance ($h_{\text{reb}} = 20$)} \\
$\mu_{\text{target}}$ & Period & PoF & TFF & traffic & \multicolumn{1}{l|}{DQ} & PoF & TFF & traffic & \multicolumn{1}{l}{DQ} \\
\hline
\hline
0.25\% & 1 & 10 (0.065) & 12 (0.110) & \colorbox{yellow}{yellow}(0.041) & 0.029 & 9 (0.137) & 12 (0.110) & \colorbox{yellow}{yellow}(0.085) & 0.036 \\
& 2 & 3(0.281)  & 383 (0.084) & \colorbox{green}{green}(0.897) & 0.813 &  3(0.281)  & 383 (0.084) & \colorbox{green}{green}(0.897) & 0.855\\
& 3 & 16 (0) & 12 (0.110) & \colorbox{red}{red}(0) & 0 & 17 (0) & 12 (0.110) & \colorbox{red}{red}(0) & 0 \\
& 4 & 12 (0.011) & 12 (0.110) & \colorbox{yellow}{yellow}(0.008)& 0.008 & 12 (0.011) & 12 (0.110) & \colorbox{yellow}{yellow}(0.008)& 0.008\\
&  5& 10 (0.065) & 12 (0.110) & \colorbox{yellow}{yellow}(0.019) & 0.016 &  12 (0.110)  & 12 (0.110) & \colorbox{yellow}{yellow}(0.008) & 0.006 \\
0.5\% & 1 & 9 (0.137) & 12 (0.110) & \colorbox{yellow}{yellow}(0.085) & 0.040 & 9 (0.137) & 12 (0.110) & \colorbox{yellow}{yellow}(0.085) & 0.042 \\
& 2 & 5 (0.909) & 295 (0.186) & \colorbox{green}{green}(0.605) & 0.439  & 3 (0.281) & 383 (0.084) & \colorbox{green}{green}(0.897) & 0.906 \\
& 3 & 22 (0) & 12 (0.110) & \colorbox{red}{red}(0) & 0 & 22 (0) & 12 (0.110) & \colorbox{red}{red}(0) & 0   \\
& 4 & 10 (0.065) & 12 (0.110) & \colorbox{yellow}{yellow}(0.041) & 0.030 & 11 (0.028) & 12 (0.110) & \colorbox{yellow}{yellow}(0.019) & 0.017  \\
& 5 & 8 (0.265) & 12 (0.110) & \colorbox{yellow}{yellow}(0.161) & 0.039 & 8 (0.265) & 12 (0.110) & \colorbox{yellow}{yellow}(0.161) & 0.046 \\
0.75\% & 1 & 10 (0.065) & 12 (0.110) & \colorbox{yellow}{yellow}(0.041) & 0.040 & 9 (0.137) & 12 (0.110) & \colorbox{yellow}{yellow}(0.085) & 0.044 \\
& 2 & 6 (0.751) & 295 (0.186) & \colorbox{green}{green}(0.430) & 0.248 & 4 (0.564) & 383 (0.084) & \colorbox{green}{green}(0.771) & 0.890\\
& 3 & 23 (0) & 12 (0.110) & \colorbox{red}{red}(0) & 0 & 23 (0) & 12 (0.110) & \colorbox{red}{red}(0) & 0    \\
& 4 & 12 (0.011) & 12 (0.110) & \colorbox{yellow}{yellow}(0.008) & 0.009 & 12 (0.011) & 12 (0.11) & \colorbox{yellow}{yellow}(0.008) & 0.01\\
&5 & 9 (0.137) & 12 (0.110) & \colorbox{yellow}{yellow}(0.085) & 0.044 & 8 (0.265) & 12 (0.110) & \colorbox{green}{green}(0.161) & 0.046 \\
    \hline
    \end{tabular}}
  \label{tab:DCC_res}%
\end{table}%

\section{Conclusions}
\label{sec:conc}

In this paper, we propose a two-stage methodology for detecting multiple
change points in both within-series and cross-correlation
structures of multivariate volatility processes. 
It first transforms the $N$-dimensional series so that
complex structural change points are detectable as change points in
the level of $N(N+1)/2$-dimensional transformed data, which provides
an input to the multiple change point detection algorithm in the second stage. 
We show the theoretical consistency of the combined methodology in
terms of the total number and locations of estimated change points, 
and verify its good performance on simulated datasets.
Also, we demonstrate the efficacy of the proposed methodology in financial risk management.
This exercise shows that,
by identifying the period of stress from the dataset of the S\&P100 constituent stocks, 
our method can serve as a formulaic approach to
calculating the (stressed) Value-at-Risk of a portfolio of risky assets.

{\small 
\bibliographystyle{apalike}
\bibliography{fbib}

\begin{thebibliography}{}

\bibitem[Andreou and Ghysels, 2002]{andreou2002}
Andreou, E. and Ghysels, E. (2002).
\newblock Detecting multiple breaks in financial market volatility dynamics.
\newblock {\em Journal of Applied Econometrics}, 17(5):579--600.

\bibitem[Andreou and Ghysels, 2003]{andreou2003}
Andreou, E. and Ghysels, E. (2003).
\newblock Tests for breaks in the conditional co-movements of asset returns.
\newblock {\em Statistica Sinica}, 13:1045--1073.

\bibitem[Aue et~al., 2009]{aue2009}
Aue, A., H{\"o}rmann, S., Horv{\'a}th, L., and Reimherr, M. (2009).
\newblock Break detection in the covariance structure of multivariate time
  series models.
\newblock {\em The Annals of Statistics}, 37(6B):4046--4087.

\bibitem[Aue and Horv{\'a}th, 2013]{aue2013}
Aue, A. and Horv{\'a}th, L. (2013).
\newblock Structural breaks in time series.
\newblock {\em Journal of Time Series Analysis}, 34:1--16.

\bibitem[Baele, 2003]{Baele2003}
Baele, L. (2003).
\newblock Did emu increase equity market correlations?
\newblock {\em Financieel Forum, Bank en Financiewezen}, 6:356--359.

\bibitem[Baillie, 1991]{Baillie1991}
Baillie, R. T.and~Bollerslev, T. (1991).
\newblock Intra-day and inter-market volatility in foreign exchange rates.
\newblock {\em The Review of Economic Studies}, 58(3):565--585.

\bibitem[{Bank of England}, 2018]{boe2018}
{Bank of England} (2018).
\newblock {\em Stress testing the UK banking system: guidance on the traded
  risk methodology for participating banks and building societies}.
\newblock
  \url{https://www.bankofengland.co.uk/-/media/boe/files/stress-testing/2018/stress-testing-the-uk-banking-system-2018-guidance.pdf}.

\bibitem[Barassi et~al., 2020]{barassi2018}
Barassi, M., Horv{\'a}th, L., and Zhao, Y. (2020).
\newblock Change-point detection in the conditional correlation structure of
  multivariate volatility models.
\newblock {\em Journal of Business \& Economic Statistics}, 38:340--349.

\bibitem[Barigozzi et~al., 2018]{barigozzi2018}
Barigozzi, M., Cho, H., and Fryzlewicz, P. (2018).
\newblock Simultaneous multiple change-point and factor analysis for
  high-dimensional time series.
\newblock {\em Journal of Econometrics}, 206(1):187--225.

\bibitem[{Basle Committee on Banking Supervision}, 1996]{BABS1996}
{Basle Committee on Banking Supervision} (1996).
\newblock {\em Supervisory Framework for the Use of `Backtesting' in
  Conjunction with the Internal Models Approach to Market Risk Capital
  Requirements}.
\newblock \url{https://www.bis.org/publ/bcbsc223.pdf}.

\bibitem[{Basle Committee on Banking Supervision}, 2010]{BABS193}
{Basle Committee on Banking Supervision} (2010).
\newblock {\em Revisions to the Basel II market risk framework}.
\newblock \url{https://www.bis.org/publ/bcbs193.pdf}.

\bibitem[Bauwens et~al., 2006]{bauwens2006}
Bauwens, L., Laurent, S., and Rombouts, J.~V. (2006).
\newblock Multivariate garch models: a survey.
\newblock {\em Journal of Applied Econometrics}, 21(1):79--109.

\bibitem[Berkes et~al., 2004]{berkes2004}
Berkes, I., Horv{\'a}th, L., and Kokoszka, P. (2004).
\newblock Testing for parameter constancy in garch(p, q) models.
\newblock {\em Statistics and Probability Letters}, 70(4):263--273.

\bibitem[Bollerslev, 1990]{bollerslev1990}
Bollerslev, T. (1990).
\newblock Modelling the coherence in short-run nominal exchange rates: a
  multivariate generalized arch model.
\newblock {\em The Review of Economics and Statistics}, pages 498--505.

\bibitem[Bollerslev et~al., 1988]{bollerslev1988}
Bollerslev, T., Engle, R.~F., and Wooldridge, J.~M. (1988).
\newblock A capital asset pricing model with time-varying covariances.
\newblock {\em Journal of Political Economy}, 96(1):116--131.

\bibitem[Boussama et~al., 2011]{boussama2011}
Boussama, F., Fuchs, F., and Stelzer, R. (2011).
\newblock Stationarity and geometric ergodicity of bekk multivariate garch
  models.
\newblock {\em Stochastic Processes and their Applications}, 121:2331--2360.

\bibitem[Cappiello et~al., 2006]{cappiello2006}
Cappiello, L., Engle, R.~F., and Sheppard, K. (2006).
\newblock Asymmetric dynamics in the correlations of global equity and bond
  returns.
\newblock {\em Journal of Financial Econometrics}, 4(4):537--572.

\bibitem[Chen and An, 1998]{chen1998}
Chen, M. and An, H.~Z. (1998).
\newblock A note on the stationarity and the existence of moments of the garch
  model.
\newblock {\em Statistica Sinica}, 8:505--510.

\bibitem[Cho, 2016]{cho2016}
Cho, H. (2016).
\newblock Change-point detection in panel data via double cusum statistic.
\newblock {\em Electronic Journal of Statistics}, 10(2):2000--2038.

\bibitem[Cho and Fryzlewicz, 2015]{cho2013}
Cho, H. and Fryzlewicz, P. (2015).
\newblock Multiple change-point detection for high-dimensional time series via
  sparsified binary segmentation.
\newblock {\em Journal of the Royal Statistical Society: Series B},
  77(2):475--507.

\bibitem[Cs{\"o}rg{\"o} and Horv{\'a}th, 1997]{csorgo1997}
Cs{\"o}rg{\"o}, M. and Horv{\'a}th, L. (1997).
\newblock {\em {Limit Theorems in Change-point Analysis}}, volume~18.
\newblock John Wiley \& Sons Inc.

\bibitem[Danielsson, 2002]{danielsson2002}
Danielsson, J. (2002).
\newblock The emperor has no clothes: Limits to risk modelling.
\newblock {\em Journal of Banking \& Finance}, 26(7):1273--1296.

\bibitem[De~Pooter and Van~Dijk, 2004]{de2004}
De~Pooter, M. and Van~Dijk, D. (2004).
\newblock Testing for changes in volatility in heteroskedastic time series -- a
  further examination.
\newblock Technical report, Econometric Institute.

\bibitem[Dette et~al., 2018]{dette2018}
Dette, H., Pan, G.~M., and Yang, Q. (2018).
\newblock Estimating a change point in a sequence of very high-dimensional
  covariance matrices.
\newblock {\em arXiv preprint, arXiv:1807.10797}.

\bibitem[Diebold and Inoue, 2001]{diebold1999}
Diebold, F. and Inoue, A. (2001).
\newblock Long range and regime switching.
\newblock {\em Journal of Econometrics}, 105(1):131--159.

\bibitem[Douc et~al., 2014]{douc2014}
Douc, R., Moulines, E., and Stoffer, D. (2014).
\newblock {\em {Nonlinear Time Series: Theory, Methods and Applications with R
  Examples}}.
\newblock Chapman and Hall/CRC.

\bibitem[Du et~al., 2011]{du2011}
Du, X., Cindy, L.~Y., and Hayes, D.~J. (2011).
\newblock Speculation and volatility spillover in the crude oil and
  agricultural commodity markets: A bayesian analysis.
\newblock {\em Energy Economics}, 33(3):497--503.

\bibitem[Duffie and Pan, 1997]{duffie1997}
Duffie, D. and Pan, J. (1997).
\newblock An overview of value at risk.
\newblock {\em The Journal of Derivatives}, 4(3):7--49.

\bibitem[Engle, 2002]{engle2002}
Engle, R. (2002).
\newblock Dynamic conditional correlation: A simple class of multivariate
  generalized autoregressive conditional heteroskedasticity models.
\newblock {\em Journal of Business \& Economic Statistics}, 20(3):339--350.

\bibitem[Engle and Kroner, 1995]{engle1995}
Engle, R.~F. and Kroner, K.~F. (1995).
\newblock Multivariate simultaneous generalized arch.
\newblock {\em Econometric Theory}, 11(01):122--150.

\bibitem[Engle and Manganelli, 2004]{engle2004caviar}
Engle, R.~F. and Manganelli, S. (2004).
\newblock {CAViaR: Conditional autoregressive value at risk by regression
  quantiles}.
\newblock {\em Journal of Business \& Economic Statistics}, 22(4):367--381.

\bibitem[{European Banking Authority}, 2012]{eba}
{European Banking Authority} (2012).
\newblock {\em Guidelines on {Stressed Value At Risk}}.
\newblock
  \url{https://www.eba.europa.eu/documents/10180/104547/EBA-BS-2012-78--GL-on-Stressed-VaR-.pdf}.

\bibitem[Ewing and Malik, 2013]{ewing2013}
Ewing, B.~T. and Malik, F. (2013).
\newblock Volatility transmission between gold and oil futures under structural
  breaks.
\newblock {\em International Review of Economics \& Finance}, 25:113--121.

\bibitem[Fan et~al., 2008]{fan2008}
Fan, J., Wang, M., and Yao, Q. (2008).
\newblock Modelling multivariate volatilities via conditionally uncorrelated
  components.
\newblock {\em Journal of the Royal Statistical Society: Series B},
  70(4):679--702.

\bibitem[Fryzlewicz and {Subba Rao}, 2011]{piotr2011}
Fryzlewicz, P. and {Subba Rao}, S. (2011).
\newblock Mixing properties of arch and time-varying arch processes.
\newblock {\em Bernoulli}, 17(1):320--346.

\bibitem[Fryzlewicz and {Subba Rao}, 2014]{piotr2013}
Fryzlewicz, P. and {Subba Rao}, S. (2014).
\newblock Multiple-change-point detection for auto-regressive conditional
  heteroscedastic processes.
\newblock {\em Journal of the Royal Statistical Society: Series B}, pages
  903--924.

\bibitem[Hamilton, 2003]{hamilton2003}
Hamilton, J.~D. (2003).
\newblock What is an oil shock?
\newblock {\em Journal of Econometrics}, 113(2):363--398.

\bibitem[Hansen and Lunde, 2005]{hansen2005}
Hansen, P.~R. and Lunde, A. (2005).
\newblock A forecast comparison of volatility models: does anything beat a
  garch(1, 1)?
\newblock {\em Journal of Applied Econometrics}, 20(7):873--889.

\bibitem[Horv{\'a}th and Hu{\v{s}}kov{\'a}, 2012]{horvath2012}
Horv{\'a}th, L. and Hu{\v{s}}kov{\'a}, M. (2012).
\newblock Change-point detection in panel data.
\newblock {\em Journal of Time Series Analysis}, 33(4):631--648.

\bibitem[J\"{a}ckel and Rebonato, 2001]{rebonato2001}
J\"{a}ckel, P. and Rebonato, R. (2001).
\newblock Valuing american options in the presence of user defined smiles and
  time-dependent volatility: Scenario analysis, model stress and lower bound
  pricing applications.
\newblock {\em Journal of Risk}, 4(1):35--61.

\bibitem[Jirak, 2015]{jirak2014}
Jirak, M. (2015).
\newblock Uniform change point tests in high dimension.
\newblock {\em The Annals of Statistics}, 43(6):2451--2483.

\bibitem[Ke et~al., 2016]{ke2016}
Ke, Y., Li, J., and Zhang, W. (2016).
\newblock Structure identification in panel data analysis.
\newblock {\em The Annals of Statistics}, 44(3):1193--1233.

\bibitem[Kokoszka and Leipus, 2000]{kokoszka2000}
Kokoszka, P. and Leipus, R. (2000).
\newblock Change-point estimation in arch models.
\newblock {\em Bernoulli}, 6(3):513--539.

\bibitem[Kokoszka and Teyssi\`{e}re, 2002]{kokoszka2002}
Kokoszka, P. and Teyssi\`{e}re, G. (2002).
\newblock Change-point detection in garch models: asymptotic and bootstrap
  tests.
\newblock Technical report, Universite Catholique de Louvain.

\bibitem[Korostelev, 1987]{korostelev1987}
Korostelev, A. (1987).
\newblock On minimax estimation of a discontinuous signal.
\newblock {\em Theory of Probability \& its Applications}, 32(4):727--730.

\bibitem[Kupiec, 1995]{kupiec1995}
Kupiec, P.~H. (1995).
\newblock Techniques for verifying the accuracy of risk measurement models.
\newblock {\em The Journal of Derivatives}, 3(2):73--84.

\bibitem[Ledoit and Wolf, 2012]{ledoit2012}
Ledoit, O. and Wolf, M. (2012).
\newblock Nonlinear shrinkage estimation of large-dimensional covariance
  matrices.
\newblock {\em The Annals of Statistics}, 40(2):1024--1060.

\bibitem[Lee et~al., 2003]{lee2003}
Lee, S., Tokutsu, Y., and Maekawa, K. (2003).
\newblock The residual cusum test for the constancy of parameters in garch(1,
  1) models.
\newblock Technical report, Seoul National University.

\bibitem[Li et~al., 2017]{li2017}
Li, J., Todorov, V., Tauchen, G., and Lin, H. (2017).
\newblock Rank tests at jump events.
\newblock {\em Journal of Business \& Economic Statistics}, pages 1--10.

\bibitem[Merlev{\`e}de et~al., 2009]{merlevede2009}
Merlev{\`e}de, F., Peligrad, M., and Rio, E. (2009).
\newblock Bernstein inequality and moderate deviations under strong mixing
  conditions.
\newblock In {\em High dimensional probability V: the Luminy volume}, pages
  273--292. Institute of Mathematical Statistics.

\bibitem[Mikosch and St\u{a}ric\u{a}, 2004]{mikosch2004}
Mikosch, T. and St\u{a}ric\u{a}, C. (2004).
\newblock Nonstationarities in financial time series, the long-range
  dependence, and the igarch effects.
\newblock {\em The Review of Economics and Statistics}, 86(1):378--390.

\bibitem[Persaud, 2000]{persaud2000}
Persaud, A. (2000).
\newblock Sending the herd off the cliff edge: the disturbing interaction
  between herding and market-sensitive risk management practices.
\newblock {\em The Journal of Risk Finance}, 2(1):59--65.

\bibitem[Pesaran and Timmermann, 2007]{pesaran2007}
Pesaran, M.~H. and Timmermann, A. (2007).
\newblock Selection of estimation window in the presence of breaks.
\newblock {\em Journal of Econometrics}, 137(1):134--161.

\bibitem[Ross, 1989]{ross1989}
Ross, S.~A. (1989).
\newblock Information and volatility: The no-arbitrage martingale approach to
  timing and resolution irrelevancy.
\newblock {\em The Journal of Finance}, 44(1):1--17.

\bibitem[Spokoiny, 2009]{spoikony2009}
Spokoiny, V. (2009).
\newblock Multiscale local change point detection with applications to
  value-at-risk.
\newblock {\em The Annals of Statistics}, 37:1405--1436.

\bibitem[Valentinyi-Endr{\'e}sz, 2004]{valentinyi2004}
Valentinyi-Endr{\'e}sz, M. (2004).
\newblock Structural breaks and financial risk management.
\newblock Technical report, Magyar Nemzeti Bank (Central Bank of Hungary).

\bibitem[Van~der Weide, 2002]{van2002}
Van~der Weide, R. (2002).
\newblock Go-garch: a multivariate generalized orthogonal garch model.
\newblock {\em Journal of Applied Econometrics}, 17(5):549--564.

\bibitem[Venkatraman, 1992]{venkatraman1992}
Venkatraman, E.~S. (1992).
\newblock Consistency results in multiple change-point problems.
\newblock {\em Technical Report No. 24, Department of Statistics, Stanford
  University}.

\bibitem[Vostrikova, 1981]{vostrikova1981}
Vostrikova, L.~J. (1981).
\newblock Detecting `disorder' in multidimensional random processes.
\newblock {\em Soviet Doklady Mathematics}, 24:55--59.

\bibitem[Vrontos et~al., 2003]{vrontos2003}
Vrontos, I.~D., Dellaportas, P., and Politis, D.~N. (2003).
\newblock A full-factor multivariate garch model.
\newblock {\em The Econometrics Journal}, 6(2):312--334.

\bibitem[Wang et~al., 2021]{wang2018}
Wang, D., Yu, Y., and Rinaldo, A. (2021).
\newblock Optimal covariance change point localization in high dimensions.
\newblock {\em Bernoulli}, 27:554--575.

\bibitem[Wang and Samworth, 2018]{wang2018s}
Wang, T. and Samworth, R.~J. (2018).
\newblock High dimensional change point estimation via sparse projection.
\newblock {\em Journal of the Royal Statistical Society: Series B}, 80:57--83.

\end{thebibliography}
}

\clearpage

\appendix
\section{Proofs of theoretical results}

\subsection*{Notation}

In addition to the notations introduced in the main text, 
$C_i, \ i=1, 2, \ldots$ are adopted
to denote fixed positive constants throughout the proofs,
which may represent different values from one usage to another.
We write $[\mathbf{a}]_i$ to denote the $i$-th element of a vector $\mathbf{a}$.
Also, $\mbf a \le \mbf b$ for two vectors $\mbf a, \mbf b \in \R^n$
indicates that $[\mbf a]_i \le [\mbf b]_i$ for all $i = 1, \ldots, n$. 
We write $\mbf 1$ to denote a vector of ones, whose dimension should be clear from the context.

\subsection{Preliminary results}

Recall that $v(t)$ denotes the index of the nearest change point among those satisfying $\eta_b < t$,
and we denote the location of such a change point by $\underline{\eta}(t)$, i.e.,
$\underline{\eta}(t) = \max\{\eta_b: \, 1 \le b \le B, \, \eta_b < t\} = \eta_{v(t)}$.
Similarly defined is $\bar{\eta}(t) = \min\{\eta_b: \, 1 \le b \le B, \, \eta_b \ge t\}$.
Also, recall the definitions of $\{\mbf r^b_t\}$ from above Proposition~\ref{prop:two}.

\begin{lem}
\label{lem:one}
Under Assumptions~\ref{eq:a0} and~\ref{eq:a2}--\ref{eq:a4}, it holds that for each $i$,
\begin{align*}
|r_{i, t}^2 - (r^{v(t)}_{i, t})^2| \le V_{i, t}
\end{align*}
where $\max_{1 \le i \le N} \E(V_{i, t}) \le c_1\rho_1^{t-\underline{\eta}(t)-1}$ with 
some fixed $\rho_1 \in (0, 1)$ and $c_1 > 0$.
Similarly, $|h_{i, t}^2 - (h^{v(t)}_{i, t})^2| \le W_{i, t}$
where $\max_{1 \le i \le N} \E(W_{i, t}) \le c_2\rho_2^{t-\underline{\eta}(t)-1}$ with 
some fixed $\rho_2 \in (0, 1)$ and $c_2>0$.
\end{lem}
\begin{proof}
The proof below is an extension of the proof given for Lemma 1 of \cite{piotr2013}.
For brevity, we omit the subscript $i$ from $r_{i, t}$ and other relevant notations
when there is no confusion.

The random recurrence equation for a GARCH($p, q$) process $\mbf r_t$ in~\eqref{eq:garch} is defined as below \citep{chen1998}:
\begin{align}
\mbf Y_t &= \mc A_t\mbf Y_{t-1} + \mbf B_t, \qquad \mbox{where} \label{eq:recur}
\\
\mbf Y_t &= (r_t^2, r_{t-1}^2, \ldots, r_{t-p+1}^2, h_t, h_{t-1}, \ldots, h_{t-q+1})^\top \nonumber
\\
\mbf B_t &= (\omega(t)\vep_t^2, 0, \ldots, 0, \omega(t), 0, \ldots, 0)^\top \nonumber
\\
\mc A_t &= \left(\begin{array}{cccccccc}
\alpha_1(t)\vep_t^2 & \cdots & \alpha_{p-1}(t)\vep_t^2 & \alpha_p(t)\vep_t^2 & \beta_1(t)\vep_t^2 & \cdots & \beta_{q-1}(t)\vep_t^2 & \beta_q(t)\vep_t^2 \\
1 & \cdot & 0 & 0 & 0 & \cdots & 0 & 0 \\
\vdots & \ddots & \vdots & \vdots & \vdots & \ddots & \vdots & \vdots \\
0 & \cdots & 1 & 0 & 0 & \cdots & 0 & 0 \\
\alpha_1(t) & \cdots & \alpha_{p-1}(t) & \alpha_p(t) & \beta_1(t) & \cdots & \beta_{q-1}(t) & \beta_q(t) \\
0 & \cdots & 0 & 0 & 1 & \cdots & 0 & 0 \\
\vdots & \ddots & \vdots & \vdots & \vdots & \ddots & \vdots & \vdots \\
0 & \cdots & 0 & 0 & 0 & \cdots & 1 & 0
\end{array}\right), \nonumber
\end{align}
where both $\mc A_t$ and $\mbf B_t$ are independent over $t$ 
and have all their elements non-negative. After iterating the equation $k$ steps backwards, we obtain
\begin{align*}
\mbf Y_t = \prod_{j=0}^{k-1}\mc A_{t-j}\mbf Y_{t-k} + \sum_{j=0}^{k-1} \prod_{l=1}^j\mc A_{t-l+1}\mbf B_{t-j} =: I + II
\end{align*}
(where $\prod_{l=1}^0\mc A_{t-l+1} = 1$).
As for $r^{v(t)}_t$,
\begin{align*}
\wt{\mbf Y}_t = \prod_{j=0}^{k-1}\wt{\mc A}_{t-j}\wt{\mbf Y}_{t-k} + \sum_{j=0}^{k-1} \prod_{l=1}^j\wt{\mc A}_{t-l+1}\wt{\mbf B}_{t-j} =: III+IV,
\end{align*}
where $\wt{\mbf Y}_t = ((r_t^{v(t)})^2, (r_{t-1}^{v(t)})^2, \ldots, (r_{t-p+1}^{v(t)})^2, h^{v(t)}_t, h^{v(t)}_{t-1}, \ldots, h^{v(t)}_{t-q+1})^\top$
and $\wt{\mc A}_t$ and $\wt{\mbf B}_t$ are defined analogously as $\mc A_t$ and $\mbf B_t$, respectively,
with the GARCH parameters associated with the stationary segment $[\underline{\eta}(t)+1, \bar{\eta}(t)]$.
By its construction, $r^{v(t)}_t$ shares the same GARCH parameters and innovation sequence as $r_t$ over $[\underline{\eta}(t)+1, \bar{\eta}(t)]$.
Therefore setting $k = t-\underline{\eta}(t)-1$ leads to $II = IV$.
As $r_t^2-(r^{v(t)}_t)^2$ is equal to the first element of $\mbf Y_t-\wt{\mbf Y}_t$, 
we have $|r_t^2-(r^{v(t)}_t)^2| \le V_t = [\prod_{j=0}^{k-1}\mc A_{t-j}\mbf Y_{t-k}]_1 + [\prod_{j=0}^{k-1}\wt{\mc A}_{t-j}\wt{\mbf Y}_{t-k}]_1$.
Since $\mc A_{t-j}, \, j=0, \ldots, k-1$ and $\mbf Y_{t-k}$ ($\wt{\mc A}_{t-j}, \, j=0, \ldots, k-1$ and $\wt{\mbf Y}_{t-k}$) are independent,
and $\mc A^* = \E(\mc A_t) = \E(\wt{\mc A}_t)$ remains the same over $t\in[\underline{\eta}(t)+1, \bar{\eta}(t)]$,
$\E(V_t) = \left[(\mc A^*)^k\E(\mbf Y_{t-k})\right]_1 + \left[(\mc A^*)^k\E(\wt{\mbf Y}_{t-k})\right]_1$.
Further, as the entries of $\mc A^*$ are non-negative, if $\mbf x$ and $\mbf y$ are $(p+q)$-vectors satisfying $\mbf 0 \le \mbf x \le \mbf y$,
we have $[(\mc A^*)^k\mbf x]_1 \le [(\mc A^*)^k\mbf y]_1$.
Under \ref{eq:a2}--\ref{eq:a3}, both $\E(\mbf Y_{t-k})$ and $\E(\wt{\mbf Y}_{t-k})$ are bounded from above by $\epsilon_2^{-1}\Xi_1\mbf 1$
and hence $\E(V_t) \le 2\epsilon_2^{-1}\Xi_1[(\mc A^*)^k\mbf 1]_1$.
Under \ref{eq:a3}, it is clear that $(\mc A^*)^{p+q}\mbf 1 \le C_1(1-\epsilon_2)\mbf 1$
for some $C_1$. Therefore, we can find for some fixed constant 
$c_1>0$ and $\rho_1\in(0, 1)$ such that
\begin{align*}
\E(V_t) \le 2\epsilon_2^{-1}\Xi_1C_1(1-\epsilon_2)^{\lfloor k/(p+q) \rfloor} 
\le c_1 \rho_1^{t-\underline{\eta}(t)-1}.
\end{align*}
Since $\rho_1$ depends on $p$, $q$, $\epsilon_2$ and $\Xi_1$ only,
the same arguments hold for all $i=1, \ldots, N$ and hence the bound on $\E(V_{i, t})$ holds uniformly in $i$ for an appropriately chosen $\rho_1$.
As for $W_t$, note that $h_t^2-(h^{v(t)}_t)^2$ is equal to the $(p+1)$-th element of $\mbf Y_t-\wt{\mbf Y}_t$.
Analogous arguments show that
there exists $c_2 > 0$, $\rho_2\in(0, 1)$ 
such that $\E(W_{i, t}) \le c_2\rho_2^{t-\underline{\eta}(t)-1}$.
\end{proof}

\subsection{Proof of Proposition~\ref{prop:one}}
\label{sec:mixing}

\cite{piotr2011} showed that time-varying, univariate ARCH processes are strong mixing at a geometric rate.
As the techniques used by the authors for the proof of their Theorem 3.1
are applicable to other Markov processes,
we extend their results by deriving the mixing rate for time-varying bivariate GARCH processes.
The arguments adopted below for showing 
the mixingness of $\mbf r_{12, t} = (r_{1, t}, r_{2, t})^\top$,
are analogously applied to showing that of any pair of $r_{i, t}$ and $r_{i', t}, \, 1 \le i < i' \le N$.

Let $\mbf X_t = (X_{1, t}, X_{2, t})^\top$ with $X_{i, t} = r_{i, t}^2$
and $\mbf Z_t = (Z_{1, t}, Z_{2, t})^\top$ with $Z_{i, t} = \vep_{i, t}^2$ for $i=1, 2$,
and $R = p \vee q$.
Recall \ref{eq:a4}, which is met by the density function of $\mbf Z_t$.
Denote the $\sigma$-algebra generated by $\mbf X_t, \ldots, \mbf X_{t+k}$ by
$\mc F_{t+k}^t = \sigma(\mbf X_t, \ldots, \mbf X_{t+k})$.
Also, for any $k>0$, let $\ubX_t^{t-k} = (\mbf X_t^\top, \ldots, \mbf X_{t-k}^\top)^\top$ and $\ubX_t^{t-k} = 0$ for $k \le 0$.
We similarly define $\ubZ_t^{t-k} = (\mbf Z_t^\top, \ldots, \mbf Z_{t-k}^\top)^\top$,
$\ubX_{i, t}^{t-k} = (X_{i, t}, \ldots, X_{i, t-k})^\top$,
$\ubh_{i, t}^{t-k} = (h_{i, t}, \ldots, h_{i, t-k})^\top$ and
$\ubZ_{i, t}^{t-k} = (Z_{i, t}, \ldots, Z_{i, t-k})^\top$.

Recall the random recurrence equation in the proof of Lemma~\ref{lem:one}.
We define $\mbf Y_{i, t}$, $\mc A_{i, t}$ and $\mbf B_{i, t}$ 
similarly as $\mbf Y_t$, $\mc A_t$ and $\mbf B_t$ of (\ref{eq:recur})
with respect to the innovations $Z_{i, t}$ and the time-varying GARCH parameters associated with $r_{i, t}$ for $i=1, 2$,
except that they are now
defined with $\alpha_{i, j}(t), \beta_{i, k}(t), \, j, k = 1, \ldots, R$
(with $\alpha_{i, j}(t) = 0, \, j > p$ and $\beta_{i, k}(t) = 0, \, k > q$
whenever relevant) such that their dimensions are
$2R$ ($\mbf Y_{i, t}$, $\mbf B_{i, t}$) and $(2R)\times(2R)$ ($\mc A_{i, t}$)
with appropriately padded zeros.
Finally, let $\mbf Y_t = (\mbf Y_{1, t}^\top, \mbf Y_{2, t}^\top)^\top$, 
$\mbf B_t = (\mbf B_{1, t}^\top, \mbf B_{2, t}^\top)^\top$ and $\mc A_t$ be the 
block diagonal matrix with $\mc A_{1, t}$ and $\mc A_{2, t}$ in its diagonal.
Then it can be shown that
\begin{align}
\label{eq:prop:one:one}
\mbf Y_{t+k+s} = \mbf B_{t+k+s} + \prod_{l=0}^{k+s-1}\mc A_{t+k+s-l}\mbf Y_t +
\sum_{l=0}^{k+s-2} \prod_{m=0}^l \mc A_{t+k+s-m}\mbf B_{t+k+s-l-1}.
\end{align}
By definition, $\mc A_t$ and $\mbf B_t$ are independent random matrices and vectors over $t$, respectively.
Let $\mc A^*_t = \E(\mc A_t)$ and $\mc A^*_{i, t} = \E(\mc A_{i, t})$.
For the proof of Proposition~\ref{prop:one}, we need the following Lemmas.

\vspace{5pt}
\begin{lem}
\label{prop:lem:one} For $s \ge 0$ and $k \ge R$, we have
\begin{align*}
X_{i, t+k+s} = Z_{i, t+k+s}\{\mc P^i_{s, k, t}(\ubV_{i, t+k+s-1}^{t+k}, \ubZ_{i, t+k-1}^{t+1}) + \mc Q^i_{s, k, t}(\ubV_{i, t}^{t-R+1}, \ubZ_{i, t+k-1}^{t+1})\},
\end{align*}
where $\ubV_{i, t}^{t - l} = ((\ubX_{i, t}^{t-l})^\top, (\ubh_{i, t}^{t-l})^\top)^\top$
for $l > 0$, and $\ubV_{i, t}^{t - l} = 0$ otherwise.

(i) When $s=0$: we have $\ubV_{i, t+k+s-1}^{t+k} = 0$ and
\begin{align*}
\mc P^i_{0, k, t}(\ubZ_{i, t+k-1}^{t+1}) &= \omega_i(t+k) + [\mc A^*_{i, t+k}\sum_{l=0}^{k-2}\prod_{m=1}^l\mc A_{i, t+k-m}\mbf B_{i, t+k-l-1}]_1,
\\
\mc Q^i_{0, k, t}(\ubV_{i, t}^{t-R+1}, \ubZ_{i, t+k-1}^{t+1}) &= [\mc A^*_{i, t+k}\prod_{l=1}^{k-1}\mc A_{i, t+k-l}\mbf Y_{i, t}]_1.
\end{align*}

(ii) When $1 \le s < R$:

\begin{align*}
& \mc P^i_{s, k, t}(\ubV_{i, t+k+s-1}^{t+k}, \ubZ_{i, t+k-1}^{t+1}) 
= \omega_i(t+k+s) + 
\\
& \qquad \qquad \sum_{j=1}^s\{\alpha_{i, j}(t+k+s)X_{i, t+k+s-j} + \beta_{i, j}(t+k+s)h_{i, t+k+s-j}\} +
\\
& \qquad \qquad \sum_{j=s+1}^R \{\alpha_{i, j}(t+k+s)Z_{i, t+k+s-j} + \beta_{i, j}(t+k+s)\}\{\omega_i(t+k+s-j) +
\\
& \qquad \qquad \qquad [\mc A^*_{i, t+k+s-j}\sum_{l=0}^{k+s-j-2}\prod_{m=1}^l\mc A_{i, t+k+s-j-m}\mbf B_{i, t+s-j-l-1}]_1\},
\\
& \mc Q^i_{s, k, t}(\ubV_{i, t}^{t-R+1}, \ubZ_{i, t+k-1}^{t+1}) = \sum_{j=s+1}^R \{\alpha_{i, j}(t+k+s)Z_{i, t+k+s-j} + \beta_{i, j}(t+k+s)\} \times
\\
& \qquad \qquad \qquad \qquad \qquad \qquad
[\mc A^*_{i, t+k+s-j}\prod_{l=1}^{k+s-j-1}\mc A_{i, t+k+s-j-l}\mbf Y_{i, t}]_1
\end{align*}

(iii) When $s \ge R$:
\begin{align*}
\mc P^i_{s, k, t}(\ubV_{i, t+k+s-1}^{t+k}, \ubZ_{i, t+k-1}^{t+1}) &= \omega_i(t+k+s) + \sum_{j=1}^R\{\alpha_{i, j}(t+k+s)X_{i, t+k+s-j} + \beta_{i, j}(t+k+s)h_{i, t+k+s-j}\},
\\
\mc Q^i_{s, k, t}(\ubV_{i, t}^{t-R+1}, \ubZ_{i, t+k-1}^{t+1}) &= 0.
\end{align*}
\end{lem}
\begin{proof}
The proof of lemma follows trivially from the representation in (\ref{eq:prop:one:one}).
\end{proof}
\vspace{5pt}

From Lemma~\ref{prop:lem:one}, the conditional density of $\mbf X_{t+k+s}$ given 
$\ubV_{t+k+s-1}^{t+k} = ((\ubV_{1, t+k+s-1}^{t+k})^\top, 
(\ubV_{2, t+k+s-1}^{t+k})^\top)^\top$,
$\ubZ_{t+k-1}^{t+1} = ((\mbf Z_{1, t+k-1}^{t+1})^\top, 
(\mbf Z_{2, t+k-1}^{t+1})^\top)^\top$ and
$\ubV_t^{t-R+1} = ((\ubV_{1, t}^{t-R+1})^\top, 
(\ubV_{2, t}^{t-R+1})^\top)^\top$,
is a function of 
\begin{align*}
Z_{i, t+k+s} = X_{i, t+k+s}\{\mc P^i_{s, k, t}(\ubV_{i, t+k+s-1}^{t+k}, \ubZ_{i, t+k-1}^{t+1}) + \mc Q^i_{s, k, t}(\ubV_{i, t}^{t-R+1}, \ubZ_{i, t+k-1}^{t+1})\}^{-1}, \, i=1, 2.
\end{align*}
That is, denoting the conditional density by $f_{s, k, t}$,
\begin{align}
& f_{s, k, t}(\mbf y | \ubV_{t+k+s-1}^{t+k}, \ubZ_{t+k-1}^{t+1}, \ubV_t^{t-R+1}) = \prod_{i=1}^2\frac{1}{\mc P^i_{s, k, t}(\ubV_{i, t+k+s-1}^{t+k}, \ubZ_{i, t+k-1}^{t+1}) + 
\mc Q^i_{s, k, t}(\ubV_{i, t}^{t-R+1}, \ubZ_{i, t+k-1}^{t+1})} \times
\nonumber
\\
& \qquad \qquad f_{1, 2}\left(\frac{y_1}
{\mc P^1_{s, k, t}(\ubV_{1, t+k+s-1}^{t+k}, \ubZ_{1, t+k-1}^{t+1}) + 
\mc Q^1_{s, k, t}(\ubV_{1, t}^{t-R+1}, \ubZ_{1, t+k-1}^{t+1})}, \right.
\nonumber \\
& \qquad \qquad \qquad \qquad \qquad \qquad
\left. \frac{y_2}
{\mc P^2_{s, k, t}(\ubV_{2, t+k+s-1}^{t+k}, \ubZ_{2, t+k-1}^{t+1}) + 
\mc Q^2_{s, k, t}(\ubV_{2, t}^{t-R+1}, \ubZ_{2, t+k-1}^{t+1})}\right).
\label{eq:prop:one:two}
\end{align}

The following lemma is adapted from Proposition~2.1 of \cite{piotr2011}.

\vspace{5pt}
\begin{lem}
\label{prop:lem:two} 
Provided its existence, denote the conditional density of $\ubX_{t+k+R_2}^{t+k}$ 
given $\ubV_{t}^{t-R_1}$,
by $f_{\ubX_{t+k+R_2}^{t+k} \vert \ubV_{t}^{t-R_1}}$.
For any $\boldsymbol{\gamma} = (\gamma_1, \ldots, \gamma_{4R_1+4})^\top \in (\R^+)^{4R_1+4}$,
define the set
$E = \{\omega:\, \ubV_t^{t-R_1}(\omega) \in \mc E\}$
where $\mc E = \{(\nu_1, \ldots, \nu_{4R_1+4})^\top:\, 
|\nu_l| \le \gamma_l \mbox{ for all } l=1, \ldots, 4R_1+4\}$.
Let $\ubW$ denote a random vector independent of $\ubV_t^{t-R_1}$
and denote its density function by $f_{\ubW}$.
Recalling the definition of $f_{s, k, t}$ in \eqref{eq:prop:one:two},
define
\begin{align*}
\mc D_{0, k, t}(\mbf y_0 | \mbf z, \mbf v) =& |f_{0, k, t}(\mbf y_0|\mbf z, \mbf v) -
f_{0, k, t}(\mbf y_0|\mbf v, \mbf 0)|,
\\
\mc D_{s, k, t}(\mbf y_s | \mbf v_{s-1}, \mbf z, \mbf v) =& 
|f_{s, k, t}(\mbf y_s|\mbf v_{s-1}, \mbf z, \mbf v) - f_{s, k, t}(\mbf y_s|\mbf v_{s-1}, \mbf z, \mbf 0)|,
\end{align*}
where $\mbf v_{s-1}$, $\mbf z$ and $\mbf v$ are vectors of appropriate dimensions.

Then, for any $R_1, R_2 \ge 0$, the followings hold:

(i) 
\begin{align*}
& \sup_{G \in \mc F_{t+k+R_2}^{t+k}, \, H \in \mc F_t^{t-R_1}} |\pr(G \cap H) - \pr(G)\pr(H)|
\\
\le&
2\sup_{\mbf v \in \mc E}\int_{\R^{2R_2+2}}\left\vert f_{\ubX_{t+k+R_2}^{t+k} \vert 
\ubV_{t}^{t-R_1}}(\mbf y|\mbf v)-
f_{\ubX_{t+k+R_2}^{t+k} \vert \ubV_{t}^{t-R_1}}(\mbf y|\mbf 0)\right\vert d{\mbf y} + 4\pr(E^c).
\end{align*}

(ii) 
\begin{align*}
& \sup_{G \in \mc F_{t+k+R_2}^{t+k}, \, H \in \mc F_t^{t-R_1}} 
|\pr(G \cap H) - \pr(G)\pr(H)|
\\
\le&
2\sum_{s=0}^{R_2}\sup_{\mbf v \in \mc E}\E_{\ubW}\left\{\sup_{\mbf v_{s-1}\in\R^{2s}} \int_{\R^2}\mc D_{s, k, t}(\mbf y_s|\mbf v_{s-1}, \ubW, \mbf v)d\mbf y_s\right\}
+ 4\pr(E^c).
\end{align*}
\end{lem}
\begin{proof}
The proof given in \cite{piotr2011} for their Proposition~2.1 is readily applicable to show the above, since
$\ubh^{t-R_1}_t \in \sigma(\ubX^{t-R_1}_t)$ and $\ubh^{t+k}_{t+k+s-1} \in \sigma(\ubX^{t+k}_{t+k+s-1})$.
\end{proof}
\vspace{5pt}

\begin{lem}
\label{prop:lem:three}
For any $\boldsymbol{\gamma} \in (\R^+)^{4R}$, we have
\begin{align}
& \sup_{G \in \mc F_{\infty}^{t+k}, \, H \in \mc F_t^{-\infty}} |\pr(G \cap H) - \pr(G)\pr(H)|
\le
2\sum_{s=0}^{R-1} \sup_{\mbf v \in \mc E} \E_{\ubW}\left\{\sup_{\mbf v_{s-1}\in\R^{2s}}
\int_{\R^2}\mc D_{s, k, t}(\mbf y_s|\mbf v_{s-1}, \ubW, \mbf v)d\mbf y_s\right\}
\nonumber \\
&
\qquad + 4\left[\sum_{i=1}^2\sum_{l=1}^R\{\pr(X_{i, t-l} \ge \gamma_{2R(i-1)+l}) + \pr(h_{i, t-l} \ge \gamma_{2R(i-1)+R+l})\}\right]
\label{prop:lem:three:eq}.
\end{align}
\end{lem}
\begin{proof}
It holds that
\begin{align*}
\sigma\{\mbf X_{t+k+R'}, \ldots, \mbf X_{t+k}\} = \sigma\{\mbf Z_{t+k+R'}, \ldots, \mbf Z_{t+k+R}, \mbf X_{t+k+R-1}, \ldots, \mbf X_{t+k}\}
\end{align*}
for any $R' \ge R = p \vee q$. 
Hence
\begin{align*}
\sup_{G \in \mc F_{\infty}^{t+k}, \, H \in \mc F_t^{-\infty}} |\pr(G \cap H) - \pr(G)\pr(H)|
= \sup_{G \in \mc F_{t+k+R-1}^{t+k}, \, H \in \mc F_t^{t-R+1}} |\pr(G \cap H) - \pr(G)\pr(H)|,
\end{align*}
and applying Lemma~\ref{prop:lem:two} with $R_1 = R_2 = R$ completes the proof.
\end{proof}
\vspace{5pt}

\begin{lem}
\label{prop:lem:five}
If \ref{eq:a4} holds, then for any positive $A_i$ and $B_i$, $i=1, 2$, we have
\begin{align*}
& \int_{\R^2} \left\vert \prod_{i=1}^2\frac{1}{A_i+B_i}f_{1, 2}\left(\frac{u}{A_1+B_1}, \frac{v}{A_2+B_2}\right) -
\prod_{i=1}^2\frac{1}{A_i}f_{1, 2}\left(\frac{u}{A_1}, \frac{v}{A_2}\right)\right\vert dudv
\\
& \qquad \qquad \qquad \qquad \le K\left(\frac{B_1}{A_1+B_1}+\frac{B_1}{A_1}+\frac{B_2}{A_2+B_2}+\frac{B_2}{A_2}\right).
\end{align*}
\end{lem}
\begin{proof}
Observe that the LHS of the above inequality is bounded from the above by
\begin{align*}
& \int_{\R^2} \prod_{i=1}^2\frac{1}{A_i+B_i}\left\vert
f_{1, 2}\left(\frac{u}{A_1+B_1}, \frac{v}{A_2+B_2}\right) -  f_{1, 2}\left(\frac{u}{A_1}, \frac{v}{A_2+B_2}\right)\right\vert dudv
\\
+& \int_{\R^2} \frac{1}{A_2+B_2}\left\vert
\left(\frac{1}{A_1+B_1}-\frac{1}{A_1}\right)f_{1, 2}\left(\frac{u}{A_1}, \frac{v}{A_2+B_2}\right)\right\vert dudv
\\
+& \int_{\R^2} \frac{1}{A_1(A_2+B_2)}\left\vert
f_{1, 2}\left(\frac{u}{A_1}, \frac{v}{A_2+B_2}\right) - f_{1, 2}\left(\frac{u}{A_1}, \frac{v}{A_2}\right)\right\vert dudv
\\
+& \int_{\R^2} \frac{1}{A_1}\left\vert
\left(\frac{1}{A_2+B_2}-\frac{1}{A_2}\right)f_{1, 2}\left(\frac{u}{A_1}, \frac{v}{A_2}\right)\right\vert dudv
=: I + II + III + IV.
\end{align*}
To bound $I$, changing the variables with $u/(A_1+B_1) = u'$ and $v/(A_2+B_2) = v'$,
\begin{align*}
I = \int_{\R^2} \left\vert f_{1, 2}(u', v') -  f_{1, 2}\left(u'\left(1+\frac{B_1}{A_1}\right), v'\right)\right\vert du'dv' \le K\frac{B_1}{A_1}.
\end{align*}
It is trivial to show that $II \le B_1/(A_1+B_1)$,
and the same arguments apply to bound $III$ and $IV$, which completes the proof.
\end{proof}
\vspace{5pt}

\begin{lem}
\label{prop:lem:four}
Under \ref{eq:a2}--\ref{eq:a4}, for all $\mbf v \in \mc E$, we have
\begin{align}
\sum_{s=0}^{R-1} \E_{\ubZ_{t+k-1}^{t+1}}\left\{\sup_{\mbf v_{s-1}\in\R^{2s}}
\int_{\R^2}\mc D_{s, k, t}(\mbf y_s|\mbf v_{s-1}, \ubZ_{t+k-1}^{t+1}, \mbf v)d\mbf y_s\right\}
&\le \frac{C_2\sum_{i=1}^2\E\{\mc Q^i_{s, k, t}(\mbf v_i, \ubZ_{i, t+k-1}^{t+1})\}}
{\min_{1 \le i \le N}\inf_{t\in\Z}\omega_i(t)} 
\nonumber \\
&\le C_3(1-\epsilon_3)^k\sum_{l=1}^{4R}\gamma_{l},
\label{prop:lem:four:ineq}
\end{align}
where $\mbf v_i, \, i = 1, 2$ denote the sub-vectors of $\mbf v$ of the equal length,
and $\epsilon_3\in(0, \epsilon_2)$ for $\epsilon_2$ defined in \ref{eq:a3}.
\end{lem}
\begin{proof}
When $\mbf v_i = \mbf 0$, we have $\mc Q^i_{s, k, t}(\mbf 0, \ubZ_{i, t+k-1}^{t+1}) = 0$ and
\begin{align*}
f_{s, k, t}(\mbf y_s|\mbf v_{s-1}, \ubZ_{t+k-1}^{t+1}, \mbf 0) = 
\prod_{i=1}^2\frac{1}{\mc P^i_{s, k, t}(\mbf v_{i, s-1}, \ubZ_{i, t+k-1}^{t+1})}
f_{1, 2}\left(\frac{y_{1, s}}{\mc P^1_{s, k, t}(\mbf v_{1, s-1}, \ubZ_{1, t+k-1}^{t+1})},
\frac{y_{2, s}}{\mc P^2_{s, k, t}(\mbf v_{2, s-1}, \ubZ_{2, t+k-1}^{t+1})}\right).
\end{align*}
Recalling (\ref{eq:prop:one:two}), we have
\begin{align*}
&& \mc D_{s, k, t}(\mbf y_s|\mbf v_{s-1}, \ubZ_{t+k-1}^{t+1}, \mbf v) = \left\vert
\prod_{i=1}^2\frac{1}{\mc P^i_{s, k, t}(\mbf v_{i, s-1}, \ubZ_{i, t+k-1}^{t+1}) 
+ \mc Q^i_{s, k, t}(\mbf v_i, \ubZ_{i, t+k-1}^{t+1})} \times \right.
\\
&& f_{1, 2}\left(\frac{y_{1, s}}
{\mc P^1_{s, k, t}(\mbf v_{1, s-1}, \ubZ_{1, t+k-1}^{t+1}) + 
\mc Q^1_{s, k, t}(\mbf v_1, \ubZ_{1, t+k-1}^{t+1})},
\frac{y_{2, s}}{\mc P^2_{s, k, t}(\mbf v_{2, s-1}, \ubZ_{2, t+k-1}^{t+1}) + 
\mc Q^2_{s, k, t}(\mbf v_2, \ubZ_{1, t+k-1}^{t+1})}\right)
\\
&& \qquad \left. - \prod_{i=1}^2\frac{1}
{\mc P^i_{s, k, t}(\mbf v_{i, s-1}, \ubZ_{1, t+k-1}^{t+1})}
f_{1, 2}\left(\frac{y_{1, s}}
{\mc P^1_{s, k, t}(\mbf v_{1, s-1}, \ubZ_{1, t+k-1}^{t+1})},
\frac{y_{2, s}}{\mc P^2_{s, k, t}(\mbf v_{2, s-1}, \ubZ_{1, t+k-1}^{t+1})}\right)\right\vert.
\end{align*}
Applying Lemma~\ref{prop:lem:five},
\begin{align*}
&& \int_{\R^2}\mc D_{s, k, t}(\mbf y_s|\mbf v_{s-1}, \ubZ_{t+k-1}^{t+1}, \mbf v)d\mbf y_s \le
\\
&& K\sum_{i=1}^2\left\{\frac{\mc Q^i_{s, k, t}(\mbf v_i, \ubZ_{i, t+k-1}^{t+1})}
{\mc P^i_{s, k, t}(\mbf v_{i, s-1}, \ubZ_{i, t+k-1}^{t+1})} +
\frac{\mc Q^i_{s, k, t}(\mbf v_i, \ubZ_{i, t+k-1}^{t+1})}
{\mc P^i_{s, k, t}(\mbf v_{i, s-1}, \ubZ_{i, t+k-1}^{t+1}) +
\mc Q^i_{s, k, t}(\mbf v_i, \ubZ_{i, t+k-1}^{t+1})}\right\}
\end{align*}
and hence
\begin{align}
\E_{\ubZ_{t+k-1}^{t+1}}\left\{\sup_{\mbf v_{s-1}\in\R^{2s}}\int_{\R^2}
\mc D_{s, k, t}(\mbf y_s|\mbf v_{s-1}, \ubZ_{t+k-1}^{t+1}, \mbf v)d\mbf y_s\right\}
\le& 2K\frac{\sum_{i=1}^2\E_{\ubZ_{t+k-1}^{t+1}}\{
\mc Q^i_{s, k, t}(\mbf v_i, \ubZ_{i, t+k-1}^{t+1})\}}
{\min_{1 \le i \le N}\inf_{t\in\Z}\omega_i(t)}
\nonumber
\\
=& 2K\frac{\sum_{i=1}^2\mc Q^i_{s, k, t}(\mbf v_i, \mbf 1)}
{\min_{1 \le i \le N}\inf_{t\in\Z}\omega_i(t)}.
\label{prop:lem:four:ineq:two}
\end{align}
Let $\bar{\gamma} = \max_{1 \le l \le 4R} \gamma_l$.
From Lemmas \ref{lem:one} and \ref{prop:lem:one},
the following holds for any $\mbf v \in \mc E$:
\begin{align*}
& \mc Q^i_{s, k, t}(\mbf 1, \mbf v) =
\sum_{j=s+1}^R \{\alpha_{i, j}(t+k+s) + \beta_{i, j}(t+k+s)\}
\left[\prod_{l=0}^{k+s-j-1}\mc A^*_{i, t+k+s-j-l}\mbf v\right]_1
\\
\le& \bar{\gamma}\sum_{j=s+1}^R \{\alpha_{i, j}(t+k+s) + \beta_{i, j}(t+k+s)\}
\left[\prod_{l=0}^{k+s-j-1}\mc A^*_{i, t+k+s-j-l}\mbf 1\right]_1
\\
\le& \bar{\gamma}(1-\epsilon_2)\sum_{j=s+1}^R(1-\epsilon_2)^{\lfloor(k+s-j)/(2R)\rfloor}
\le C_4(1-\epsilon_3)^k\sum_{l=1}^{4R}\gamma_{l}
\end{align*}
for some $C_4>0$ and $\epsilon_3\in(0, \epsilon_2)$ which,
when plugged into (\ref{prop:lem:four:ineq:two}), yields (\ref{prop:lem:four:ineq}).
\end{proof}
\vspace{5pt}

Now we are fully equipped to prove Proposition~\ref{prop:one}.
Lemma~\ref{prop:lem:four} provides the upper bound on the first term in (\ref{prop:lem:three:eq}).
Therefore it only remains to bound the probability terms in (\ref{prop:lem:three:eq}).
Using Markov's inequality,
\begin{align*}
\pr(X_{i, t-l} \ge \gamma) \le \gamma^{-1}\E(X_{i, t-l}) \le \gamma^{-1}\sup_{t\in\Z}
\frac{\omega_i(t)}{1-\sum_{j=1}^R\{\alpha_{i, j}(t)+\beta_{i, j}(t)\}} 
< \frac{C_5}{\gamma}
\end{align*}
for some fixed $C_5>0$, and $\pr(h_{i, t-l} \ge \gamma)$ is similarly bounded.
Thus,
\begin{align*}
\sup_{G \in \mc F_{\infty}^{t+k}, \, H \in \mc F_t^{-\infty}} |\pr(G \cap H) - \pr(G)\pr(H)| \le C_3(1-\epsilon_3)^k\sum_{l=1}^{4R}\gamma_l + 2C_5\sum_{l=1}^{4R}\frac{1}{\gamma_l}.
\end{align*}
Setting $\gamma_l = (1-\epsilon_3)^{-k/2}$ for all $l=1, \ldots, 4R$, we obtain
\begin{align*}
\sup_{t\in\Z}\sup_{G \in \mc F_{\infty}^{t+k}, \, H \in \mc F_t^{-\infty}} |\pr(G \cap H) - \pr(G)\pr(H)| \le M\alpha^k
\end{align*}
for some $\alpha \in [\sqrt{1-\epsilon_3}, 1)$ and fixed $M>0$.
Since $\alpha$ does not depend on the index of $\mbf r_{12}$, the same arguments apply to any $\mbf r_{ii', t}, \, 1 \le i < i' \le N$
and hence the proposition is proved. \hfill $\square$

\subsection{Proof of Proposition~\ref{prop:two}}
\label{pf:lem:one}

By its definition, $g_0$ is bounded and
and is Lipschitz continuous in its squared arguments,
i.e.\ there exist $\bar{g}, C_g \in (0, \infty)$ such that
\begin{align}
\label{assum:a5}
|g_0| \le \bar{g} \quad \text{and} \quad
|g_0(z_0, \ldots, z_{p+q}) - g_0(z_0', \ldots, z_{p+q}')| \le C_g\sum_{k=0}^{p+q}|z_k^2-(z_k')^2|.
\end{align}

Part (i) follows trivially from the definition of $\wt{U}^{v(t)}_{i, t}$ and $\wt{U}^{v(t)}_{ii', t}$.
More specifically, since $r^{v(t)}_{i, t}$ `jumps' from one stationary process $r^b_{i, t}$ to another $r^{b+1}_{i, t}$
without any boundary effect, we have the expectations of 
$(\wt{U}^{v(t)}_{i, t})^2$ and $\wt{U}^{v(t)}_{ii', t}$ exactly piecewise constant with their change points
coinciding with $\mc B$.
For the proof of (ii), we adopt the arguments similar to those used in the proof of Lemma 4 in \cite{piotr2013}.

Recall that $v(t)$ denotes the index of the nearest change point among those satisfying $\eta_b < t$,
and the definition of $z_{j, t}$ in the context of time series segmentation,
namely $z_{j, t} = U_{i, t}^2 - \wt{g}_{i, t}$
or $z_{j, t} = U_{ii', t} - \wt{g}_{ii', t}$ for some $i, i'\in\{1, \ldots, N\}$.
Note that the claim of the proposition follows when $\pr(\mc E_1 \cap \mc E_2) \to 1$, where
\begin{align*}
\mc E_1 =& \left\{\max_{1 \le i \le N}\max_{1 \le s < e \le T} \frac{1}{\sqrt{e-s+1}}\left\vert
\sum_{t=s}^e (U_{i, t}^2 - \wt{g}_{i, t}) \right\vert \le c'\sqrt{\log(T)}  \right\}, \quad \mbox{and}
\\
\mc E_2 =& \left\{\max_{1 \le i < i' \le N}\max_{1 \le s < e \le T} \frac{1}{\sqrt{e-s+1}}\left\vert
\sum_{t=s}^e (U_{ii', t} - \wt{g}_{ii', t}) \right\vert \le c''\sqrt{\log(T)}  \right\}
\end{align*}
for some fixed $c', c''>0$.

We first focus on the case when $z_{j, t} = U_{i, t}^2 - \wt{g}_{i, t}$.
We investigate the probability of the following event:
\begin{align}
\label{eq:lem:two:one}
\frac{1}{\sqrt{e-s+1}} \left\vert\sum_{t=s}^e z_t\right\vert > \lambda_T
\end{align}
for $\lambda_T \asymp \sqrt{\log(T)}$.
Let $\nu = e-s+1$.

\vspace{5pt}
(a) When $\nu < \lambda_T^2/(4\bar{g}^4)$: 
Note that $|z_{j, t}| \le |U_{i, t}^2| + |\wt{g}_{i, t}| \le 2\bar{g}^2$ from \eqref{assum:a5}.
Hence $\nu^{-1/2}|\sum_{t=s}^e z_t| \le \nu^{1/2}2\bar{g}^2 < \lambda_T$.

\vspace{5pt}
(b) When $\nu > \lambda_T^2/(4\bar{g}^4)$:
Decompose $z_t$ as
\begin{align*} 
z_{j, t} = U_{i, t}^2 - \wt{g}_{i, t} = \{U_{i, t}^2 - \E(U_{i, t}^2)\} + \{\E(U_{i, t}^2) - \wt{g}_{i, t}\} 
=: \xi_{i, t} + \zeta_{i, t}.
\end{align*}
Also from~\eqref{assum:a5},
\begin{align*}
|\zeta_{i, t}| =& |\E(U_{i, t}^2) - \E(\wt{U}_{i, t}^2)| \le 
2\bar{g} \, C_g\left\{\sum_{k=0}^p\E|r_{i, t-k}^2-(r^{v(t)}_{i, t-k})^2| + 
\sum_{l=1}^q\E|h_{i, t-l}^2-(h^{v(t)}_{i, t-l})^2|\right\}
\\
\le& 2\bar{g} \, C_g\left\{\sum_{k=0}^p\E(V_{i, t-k}) + 
\sum_{l=1}^q\E(W_{i, t-l})\right\} \le C_1(p+q)\rho^{t-\underline{\eta}(t)-1}
\end{align*}
for $\rho = \rho_1 \vee \rho_2 \in (0, 1)$ and fixed $C_1>0$ thanks to Lemma~\ref{lem:one}.
Therefore
\begin{align*}
\left\vert \sum_{t=s}^e \zeta_{i, t} \right\vert \le& \sum_{t=s}^e |\zeta_{i, t}| \le \sum_{t=1}^T |\zeta_{i, t}| 
\le C_1(p+q)\sum_{t=1}^T \rho^{t-\underline{\eta}(t)-1}
= C_1(p+q)\sum_{b=0}^B\sum_{t=\eta_b+1}^{\eta_{b+1}} \rho^{t-\eta_b-1}
\\
\le& \frac{C_2(p+q)(B+1)}{1-\rho} \le C_2(p+q)B
\end{align*}
for some $C_2>0$.
Then, the probability of the event in (\ref{eq:lem:two:one}) is bounded from the above by
$\pr\{\nu^{-1/2}|\sum_{t=s}^e \xi_{i, t}| \ge \wt{\lambda}_T\}$ 
with $\wt{\lambda}_T = \lambda_T-C_2(p+q)B\nu^{-1/2}$.
From Proposition~\ref{prop:one}, $U_{i, t}$ is a bounded, 
strong mixing process with its $\alpha$-mixing coefficient $\alpha(k) \asymp \alpha^k$
for some $\alpha \in (0, 1)$.
Combining the mixing property of $\xi_{i, t}$ with the fact that $|\xi_{i, t}| \le 2\bar{g}^2$, 
we apply Theorem~1 of \cite{merlevede2009} and derive
\begin{align}
\label{eq:bosq}
\pr\left(\frac{1}{\sqrt\nu} \left\vert \sum_{t=s}^e \xi_{i, t}\right\vert \ge \wt{\lambda}_T\right) \le
\exp\left(- \frac{C_2\nu\wt\lambda_T^2}{4\bar{g}^4\nu + 2\bar{g}^2\nu^{1/2}\wt\lambda_T \cdot \log(T) \,
\log\log(T)}\right),
\end{align}
where $C_2$ depends only on $\alpha$ and hence not on $j$ ($i$), $s$ or $e$.
Since $\wt{\lambda}_T \to \infty$ as $T \to \infty$, while $\wt{\lambda}_T < \lambda_T$ and 
$\nu \ge \lambda_T^2/(4\bar{g}^4)$, 
we have the RHS of \eqref{eq:bosq} bounded by
$C_3\exp(-C_4\wt{\lambda}_T^2)$ for some fixed $C_3, C_4 > 0$.
Summarising (a)--(b) above, it can be shown that
\begin{align*}
\pr(\mc E_1) \ge 1 - C_3NT^2\exp(-C_4\wt{\lambda}_T^2) \to 1
\end{align*}
as $T \to \infty$ under \ref{eq:a1} and \ref{eq:b3}, for a sufficiently chosen $c'$.

We now turn our attention to $\pr(\mc E_2 | \mc E_1)$.
Let $z_{j, t} = U_{ii', t} - \wt{g}_{ii', t}$.
By construction,
\begin{align*}
\frac{1}{\sqrt\nu} \Big\vert\sum_{t=s}^e z_{j ,t} \Big\vert =& 
\frac{1}{\sqrt\nu} \sum_{t=s}^e \left\vert (U_{i, t} + s_{i, i'}U_{i', t})^2 - 
\E\{\wt{U}^{v(t)}_{i, t} + s_{i, i'}\wt{U}^{v(t)}_{i', t}\}^2 \right\vert
\\
\le& \frac{1}{\sqrt\nu} \Big\vert \sum_{t=s}^e (U^2_{i, t}-\wt{g}_{i, t}) \Big\vert + 
\frac{1}{\sqrt\nu} \Big\vert \sum_{t=s}^e (U_{i', t}^2-\wt{g}_{i', t}) \Big\vert
+ \frac{2}{\sqrt\nu} \Big\vert
\sum_{t=s}^e [U_{i, t}U_{i', t}-\E\{\wt{U}^{v(t)}_{i, t}\wt{U}^{v(t)}_{i', t}\}] \Big\vert.
\end{align*}
Given $\mc E_1$, we only need to investigate the boundedness of the last term.
Observe that
\begin{align*}
& U_{i, t}U_{i', t}-\E\{\wt{U}^{v(t)}_{i, t}\wt{U}^{v(t)}_{i', t}\}
\\
=& \{U_{i, t}U_{i', t}-\E(U_{i, t}U_{i', t})\} +
[\E(U_{i, t}U_{i', t})-\E\{\wt{U}^{v(t)}_{i, t}\wt{U}^{v(t)}_{i', t}\}]
=: \xi'_{ii', t} + \zeta'_{ii', t}.
\end{align*}
Since $g_3$ (defined in Appendix \ref{sec:pf:g2}) is also Lipschitz continuous in its squared arguments,
we can bound $\nu^{-1/2}|\sum_{t=s}^e \zeta'_{ii', t}|$ similarly as $\nu^{-1/2}|\sum_{t=s}^e \zeta_{i', t}|$.
As for $\nu^{-1/2}|\sum_{t=s}^e \xi'_{ii', t}|$
since $\xi'_{ii', t}$ is a zero-mean, strong mixing process from Proposition~\ref{prop:one},
its boundedness analogously follows as that of $\nu^{-1/2}|\sum_{t = s}^2 \xi_{i, t}|$,
and $\pr(\mc E_2 | \mc E_1) \to 1$, which completes the proof.


\subsection{Proof of Theorem~\ref{thm:dcbs}}
\label{sec:pf:one}

\begin{lem}
\label{lem:three}
Under the conditions of Proposition~\ref{prop:two},
\begin{align*}
\max_{1 \le j \le d} \max_{1 \le s \le c < e \le T} |\mc Z^j_{s, c, e}| = O_p(\sqrt{\log(T)}),
\end{align*}
where
\begin{align*}
\mc Z^j_{s, c, e}
= \sqrt{\frac{e-c}{(e-s+1)(c-s+1)}} \sum_{t=s}^c z_{j, t} -
\sqrt{\frac{c-s+1}{(e-s+1)(e-c)}} \sum_{t=c+1}^e z_{j, t}.
\end{align*}
\end{lem}
\begin{proof}
Note that
\begin{align*}
|\mc Z^j_{s, c, e}|
\le \sqrt{\frac{e-c}{(e-s+1)(c-s+1)}}\left\vert\sum_{t=s}^c z_{j, t} \right\vert +
\sqrt{\frac{c-s+1}{(e-s+1)(e-c)}}\left\vert\sum_{t=c+1}^e z_{j, t} \right\vert =: V+VI.
\end{align*}
From Lemma 1,
\begin{align*}
V = \sqrt{\frac{e-c}{e-s+1}} \cdot \frac{1}{\sqrt{c-s+1}}\left\vert \sum_{t=s}^c z_{j, t} \right\vert
= O_p(\sqrt{\log(T)}),
\end{align*}
and similarly $VI = O_p(\sqrt{\log(T)}),$, 
both of which hold uniformly in $j$ and $(s, c, e)$. 
Therefore, for some fixed $c > 0$,
\begin{align*}
\max_{1 \le j \le N}\max_{1 \le s \le c < e \le T} |\mc Z^j_{s, c, e}|
\le \max_{1 \le s \le c < e \le T} \left\{\sqrt{\frac{c-s+1}{e-s+1}}+\sqrt{\frac{e-c}{e-s+1}}\right\}
\cdot c\sqrt{\log(T)}
\le c\sqrt{2\log(T)}
\end{align*}
with probability tending to one as $T \to \infty$.
\end{proof}

\vspace{5pt}
By the definition of $g_0$ and the discussion in Appendix~\ref{sec:pf:g2},
there exists a fixed constant $\bar{f} > 0$ such that
$\max_{1 \le j \le d} \max_{1 \le t \le T} |f_{j, t}| \le \bar{f}$.
Proposition~\ref{prop:two} and Lemma~\ref{lem:three} place a logarithmic bound on
the partial sums and the cumulative sums of $z_{j, t}$.
With these results in place of Lemmas~1--2 of \cite{cho2016},
we are able to apply the arguments identical to those employed there to prove 
Theorem~\ref{thm:dcbs} under \ref{eq:b2}--\ref{eq:b4}.

\subsection{Properties of $g_2$}
\label{sec:pf:g2}

We show that from~\eqref{assum:a5}, 
$g_2$ is bounded and Lipschitz continuous in its squared arguments.
Firstly, note that
\begin{align}
g_2(\mathbf{z}, \mathbf{z}') =& g_1(\mathbf{z})+g_1(\mathbf{z}') \pm 2 g_0(\mathbf{z})g_0(\mathbf{z}'),
\label{eq:gtwo:new}
\end{align}
and hence $|g_2(\mathbf{z}, \mathbf{z}')| \le |g_1(\mathbf{z})| + |g_1(\mathbf{z}')| + 2|g_0(\mathbf{z})g_0(\mathbf{z}')| \le 4\bar{g}$
for any $\mathbf{z}, \mathbf{z}' \in \R^{p+q+1}$.
From \eqref{eq:gtwo:new}, we only need to establish the Lipschitz continuity of
$g_3(\mathbf{z}, \mathbf{z}') = g_0(\mathbf{z})g_0(\mathbf{z}')$.
Further introducing the notations $\mathbf{w}, \mathbf{w}' \in \R^{p+q+1}$,
\begin{align*}
& |g_3(\mathbf{z}, \mathbf{w}) - g_3(\mathbf{z}', \mathbf{w}')| =
|g_0(\mathbf{z})g_0(\mathbf{w}) - g_0(\mathbf{z}')g_0(\mathbf{w}')|
\\
\le& |g_0(\mathbf{z})\{g_0(\mathbf{w}) - g_0(\mathbf{w}')\}|
+ |g_0(\mathbf{w}')\{g_0(\mathbf{z}) - g_0(\mathbf{z}')\}|
\le \bar{g}\{|g_0(\mathbf{w}) - g_0(\mathbf{w}')| + |g_0(\mathbf{z}) - g_0(\mathbf{z}')|\},
\end{align*}
and thus follows the Lipschitz continuity of $g_2$ in its squared arguments.

\clearpage

\section{Additional simulation results}
\label{sec:sim:add}

\subsection{Power of the test}

In this section, we provide additional simulation results
complementing the simulation studies conducted in Section~\ref{sec:sim} of the main text. 

\begin{enumerate}[label = (M\arabic*), start = 4]
\setlength\itemsep{0em} 
\item \label{m:1} \textbf{tv-MGARCH ($1$, $1$) processes with a single change point.} 
Change points are introduced to GARCH parameters in~\eqref{eq:garch}
at $\eta_1 \in \{[T/2], [9T/10]\}$, where $T = 1000$ and $N \in \{50, 100\}$. 
For a randomly chosen $\mathcal{S}_1 \subset \{1, \ldots, N\}$,
GARCH parameters $\omega_i(t), \alpha_{i, 1}(t)$ and $\beta_{i, 1}(t)$ 
for $i \in \mathcal{S}_1$ change at $t = \eta_1$ as
$\omega_i(t) = \omega^{(1)}\mathbb{I}(t \le \eta_1) + \omega^{(2)}\mathbb{I}(t > \eta_1) + \delta_{\omega, i}$, 
$\alpha_{i, 1}(t) = \alpha^{(1)}_1\mathbb{I}(t \le \eta_1) + \alpha^{(2)}_1\mathbb{I}(t > \eta_1) + \delta_{\alpha, i}$ 
and $\beta_{i, 1}(t) = \beta^{(1)}_1\mathbb{I}(t \le \eta_1) + \beta^{(2)}_1\mathbb{I}(t > \eta_1)  + \delta_{\beta, i}$, 
where $|\mathcal{S}_1| = [\varrho N]$ with
$\varrho \in \{1, 0.75, 0.5, 0.25\}$ controlling the `sparsity' of
the change point. 
We have $\delta_{\cdot, i} \sim_{\iid} \mc U(-\Delta, \Delta)$ is as in (M0), 
and $\bvep_t \sim_{\iid} \mc N(\mbf 0, \bm\Sigma_\vep(t))$ with 
$\bm\Sigma_\vep(t) = \bm\Sigma_\vep$ defined in (M0).
\begin{enumerate}[label = (M4.\arabic*), start = 1]
\setlength\itemsep{0em} 
\item \label{m:1:1} $(\omega, \alpha_1, \beta_1):$ $(0.4, 0.1, 0.5)$ $\to$ $(0.4, 0.1, 0.6)$. 
\item \label{m:1:2} $(\omega, \alpha_1, \beta_1):$ $(0.4, 0.1, 0.5)$ $\to$ $(0.4, 0.1, 0.8)$.
\item \label{m:1:3} $(\omega, \alpha_1, \beta_1):$ $(0.1, 0.1, 0.8)$ $\to$ $(0.1, 0.1, 0.7)$. 
\item \label{m:1:4} $(\omega, \alpha_1,\beta_1):$ $(0.1, 0.1, 0.8)$ $\to$ $(0.1, 0.1, 0.4)$.
\item \label{m:1:5} $(\omega, \alpha_1, \beta_1):$ $(0.4, 0.1, 0.5)$ $\to$ $(0.5, 0.1, 0.5)$. 
\item \label{m:1:6} $(\omega, \alpha_1, \beta_1):$ $(0.4, 0.1, 0.5)$ $\to$ $(0.8, 0.1, 0.5)$.
\item \label{m:1:7} $(\omega, \alpha_1, \beta_1):$ $(0.1, 0.1, 0.8)$ $\to$ $(0.3, 0.1, 0.8)$. 
\item \label{m:1:8} $(\omega, \alpha_1, \beta_1):$ $(0.1, 0.1, 0.8)$ $\to$ $(0.5, 0.1, 0.8)$.
\end{enumerate}
\end{enumerate}

Tables~\ref{table:sim:one:50}--\ref{table:sim:one:100:skew}
present the results on the power of the test 
from conducting a single iteration of the DCBS algorithm.
As expected, the test achieves higher power when the change point
is cross-sectionally dense (with larger $\varrho$) and located centrally, 
and the same applies to the localisation accuracy of the change point estimator.
For most GARCH parameter configurations, the test attains power above $0.9$ even when the
change point is relatively sparse ($\varrho = 0.25$)
except for \ref{m:1:2}, \ref{m:1:4} and \ref{m:1:8}.
Even when the location of the change point is skewed ($\eta_1 = [9T/10]$), 
our method generally attains high power and localisation accuracy if the change point is not too sparse, 
in most settings where it shows good performance
when the change point is centrally located ($\eta_1 = [T/2]$).

\begin{table}[htbp]
\caption{\ref{m:1} Power of the change point test at $\alpha = 0.05$
and the accuracy of $\heta_1$ (\% of $|\heta_1 - \eta_1| < \log^2T$) 
when $N = 50$, $T = 1000$ and $\eta_1 = [T/2]$.}
\label{table:sim:one:50} \centering
\small{\begin{tabular}{c|cc|cc|cc|cc} 
\hline
&   \multicolumn{2}{c}{\ref{m:1:1}} &        \multicolumn{2}{c}{\ref{m:1:2}} &        \multicolumn{2}{c}{\ref{m:1:3}} &        \multicolumn{2}{c}{\ref{m:1:4}}      \\
$\varrho$ & power & accuracy (\%) & power & accuracy (\%) & power
& accuracy (\%) & power & accuracy (\%)   \\  
\hline\hline
1 & 1.00 &  100 &   1.00 &  100 &   1.00 &  100 &   0.96 &  84  \\
0.75 &  1.00 &  100 &   1.00 &  100 &   1.00 &  100 &   0.91 &  79  \\
0.5 &   1.00 &  100 &   1.00 &  99 &    1.00 &  100 &   0.57 &  47  \\
0.25 &  0.96 &  91 &    0.83 &  73 &    0.99 &  97 &    0.12 &  7
\\  \hline
&   \multicolumn{2}{c}{\ref{m:1:5}} &        \multicolumn{2}{c}{\ref{m:1:6}} &        \multicolumn{2}{c}{\ref{m:1:7}} &        \multicolumn{2}{c}{\ref{m:1:8}}      \\
$\varrho$ & power & accuracy (\%) & power & accuracy (\%) & power
& accuracy (\%) & power & accuracy (\%)   \\  
\hline\hline
1 & 1.00 &  100 &   1.00 &  100 &   1.00 &  100 &   1.00 &  98  \\
0.75 &  1.00 &  100 &   1.00 &  100 &   1.00 &  100 &   1.00 &  98  \\
0.5 &   1.00 &  100 &   1.00 &  100 &   1.00 &  100 &   0.98 &  93  \\
0.25 &  0.99 &  93 &    1.00 &  98 &    1.00 &  99 &    0.54 &  43
\\  \hline
\end{tabular}}
\end{table}

\begin{table}[htbp]
\caption{\ref{m:1} Power of the change point test at $\alpha = 0.05$
and the accuracy of $\heta_1$ (\% of $|\heta_1 - \eta_1| <
\log^2T$) when $N = 50$, $T = 1000$ and $\eta_1 = [0.9T]$.}
\label{table:sim:one:50:skew} \centering
\small{\begin{tabular}{c|cc|cc|cc|cc} 
\hline
&   \multicolumn{2}{c}{\ref{m:1:1}} &        \multicolumn{2}{c}{\ref{m:1:2}} &        \multicolumn{2}{c}{\ref{m:1:3}} &        \multicolumn{2}{c}{\ref{m:1:4}}      \\
$\varrho$ & power & accuracy (\%) & power & accuracy (\%) & power
& accuracy (\%) & power & accuracy (\%)   \\ \hline\hline
1 & 1.00 &  100 &   1.00 &  99 &    1.00 &  100 &   0.43 &  37  \\
0.75 &  1.00 &  100 &   1.00 &  99 &    1.00 &  100 &   0.27 &  20  \\
0.5 &   1.00 &  100 &   0.89 &  84 &    1.00 &  100 &   0.09 &  6   \\
0.25 &  0.92 &  90 &    0.32 &  28 &    0.96 &  95 &    0.08 &  2
\\  \hline
&   \multicolumn{2}{c}{\ref{m:1:5}} &        \multicolumn{2}{c}{\ref{m:1:6}} &        \multicolumn{2}{c}{\ref{m:1:7}} &        \multicolumn{2}{c}{\ref{m:1:8}}      \\
$\varrho$ & power & accuracy (\%) & power & accuracy (\%) & power
& accuracy (\%) & power & accuracy (\%)   \\ 
\hline\hline
1 & 1.00 &  100 &   1.00 &  100 &   1.00 &  100 &   0.96 &  84  \\
0.75 &  1.00 &  100 &   0.99 &  97 &    1.00 &  100 &   0.89 &  79  \\
0.5 &   1.00 &  100 &   0.87 &  86 &    1.00 &  100 &   0.61 &  50  \\
0.25 &  0.63 &  62 &    0.10 &  8 & 1.00 &  100 &   0.16 &  10  \\
\hline
\end{tabular}}
\end{table}

\begin{table}[htbp]
\caption{\ref{m:1} Power of the change point test at $\alpha = 0.05$
and the accuracy of $\heta_1$ when $N = 100$, $T = 1000$ and
$\eta_1 = [T/2]$.} \label{table:sim:one:100} \centering
\small{\begin{tabular}{c|cc|cc|cc|cc}
\hline
&   \multicolumn{2}{c}{\ref{m:1:1}} &        \multicolumn{2}{c}{\ref{m:1:2}} &        \multicolumn{2}{c}{\ref{m:1:3}} &        \multicolumn{2}{c}{\ref{m:1:4}}      \\
$\varrho$ & power & accuracy (\%) & power & accuracy (\%) & power
& accuracy (\%) & power & accuracy (\%)   \\  
\hline\hline
1 & 1.00 &  100 &   1.00 &  100 &   1.00 &  100 &   1.00 &  96  \\
0.75 &  1.00 &  100 &   1.00 &  100 &   1.00 &  100 &   1.00 &  93  \\
0.5 &   1.00 &  100 &   1.00 &  100 &   1.00 &  100 &   0.94 &  83  \\
0.25 &  1.00 &  98 &    1.00 &  90 &    1.00 &  100 &   0.46 &  36
\\  
\hline
&   \multicolumn{2}{c}{\ref{m:1:5}} &        \multicolumn{2}{c}{\ref{m:1:6}} &        \multicolumn{2}{c}{\ref{m:1:6}} &        \multicolumn{2}{c}{\ref{m:1:8}}      \\
$\varrho$ & power & accuracy (\%) & power & accuracy (\%) & power
& accuracy (\%) & power & accuracy (\%)   \\  
\hline\hline
1 & 1.00 &  100 &   1.00 &  100 &   1.00 &  100 &   1.00 &  100 \\
0.75 &  1.00 &  100 &   1.00 &  100 &   1.00 &  100 &   1.00 &  100 \\
0.5 &   1.00 &  100 &   1.00 &  100 &   1.00 &  100 &   1.00 &  99  \\
0.25 &  1.00 &  100 &   1.00 &  100 &   1.00 &  100 &   0.98 &  93
\\  \hline
\end{tabular}}
\end{table}

\begin{table}[htbp]
\caption{\ref{m:1} Power of the change point test at $\alpha = 0.05$
and the accuracy of $\heta_1$ when $N = 100$, $T = 1000$ and
$\eta_1 = [9T/10]$.} \label{table:sim:one:100:skew} \centering
\small{\begin{tabular}{c|cc|cc|cc|cc} 
\hline
&   \multicolumn{2}{c}{\ref{m:1:1}} &        \multicolumn{2}{c}{\ref{m:1:2}} &        \multicolumn{2}{c}{\ref{m:1:3}} &        \multicolumn{2}{c}{\ref{m:1:4}}      \\
$\varrho$ & power & accuracy (\%) & power & accuracy (\%) & power
& accuracy (\%) & power & accuracy (\%)   \\  
\hline\hline
1 & 1.00 &  100 &   1.00 &  100 &   1.00 &  100 &   0.87 &  78  \\
0.75 &  1.00 &  100 &   1.00 &  100 &   1.00 &  100 &   0.65 &  57  \\
0.5 &   1.00 &  100 &   1.00 &  100 &   1.00 &  100 &   0.18 &  15  \\
0.25 &  1.00 &  100 &   0.81 &  76 &    1.00 &  100 &   0.07 &  1
\\  \hline
&   \multicolumn{2}{c}{\ref{m:1:5}} &        \multicolumn{2}{c}{\ref{m:1:6}} &        \multicolumn{2}{c}{\ref{m:1:7}} &        \multicolumn{2}{c}{\ref{m:1:8}}      \\
$\varrho$ & power & accuracy (\%) & power & accuracy (\%) & power
& accuracy (\%) & power & accuracy (\%)   \\  
\hline\hline
1 & 1.00 &  100 &   1.00 &  100 &   1.00 &  100 &   1.00 &  98  \\
0.75 &  1.00 &  100 &   1.00 &  100 &   1.00 &  100 &   1.00 &  97  \\
0.5 &   1.00 &  100 &   1.00 &  100 &   1.00 &  100 &   0.93 &  84  \\
0.25 &  1.00 &  100 &   0.52 &  52 &    1.00 &  100 &   0.31 &  26
\\  \hline
\end{tabular}}
\end{table}

\subsection{Localisation accuracy}

\begin{figure}[htbp]
\centering
\includegraphics[width = 1\textwidth]{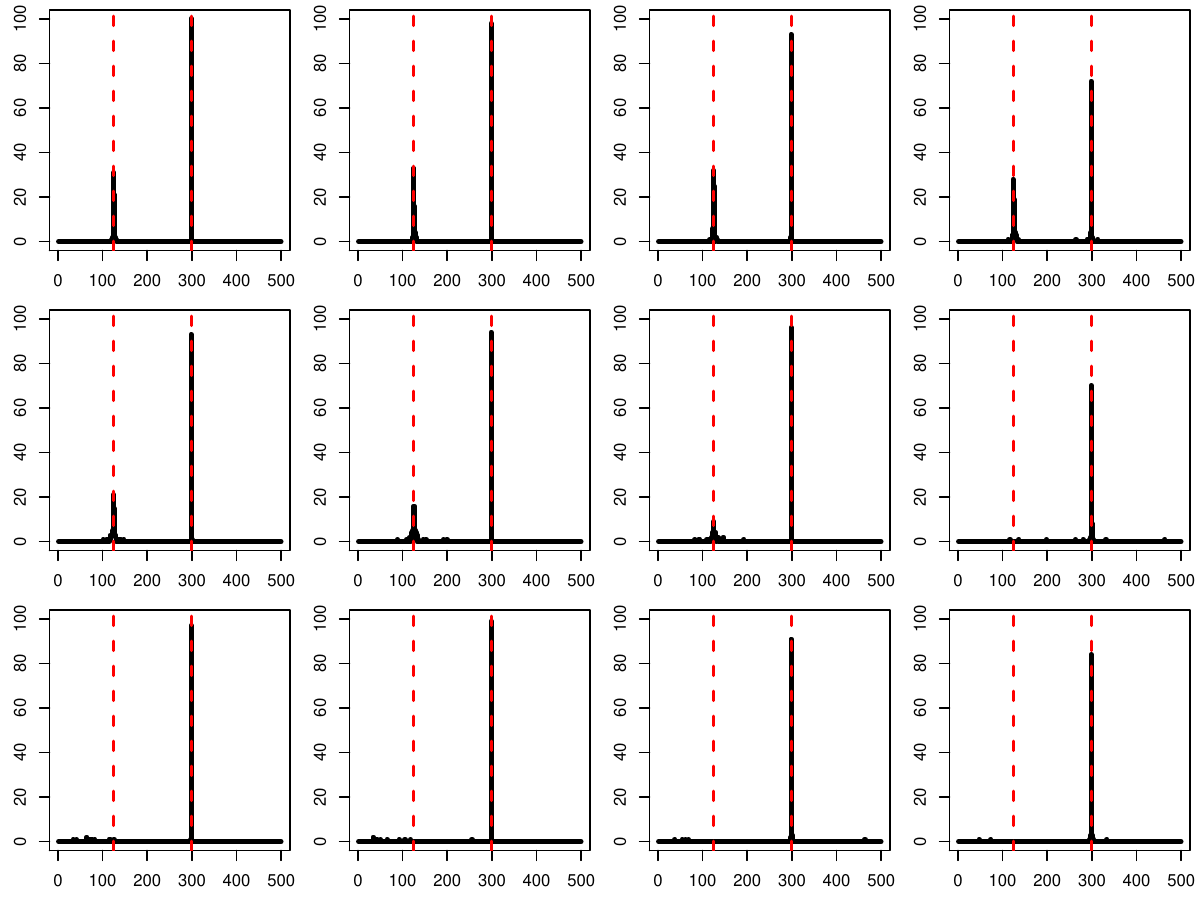}
\caption{\ref{m:2} Summary of estimated change points over $100$ realisations when $N = 50$,
for \ref{m:2:1}--\ref{m:2:3} (top to bottom) with $\varrho \in \{1, 0.75, 0.5, 0.25\}$ (left to right)
and Gaussian innovations;
$\eta_1$ and $\eta_2$ are indicated by vertical broken lines.}
\label{fig:sim:two:50}
\end{figure}

\begin{figure}[htbp]
\centering
\includegraphics[width = 1\textwidth]{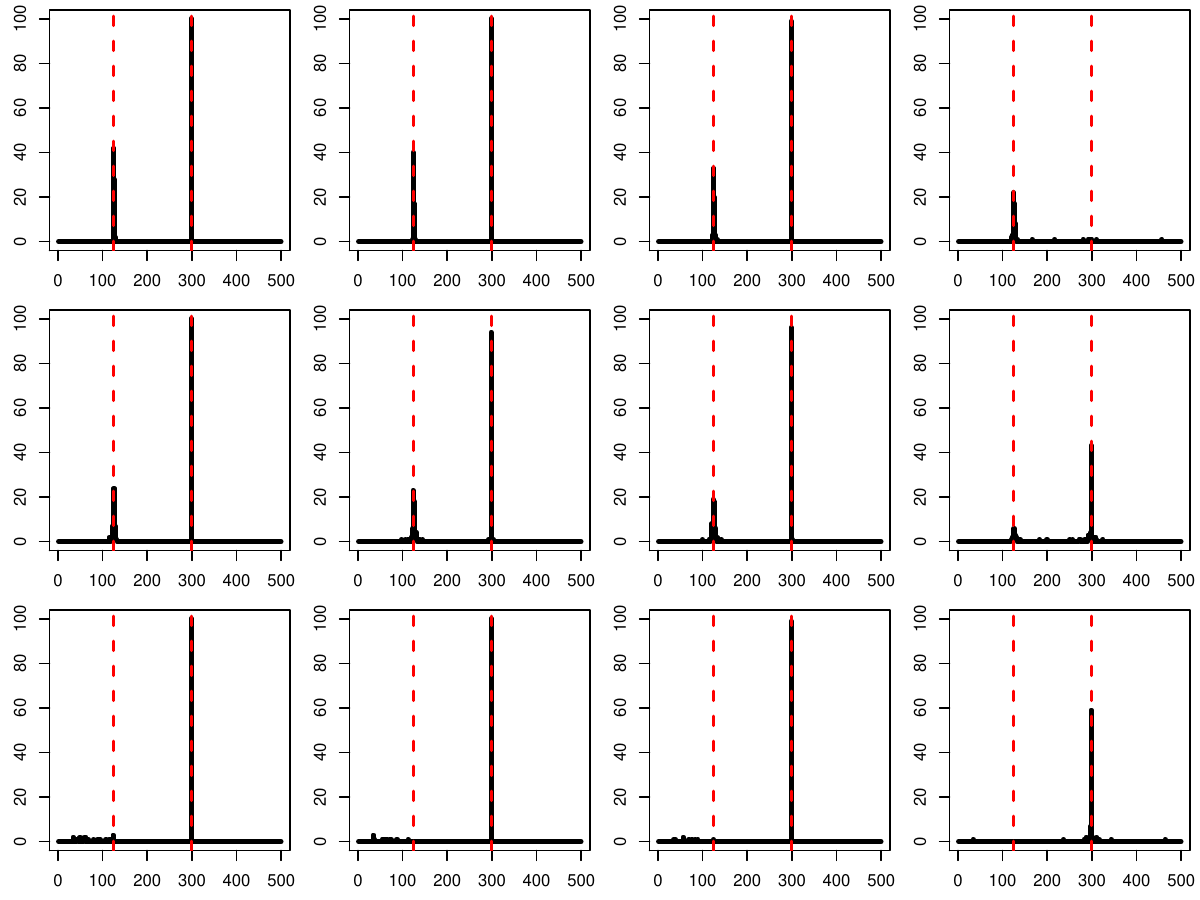}
\caption{\ref{m:2} Summary of estimated change points over $100$
realisations when $N = 100$.}
 \label{fig:sim:two:100}
\end{figure}

\begin{figure}[htbp]
\centering
\includegraphics[width = 1\textwidth]{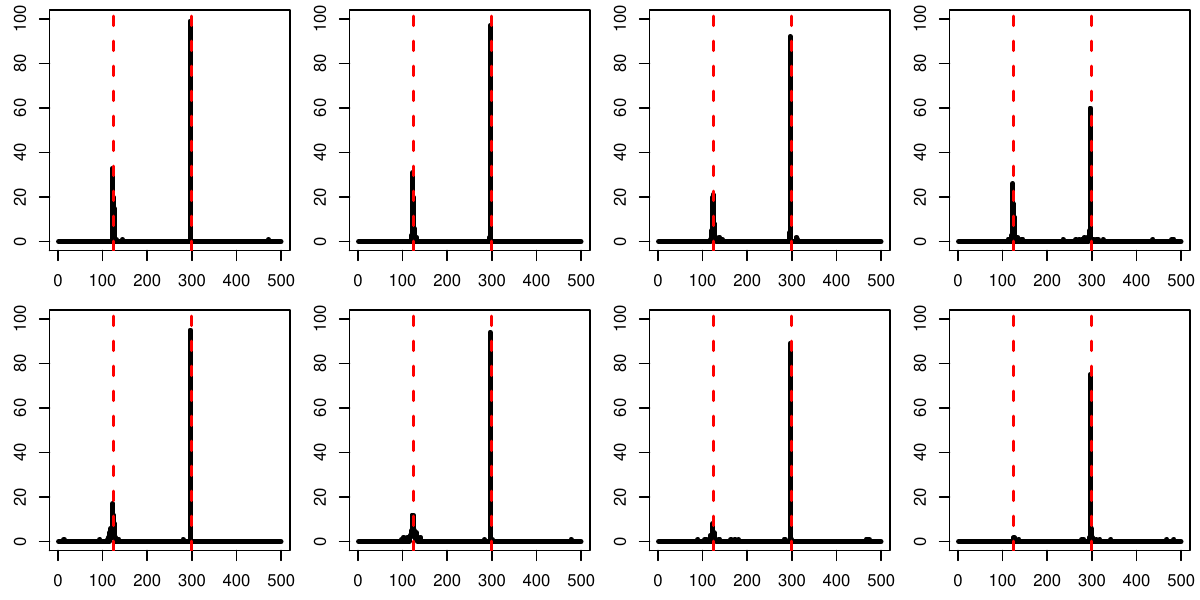}
\caption{\ref{m:3:1} Summary of estimated change points over $100$ realisations when $N = 50$,
for (M2.1.1)--(M2.1.2) (top to bottom) with $\varrho \in \{1, 0.75, 0.5, 0.25\}$ (left to right).}
\label{fig:sim:three:one:50}
\end{figure}

\begin{figure}[htbp]
\centering
\includegraphics[width = 1\textwidth]{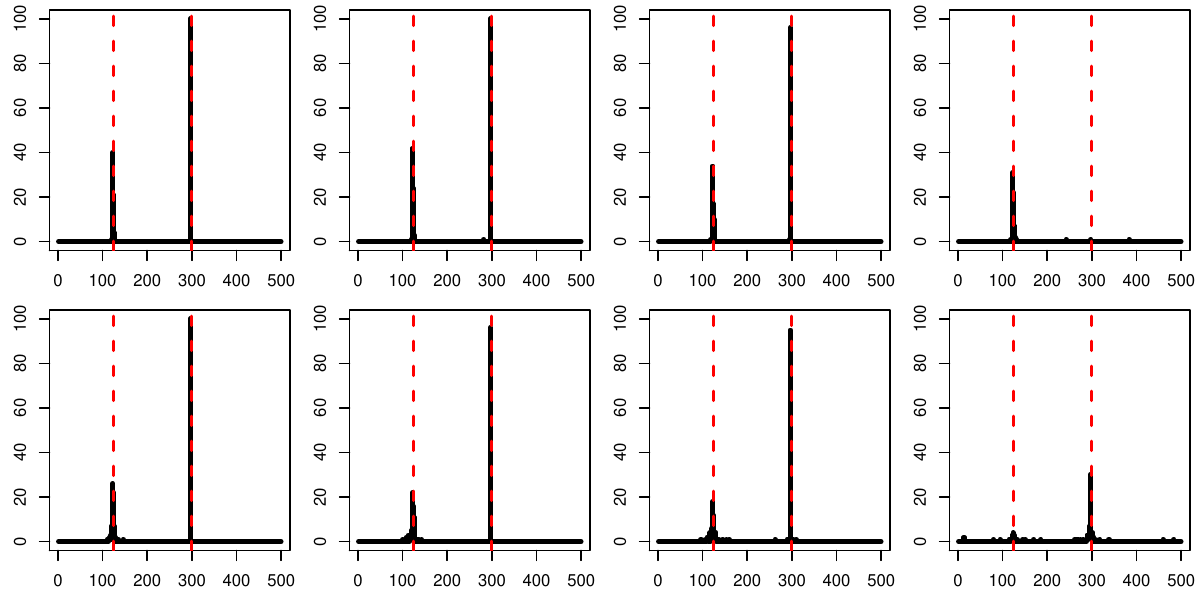}
\caption{\ref{m:3:1} Summary of estimated change points over $100$ realisations when $N = 100$.}
\label{fig:sim:three:one:100}
\end{figure}

\begin{figure}[htbp]
\centering
\includegraphics[width = 1\textwidth]{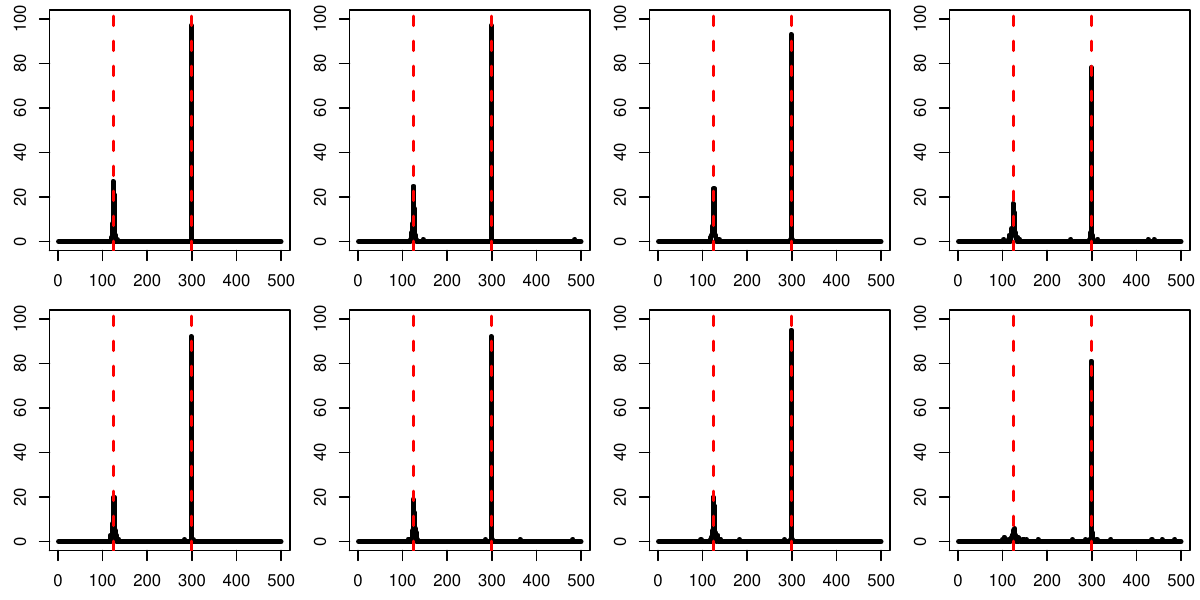}
\caption{\ref{m:3:2} Summary of estimated change points over $100$ realisations when $N = 50$,
for \ref{m:3:2:1}--\ref{m:3:2:2} (top to bottom) with $\varrho \in \{1, 0.75, 0.5, 0.25\}$ (left to right).}
\label{fig:sim:three:two:50}
\end{figure}

\begin{figure}[htbp]
\centering
\includegraphics[width = 1\textwidth]{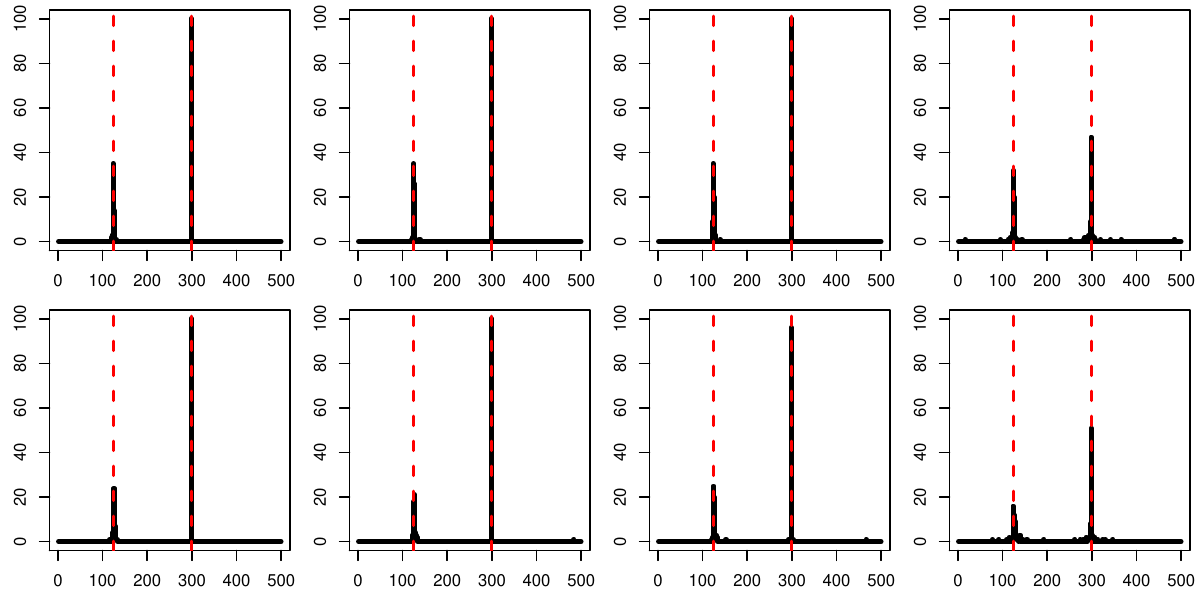}
\caption{\ref{m:3:2} Summary of estimated change points over $100$
realisations when $N = 100$.} 
\label{fig:sim:three:two:100}
\end{figure}

\begin{figure}[htbp]
\centering
\includegraphics[width = 1\textwidth]{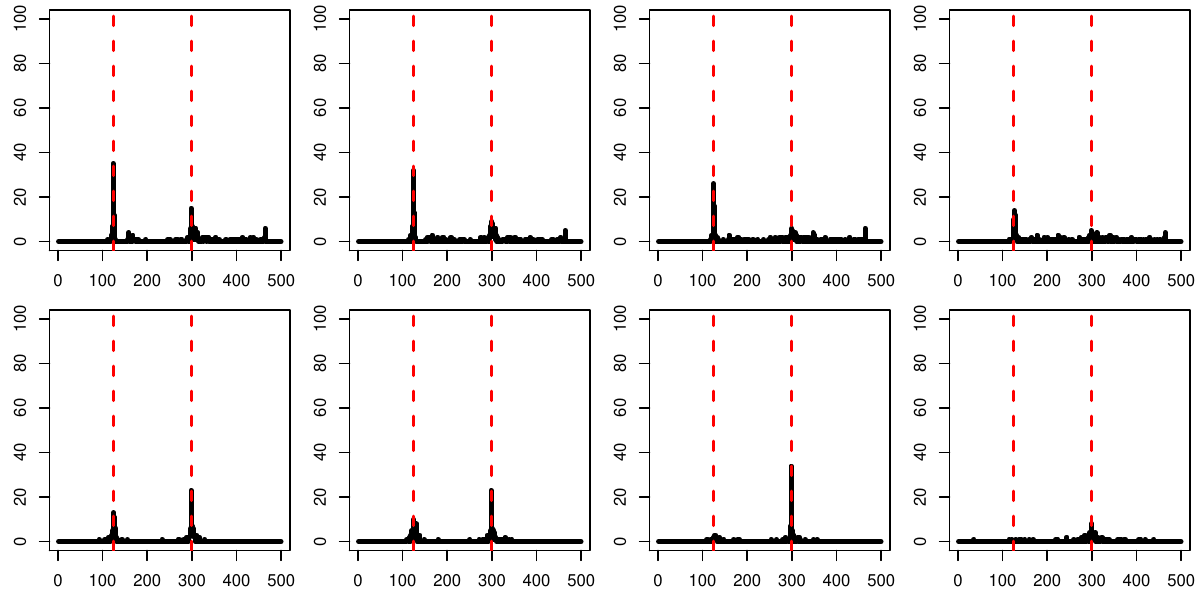}
\caption{\ref{m:4} Summary of estimated change points over $100$ realisations when $N = 50$,
for \ref{m:4:1}--\ref{m:4:2} (top to bottom) with $\varrho \in \{1, 0.75, 0.5, 0.25\}$ (left to right).}
\label{fig:sim:four:50}
\end{figure}

\begin{figure}[htbp]
\centering
\includegraphics[width = 1\textwidth]{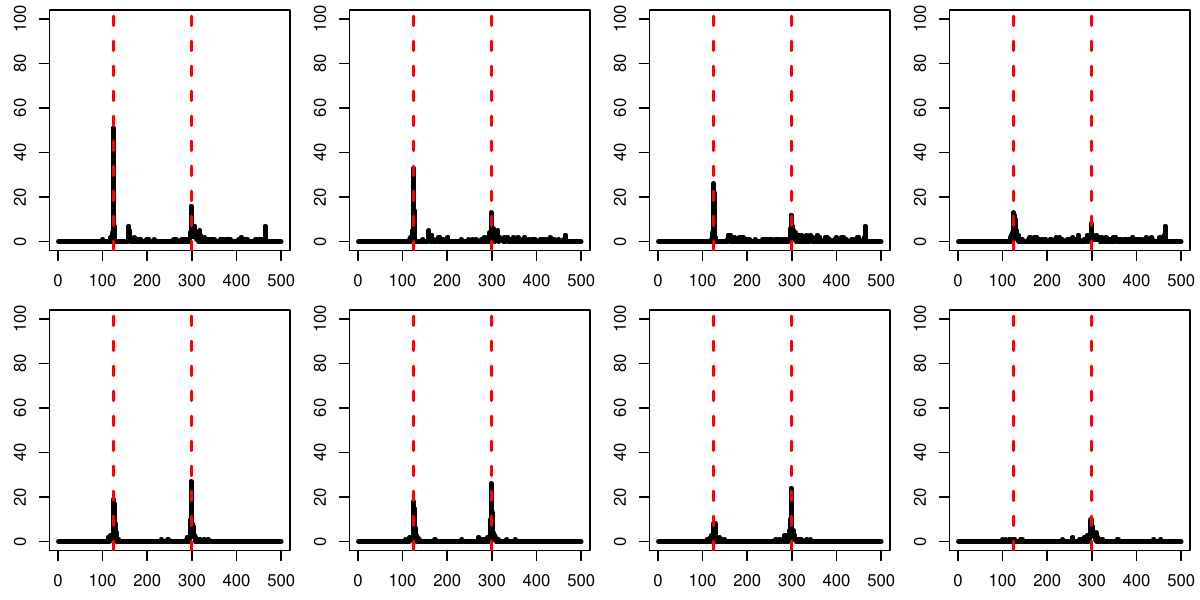}
\caption{\ref{m:4} Summary of estimated change points over $100$
realisations when $N = 100$.} \label{fig:sim:four:100}
\end{figure}

Figures~\ref{fig:sim:two:50}--\ref{fig:sim:four:100}
illustrate the locations of estimated change points
from the simulation studies conducted in Section~\ref{sec:sim} of the main text.

\end{document}